%% file: VIMOS_paper1_final_MNRAS.tex
\documentclass[usenatbib,usegraphicx,useAMS]{mn2e}
 \usepackage[T1]{fontenc}
  \usepackage{multirow}
 \usepackage{color}
 \usepackage{times}
 \newcommand{\Fe}{\textsc{feii}}
 \newcommand{\Ox}{\textsc{[oiii]}}
 \newcommand{\Ni}{\textsc{[nii]}}
 \newcommand{\Su}{\textsc{[sii]}}
 \newcommand{\Hb}{\mbox{H$\beta$}}
 \newcommand{\Ha}{\mbox{H$\alpha$}}
 \newcommand{\HII}{\textsc{hii}}
 \newcommand{\QDeb}{\textsc{qdeblend${}^{\mathrm{3D}}$}}

\begin{document}

\title[Integral field spectroscopy of nearby QSOs I.]{Integral field spectroscopy of nearby QSOs:\\ I. ENLR size\,--\,luminosity relation, 
ongoing star formation \& resolved gas-phase metallicities \thanks{Based on observations made with VIMOS integral field spectrograph mounted 
to the Melipal VLT telescope at ESO-Paranal Observatory (programs 072B-0550 and 083B-0801; PI: K. Jahnke)}}

\author[Husemann et al.]{B.~Husemann$^{1,2}$\thanks{ESO fellow, bhuseman@eso.org}, K.~Jahnke,$^3$, S.~F.~S\'anchez$^{4,5,6}$, L.~Wisotzki$^{2}$, \newauthor
D.~Nugroho$^3$, D.~Kupko$^2$,  M.~Schramm$^7$\\
$^1$European Southern Observatory, Karl-Schwarzschild-Str. 2, 85748 Garching b. M\"unchen, Germany\\
$^2$Leibniz-Institut f\"ur Astrophysik Potsdam, An der Sternwarte 16, 14482 Potsdam, Germany\\
$^3$Max-Planck-Institut f\"ur Astronomie, K\"onigstuhl 17, D-69117 Heidelberg, Germany\\
$^4$Instituto de Astronom\'\i a,Universidad Nacional Auton\'oma de Mexico, A.P. 70-264, 04510, M\'exico,D.F.\\
$^5$Instituto de Astrof\'isica de Andaluc\'ia (IAA/CSIC), Glorieta de la Astronom\'{\i}a s/n Aptdo. 3004, E-18080 Granada, Spain\\
$^6$Centro Astron\'omico Hispano-Alem\'an, Calar Alto (CSIC-MPG), C/Jes\'us Durb\'an Rem\'on 2-2, E-04004 Almeria, Spain\\
$^7$Kavli Institute for the Physics and Mathematics of the Universe (WPI), Todai Institutes for Advanced Study, the University of Tokyo, Kashiwa, Japan 277-8583
}
\maketitle
\begin{abstract}
We present optical integral field spectroscopy for a flux-limited sample of 19 quasi-stellar objects (QSOs) at low redshift ($z<0.2$) and 
spatially resolve their ionized gas properties at a physical resolution of 2--5\,kpc. Extended ionized gas exists in all QSO host galaxies irrespective of their morphological 
types. The extended narrow-line regions (ENLRs), photoionized by the radiation of active galactic nuclei (AGN), have sizes of up to several kpc and correlate more strongly with the 
QSO continuum luminosity at 5100\AA\ than with the integrated \Ox\ luminosity.  We find a relation of the form 
$\log r \propto (0.46\pm0.04)\log L_{5100}$, reinforcing the picture of an approximately constant ionization parameter for the ionized clouds 
across the ENLR. Besides the ENLR, we also find gas ionized by young massive stars in more than 50 per cent of the galaxies on kpc scales.
In more than half of the sample, the specific star formation rates based on the extinction-corrected H$\alpha$ luminosity
are consistent with those of inactive disc-dominated galaxies, even for some bulge-dominated QSO hosts.  
Enhanced star formation rates of up to $\sim$70\,$\mathrm{M}_{\sun}\,\mathrm{yr}^{-1}$ are rare and always associated with signatures of major mergers. 
Comparison with the star formation rate based on the 60$\mu$m+100$\mu$m FIR luminosity suggests that the FIR luminosity is systematically contaminated by AGN emission and H$\alpha$ appears 
to be a more robust and sensitive tracer for the star formation rate.
Evidence for efficient AGN feedback is scarce in our sample, but some of our QSO hosts lack signatures of ongoing star formation leading to a reduced specific star formation rate with respect to the 
main sequence of galaxies. Whether this is causally linked to the AGN or simply caused by gas depletion remains an open question. 
Based on 12 QSOs where we can make measurements, we find that on average bulge-dominated QSO host galaxies tend to fall below the mass-metallicity relation compared to their disc-dominated 
counterparts.  While not yet statistically significant for our small sample, this may provide a useful diagnostic for future large surveys if this metal dilution can be shown to be linked to recent 
or ongoing galaxy interactions.
\end{abstract}

\section{Introduction}
Active galactic nuclei (AGN) are powered by gas accretion on to super-massive black holes at the centre of galaxies and have been thought to 
significantly affect the evolution of their host galaxies through quenching of star formation in the most massive systems.  Very luminous AGN, 
recognized as quasi-stellar objects (QSOs), were often found in major mergers, which have been thought to be the main triggering mechanism for 
a QSO phase. This has led to an evolutionary scenario for the formation of bulge-dominated galaxies from gas-rich major merger with enhanced 
star formation that is followed by a QSO phase to quench star formation \citep[e.g.][]{Sanders:1988a,Sanders:1988b,Hutchings:1992,Canalizo:2001,
Hopkins:2006a} and resulting in the passive spheroidal galaxies we observe at present. However, the AGN-merger connection is strongly 
debated and an unsolved issue. On one hand, several studies found an increased fraction of AGN in galaxies with close companions, ongoing mergers or
in post-merger systems \citep{Koss:2010,RamosAlmeida:2010,Ellison:2011,Bessiere:2012,Cotini:2013,Sabater:2013}. On the other hand, many studies claimed to find 
no significant excess of mergers in AGN hosts \citep{Dunlop:2003,Sanchez:2004b,Grogin:2005,Li:2008b,Gabor:2009,Tal:2009,
Cisternas:2011,Kocevski:2012,Schawinski:2012, Boehm:2013}. These studies report that a significant fraction of AGN appear to reside in isolated 
disc-dominated galaxies for which internal processes are likely responsible for fuelling their active nuclei. Both findings could be in agreement when the AGN-merger connection is a 
function of AGN luminosity as reported by \citet{Treister:2012}, considering the large
number of disturbed host galaxies for the most luminous QSOs in the local Universe \citep[e.g.][]{Veilleux:2009}. On the other hand, the morphological analysis of  
X-ray selected AGN host galaxies at $0.5<z<0.8$ indicate no significant luminosity dependence of the AGN-merger connection \citep{Villforth:2014}. 
A potential problem in the interpretation of the current data is that the faint signatures of recent interactions can easily be missed when the image depth is too shallow 
\citep[e.g.][]{Bennert:2008} or that there could be a significant time delay between the onset of AGN activity and the merger event \citep[e.g.][]{Wild:2010}.

The exceptionally blue colours of bulge-dominated QSO host galaxies compared to their passive inactive counterparts 
\citep[e.g.][]{Kauffmann:2003,Jahnke:2004c,Jahnke:2004,Sanchez:2004b,Zakamska:2006,Schramm:2008} imply the presence of a significant population 
of young stars. These QSO hosts were therefore thought to be in a transition phase during which continued star formation is suppressed as a 
consequence of AGN feedback supporting the merger-induced evolution. The presence of an intermediate-age stellar population (1-2Gyr) has 
been subsequently confirmed with deep long-slit spectroscopy \citep{Canalizo:2001,Jahnke:2007,Wold:2010, Canalizo:2013}. However, it is 
observationally difficult to reliably quantify the amount of ongoing star formation in these luminous QSOs and results based on [\textsc{oii}] 
line strength from unresolved spectroscopy \citep{Ho:2005,Silverman:2009} and infrared diagnostics \citep[e.g.][]{Schweitzer:2006,Lacy:2007,
Santini:2012,Rosario:2012,Urrutia:2012} are strongly debated. Those studies lead to inconsistent results on the enhancement or suppression of 
ongoing star formation in AGN hosts.  

One particular problem in estimating the ongoing star formation from optical line luminosities is the coupling between QSO radiation and the 
interstellar gas of the host galaxy. It is well established that AGN can ionize the gas out to several kpc, the so-called extended narrow-line 
region \citep[(ENLR),][]{Unger:1987}, based on narrow-band imaging of the \Ox\ $\lambda5007$ emission line \citep[e.g.][]{McCarthy:1987,
Stockton:1987,Bennert:2002,Schmitt:2003b,Villar-Martin:2010}. Only spatially resolved quantitative spectroscopy is able to separate the relative 
contribution of \HII\ regions powered by young massive stars and AGN photoionization across the host galaxies.  A severe problem for studying 
luminous type 1 (unobscured) QSO from the ground is the seeing that smears out the light of the bright nucleus. Thus, long-slit and integral 
field spectroscopic studies often focused on type 2 (obscured) QSOs to study the ENLR \citep[e.g.][]{Humphrey:2010,Greene:2011,Villar-Martin:2011a,
Liu:2013}. The light of the nucleus is blocked by obscuring material along our line of sight \citep{Antonucci:1993} which minimizes any host 
galaxy contamination for type 2 QSOs. However, the intrinsic properties of the AGN, like the accretion rate, black hole mass and AGN luminosity 
can only be indirectly estimated for those QSOs and are subject to systematic uncertainties.

In this article, we present rest-frame optical integral field spectroscopy observations of a flux-limited sample of 19 nearby ($z<0.2$) 
type 1 QSOs, which correspond to the most luminous AGN at their respective redshifts. We use a dedicated algorithm to decompose the QSO and 
host galaxy light in the three-dimensional data which we have already successfully applied to similar observations of QSOs \citep{Sanchez:2004a,
Christensen:2006,Husemann:2008,Husemann:2010,Husemann:2011,Husemann:2013a}. We leave a detailed analysis of the stellar populations as well as 
gas and stellar kinematics for future papers in the series and focus solely on the spatially resolved characterization of the ionized gas via 
standard emission-line diagnostics.  Specifically, we separate \HII-like regions and the ENLR contribution to study the ENLR size-luminosity 
relation for type 1 AGN  and  estimate specific star formation rates (SSFRs) from \HII-like regions in comparison to the overall population of 
normal galaxies. Furthermore, we measure the gas-phase oxygen abundance as a diagnostic for the origin of the gas, which can be used to 
distinguish between internal processes and galaxy interactions as AGN triggering mechanisms in different morphological types.

In Section~\ref{sect:sample_obs}, we present the QSO sample, and describe the IFU observations and data reduction. In Section~\ref{sect:debl}, 
we outline the QSO-host galaxy deblending process and provide a detailed description of the emission-line measurements together with the 
corresponding ionized gas diagnostics in Section~\ref{sect:emlines}. Our main results are presented and discussed in Section~\ref{sect:results} 
followed by our summary and conclusions (Section~\ref{sect:conclusions}). Throughout the paper we assume a cosmological model with 
$H_0=70\,\mathrm{km}\,\mathrm{s}^{-1}\,\mathrm{Mpc}^{-1}$, $\Omega_\mathrm{m}=0.3$, and $\Omega_\Lambda=0.7$.

\section{QSO sample and observations}\label{sect:sample_obs}
\subsection{Sample characteristics}
Our flux-limited QSO sample is drawn from the Hamburg/ESO survey \citep[HES, ][]{Reimers:1996,Wisotzki:1996,Wisotzki:2000} and  
consists of the brightest QSOs above well defined flux limits within an area of $611\,\mathrm{deg}^2$ at $0.027<z<0.2$. It is a low-redshift 
subset of the sample defined by \citet{Koehler:1997} to study the local QSO luminosity function where details of the sample selection can be found. 
The QSOs have total apparent magnitudes in the range of $13.7<V_\mathrm{total}<16.8$, and host magnitudes of $14.5<V_\mathrm{host}<18.0$ with 
corresponding host luminosities of $-25.6<M_K<-23.2$ in the $K$ band. We summarize their main characteristics in Table~\ref{tbl:sample} and 
describe them below in more detail.

\begin{table*}
\begin{minipage}{150mm}
\begin{footnotesize}
\caption{Overview of the sample characteristics}
\label{tbl:sample}
\input{tab1.tex}
\end{footnotesize}
\end{minipage}
\end{table*}

An extensive set of ground-based multi-colour $BVRIJHK$ imaging observation for this sample is available and was presented by \citet{Jahnke:2004b}.
It contains 9 bulge-dominated QSO host galaxies, 8 disc-dominated QSO host galaxies, and 2 cases of ongoing major mergers.  
In addition, a few objects have confirmed close companion galaxies with or without signatures of ongoing interactions.  
High-resolution \textit{HST} images  were obtained for 9 objects in different programs\footnote{
\textit{HST} images from the following programs: ``The nature of quasar host galaxies: combining ACS imaging and VLT Integral 
Field Spectroscopy'' (Proposal 10238, PI: F. Courbin), ``Subarcsecond structure in nearby AGN'' (Proposal 5479, PI: M. Malkan),
``WFC imaging of nearby bright Quasars'' (Proposal 5434, PI: J. Bahcall), ``High-resolution imaging of X-ray selected AGN'' (Proposal 6361, 
PI: B. Boyle)}, some of which are unpublished so far. We retrieved all archival \emph{HST} images from the Hubble Legacy Archive
\footnote{Website of the Hubble Legacy Archive: http://hla.stsci.edu} and performed a deblending of the QSO and host components with 
\textsc{galfit} \citep{Peng:2002,Peng:2010} using either a dedicated PSF star observation or, alternatively, a PSF model created with 
\textsc{tinytim} \citep{Krist:1995}. The QSO-subtracted \textit{HST} or ground-based images (mainly $V$-band) of our targets are shown 
in the overview figures (Fig.~\ref{fig:data_overview}) for each object later on.

We used the broad-band photometric information available in either 6 or 7 different bands to infer the stellar mass for each host galaxy
by estimating their mass-to-light ratio via a SED template fitting approach. We generated a set of template SEDs with the \citet{Bruzual:2003} population synthesis code 
assuming Padova 1994 evolutionary tracks and a Chabrier initial mass function (IMF). Furthermore, we adopted a solar metallicity
for all template SEDs. Since our host galaxies exhibit bluer colours than the inactive ones \citep{Jahnke:2004b}, we account for 
signs of ongoing or recent star formation by creating a library of 2190 composite stellar population (CSP) models. The CSP models are two 
component models with exponential declining star formation histories. The e-folding times are chosen to be $\tau=$10\,Myr and 
$100$\,Myr, respectively, for the recent burst components and $\tau=1$\,Gyr for the old stellar population.  Finally, we estimate the 
stellar masses of  each host galaxy using a template-fitting algorithm where the redshift of the templates is fixed to the known redshift
 of the QSO during the $\chi^2$ minimization. To assess the uncertainty of our stellar mass estimates, we varied the observed flux 
 in each bandpass according to the Gaussian distributed $\sigma$ flux error. Here, we also included the effect of contamination from 
 emission lines falling into the bandpass. After fitting 100 of those mock SEDs we defined the confidence interval such that it covers 99\% of 
 the range in stellar masses around the best fit value.
 
To characterize the radio properties of our sample, we classified the QSOs into radio-loud, radio-intermediate and radio-quiet based on the 
$R$ parameter \citep[e.g.][]{Kellermann:1989}, which is defined as the ratio of the flux density at 6\,cm (5\,GHz) over that at 4400\AA. 
Follow-up observations for HES QSOs at 5\,GHz with the Very Large Array (VLA) were only done for 7 QSOs of our sample. For the other objects 
we used the measurements and upper limits from the NRAO VLA Sky Survey (NVSS) at 1.4\,GHz \citep{Condon:1998} as a surrogate for the radio 
flux at 5\,GHz assuming a power-law radio spectral index of $\alpha_r=-0.5$ at the dividing line between steep and flat-spectrum radio source. 
The vast majority (12/19) of the QSOs in the sample  are radio-quiet QSOs ($R<1$), including all objects with an upper limit in $R$ close to 1. 
Five QSOs exhibit an intermediate level of radio emission with $R$ parameters in between $1\leq R\leq 10$ and two QSOs in our sample 
($\sim$10 per cent), HE~1020$-$1022 and HE~1434$-$1600, are clearly radio-loud QSOs ($R>10$).

In addition, we collected the IRAS 60$\mu$m and 100$\mu$m fluxes from the IRAS Faint Source Catalogue \citep{Moshir:1990}. Only four objects
in our sample have been detected by IRAS. For the rest of the sample, we adopt upper limits of 0.2\,Jy and 1.0\,Jy, respectively, which were estimated by
\citet{Moshir:1990} for the catalogue.

\subsection{VIMOS integral field spectroscopy}
Integral field spectroscopy of all 19 QSOs in the sample was obtained with the VIsible MultiObject Spectrograph \citep[VIMOS,][]{LeFevre:2003} 
mounted on UT3 (Melipal) of the ESO Very Large Telescope in Chile. The observations were carried out in service mode in period 72 and 83 during 
dark time. We used the high resolution grisms (HR grisms) in order to be able to obtain accurate kinematic information from emission and 
absorption lines as well as to detect and deblend kinematically different emission-line components. The spectral resolution of the HR blue,
HR orange, and HR red grisms are $\lambda~\Delta \lambda^{-1}\sim$2550, $\sim$2650 and $\sim$3100, respectively.  
Because the wavelength coverage is limited for the HR grisms, observations in two different instrumental setups were often taken to cover the 
important spectral regions around H$\beta$ and H$\alpha$. 

Depending on the apparent angular sizes of the QSO host galaxies, we selected different magnifications of 0\farcs33\,$\times$\,0\farcs33 or
0\farcs67\,$\times$\,0\farcs67 per spaxel\footnote{spectral pixel, a single spatial resolution element containing a spectrum along 
the entire wavelength range}. This yields a field-of-view (FoV) of 13\arcsec\,$\times$\,13\arcsec\ or 27\arcsec\,$\times$\,27\arcsec\ for the 
1600 fibres arranged in a rectangular grid of 40\,$\times$\,40 spaxels. The total integration time per instrumental setup ranged between 900\,s 
and 3150\,s, split into at least 3 exposures. A dithering scheme allowed the rejection of dead fibres during data combination. The median seeing 
of all observations was 1\farcs3 and the median airmass was 1.2. Details on the individual observations are given in Table~\ref{tbl:obslog}.

\begin{table*}
\begin{minipage}{136mm}
\begin{footnotesize}
\caption{VIMOS observational log.}
\label{tbl:obslog}
\input{tab2.tex}

\end{footnotesize}
\end{minipage}
\end{table*}

Unfortunately, a few QSOs were not properly centred in the VIMOS FoV because of the blind acquisition procedure, which affected observations 
in the high magnification mode ($0\farcs33$ spaxels) more severely. Thus, HE~1201$-$2408 and HE~1434$-$1600 are only partially covered with 
our targeted VIMOS FoV and HE~1335$-$0847 is not covered at all with the HR orange grism setup. Observation of HE~1020$-$1022 suffered from 
poor photometric conditions resulting in an exceptionally bad  spectrophotometry. The spectra of the HR orange and HR red observations do not
match in the overlapping wavelength range, neither in slope nor in absolute calibration, so that we rejected this object from any detailed 
analysis in this paper.

\subsection{Data reduction}
We used a completely self-made reduction pipeline for the complex VIMOS data reduction, which is based on the reduction pipeline written 
in Python for the data of the Calar Alto Legacy Integral-field Area (CALIFA) survey \citep{Sanchez:2012a,Husemann:2013b}. The CALIFA pipeline can 
almost directly be applied to the VIMOS data with just a few dedicated modifications, because the CALIFA survey also uses a fibre-fed IFU similar to VIMOS. 
A key feature of the pipeline is the use of pixel tables and only applies resampling steps for the wavelength calibration
and spatial resampling of the final cubes. The pipeline performs all basic reduction steps: Bias subtraction, automatic fibre identification 
with rejection of bad fibres, cosmic ray detection/rejection with PyCosmic \citep{Husemann:2012a}, straylight subtraction, fibre tracing, flexure correction, spectral extraction, 
wavelength calibration, fibre flat-fielding and flux calibration. We provide a description of the different reduction steps below, including 
our special treatment for the VIMOS instrument and how the final data cubes are created from multiple exposures.

A master bias frame was created for each of the four independent spectrograph CCDs as the median of 5 bias frames regularly taken each day. 
These master bias frames were subsequently subtracted from all raw images. Automatic fibre identification often fails for VIMOS.
We take the known low transmission (bad) fibres and the flexure offsets into account when automatically cross-matching the 
fibre peak positions with a VIMOS fibre position template. The fibre peaks were then traced along the dispersion axis. This process was 
robust except for the HR blue observations where the tracing was lost for a few fibres at the blue end ($\lambda<4300\AA$) caused by their 
low transmission. We ignored that effect because that spectral region is unimportant for our scientific analysis.

Strong flexure is an important effect that varies with the position of the telescope. Thus, the traces of fibres in the continuum lamp exposure 
do not necessarily match with that of the science frames taken at slightly different telescope positions. To estimate the relative offsets 
between the traces in the science frames and the continuum exposure, we measured the fibre positions directly in the science frame by co-adding 
the light of 200 pixels along the dispersion direction at 5--6 locations distributed along the dispersion axis. Afterwards we extrapolated the 
measured offsets on to the entire dispersion axis using a Legendre polynomial of 2nd order. We find flexure offsets up to $\pm2.5$\,pixels in 
cross-dispersion for the observations from 2004 and only $\pm 0.5$\,pixels for the observations from 2009 following an upgrade of the instrument. 
Flexure offsets are also expected to occur in the dispersion direction. We estimated those flexure offset from the difference between the 
measured and expected wavelength of prominent night sky emission lines. The offsets were then extrapolated to the entire wavelength range with a Legendre polynomial 
of 2nd order. However, only one sky line in the HR blue setup is bright enough to estimate the flexure offsets, so that we applied a 0th-order correction
in that case. In addition, we correct the wavelength calibration for the heliocentric velocity shift of the target at the time of observation.

Each spectrograph of VIMOS covers 400 fibres densely projected on to each CCD so that cross-talk between fibres is an issue. 
We use an optimal extraction algorithm \citep{Horne:1986} assuming a Gaussian profile in cross-dispersion for each fibre with fixed
position and FWHM individually. The FWHM fibre profiles were determined independently for blocks of 20 fibres simultaneously at every 50th pixel 
in dispersion direction which is extrapolated along the dispersion axis with a lower order polynomial. Bad CCD pixels and cosmics ray hits are 
masked during the extraction process. The value of the extracted spectrum is flagged as bad, if the brightest 3 pixels of a fibre at a certain 
spectral pixel are bad. The error of each spectrum is computed with the optimal extraction algorithm based on the Poisson and read-out noise of each CCD pixel. 
At the end of the spectra extraction process, we store the data as a row-stacked spectra (RSS) file containing the spectra, wavelength, error and bad pixel masked
as different extensions.

\begin{figure*}
 \centering
 \includegraphics[width=\textwidth]{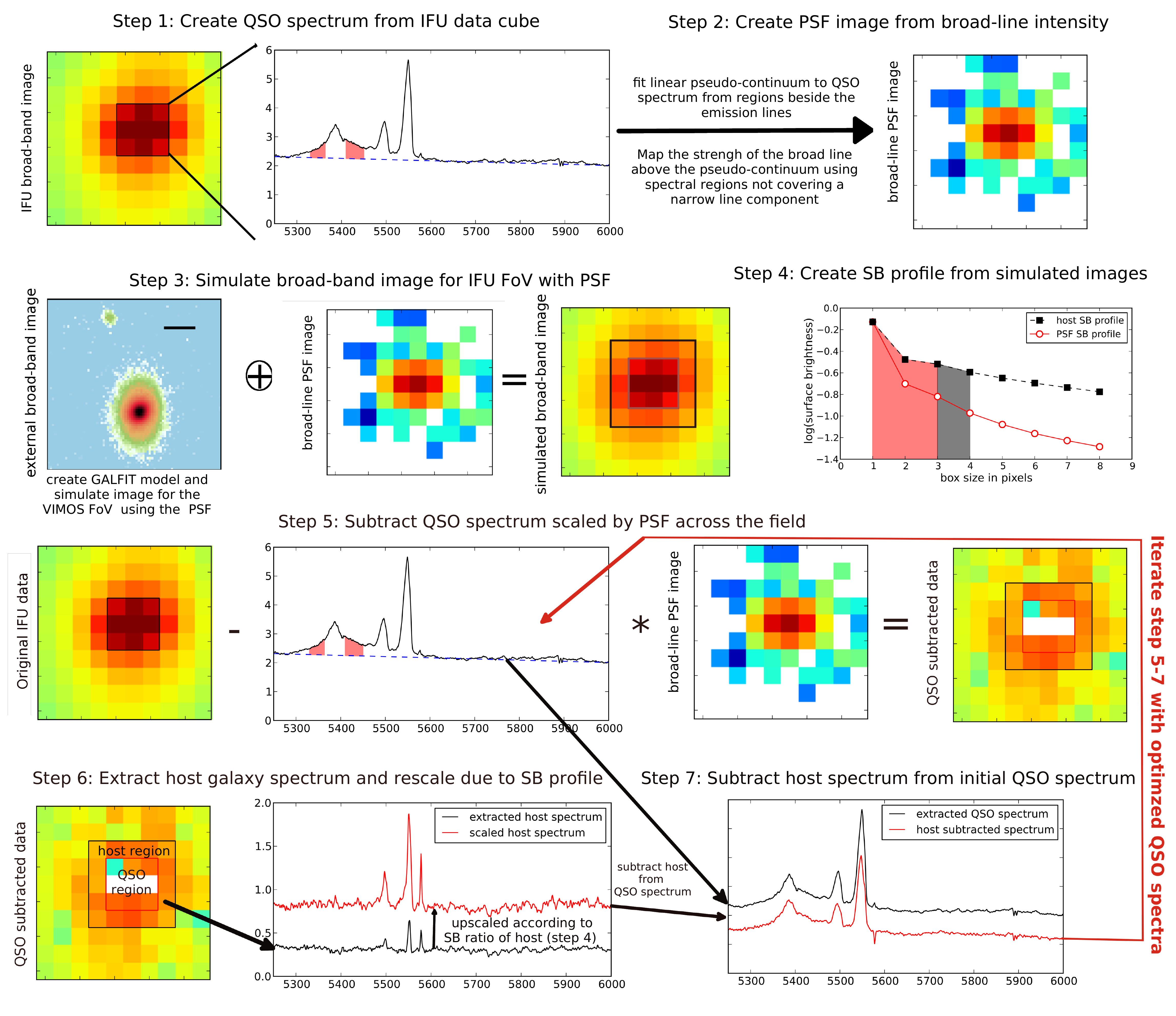}
 \caption{Sketch to illustrate the various steps of the iterative algorithm to deblend the QSO from the host galaxy emission with \QDeb.}
 \label{fig:QDeb_algo}
 \end{figure*}
Afterwards, we adaptively smoothed the spectra to a common spectral resolution of 3\AA\ (FWHM) that were estimated from the FWHM of the 
lines in the arc lamp frame. This is important for modelling the continuum over the entire wavelength range and interpreting line width properly.
All spectra are resampled to a common wavelength solution for each instrumental setup. We chose the sampling points
for the different setups such that they are consistent in their overlapping wavelength range to allow a simple combination. Fibre-flats were created from the corresponding continuum lamp exposures 
for each object to correct differences in the 
fibre-to-fibre transmission. We find residual fibre-to-fibre variations of $\pm5$ per cent across the field as measured by the flux of the 
[\textsc{oi}]\,$\lambda 5577$ night-sky line in blank sky fields. We note that those variations are significantly smaller than reported for 
other VIMOS IFU observations, e. g. up to $\pm30$ per cent \citep{Arribas:2008}, which is most likely related to the fact that other VIMOS
pipelines do not properly take into account the significant cross-talk.

Flux-calibration was performed based on spectroscopic standard star observations that were processed along the same steps outlined above. 
We computed a master sensitivity function for a given instrumental setup by averaging the sensitivity curves of all available standard star 
observations close to our science targets. The master sensitivity function was then used to perform a relative flux calibration of the science 
data.

We subsequently applied the following post-processing to create science ready datacubes. A mean sky spectrum was extracted from four 
spatial regions at the edges of the VIMOS FoV that were free from host galaxy emission, and they were subtracted them from all the spectra in 
each of the four quadrants separately to remove the background signal.  We traced the positional change of the bright point-like QSO as a 
function of wavelength caused by the atmospheric dispersion. The different dithered science exposures were then registered with respect 
to the QSO position at a given wavelength and combined in a single step with the drizzle algorithm \citep{Fruchter:2002}. During the combination of dithered
exposures bad pixels and bad fibres are masked. Thereby, we limit the spatial resampling steps to one while correcting for atmospheric dispersion
at the same time. This allows the propagate of errors with least correlated noise.

To establish an absolute flux calibration, we matched the synthetic $V$-band photometry of the VIMOS data with the $V$-band photometry of the ground-based images taken
from \citet{Jahnke:2004b}. The spectra were also corrected for Galactic extinction assuming the attenuation law of \citet{Cardelli:1989} together 
with the corresponding $V$ band extinction ($A_V$) along the line-of-sight for each object measured by \citet{Schlegel:1998}.  Finally, we 
removed the telluric absorption bands with a normalized absorption template generated from the standard star observations.

\section{Spectral QSO-host deblending in 3D}\label{sect:debl}
\subsection{Applying \QDeb\ to VIMOS data}
Studying the properties of QSO host galaxies, and type 1 AGN hosts in general, requires a robust deblending of the AGN and host galaxy light. 
In the case of 3D spectroscopy, this deblending needs to be done in the spatial \emph{and} spectral dimensions, for which a dedicated algorithm 
is needed.  We presented a dedicated software tool \QDeb\ for this task\footnote{available for download at 
http://sourceforge.net/projects/qdeblend/} in \citet{Husemann:2013a}, which is an improved version of the iterative algorithm initially 
presented by \citet{Christensen:2006} to detect extended Ly$\alpha$ emission in IFU data of high-redshift QSOs. The basic concept of \QDeb\ 
is that the spectrum in each spaxel is a superposition of the host galaxy spectrum at this position and of the AGN spectrum modulated in 
absolute flux according to the point-spread function (PSF) of the observation. Although current integral field spectrographs usually do 
not capture stars simultaneously with the target, given their small FoV, type 1 AGN offer the opportunity to self-calibrate a PSF based on 
their broad emission lines \citep{Jahnke:2004}.  More details about \QDeb\ and its algorithm can be found in \citet{Husemann:2013a} or in the 
user manual of \QDeb. Here, we briefly outline the process and show a sketch of the different steps in Fig.~\ref{fig:QDeb_algo}
to illustrate the iterative algorithm of \QDeb\ applied to our QSO observations.

In the first iteration, a high S/N co-added QSO spectrum is extracted from a 3\,$\times$\,3 spaxels region centred on the brightest 
QSO spaxels (step 1). The broad emission-line originate from the QSO broad-line regions on scales of a few pc or less and will appear as a point source in our observations.
We measure the relative brightness of the broad-line wings against the local pseudo-continuum to reconstruct the PSF of the IFU observations (step 2). 
We simulate a broad-band image for the exact VIMOS FoV and PSF based on a host galaxy model obtained from a 
nucleus-to-host decomposition with GALFIT of an available broad-band image (step 3). Then we create a SB profile of the host galaxy (step 4) that we use later in the iterative process.
A pure QSO datacube is constructed by scaling the high S/N QSO spectrum to match the flux in the broad-line wings in each spaxel. The QSO datacube is
then subtracted from the original cube and is supposed to contain only host galaxy emission (step 5). However, the initially extracted QSO spectrum is 
inevitably contaminated by some host galaxy light. We iteratively remove that host galaxy contribution with \QDeb. In four subsequent iterations, we extract a host galaxy spectrum from a single 
spaxel wide annulus \emph{around} the central QSO region (3$\times$3 spaxels) from the QSO-subtracted datacube (step 6). Since the radial SB gradient of the PSF is much steeper than that of 
the host galaxy, 
the annulus is dominated by host galaxy light after the QSO subtraction. We rescale the host galaxy spectrum to match the expected host 
galaxy surface brightness \emph{within} the QSO region as determined from the 2D SB profile (estimated in step 4) before we subtract it from the initial QSO spectrum to be
used for the next iteration (step 7).

Although the FWHM of the PSF changes slowly with wavelength, we could only estimate a PSF at the observed wavelength of \Hb\ and \Ha. We 
therefore split the datacubes when both lines were covered in a single observation. The separated datacubes were then individually processed 
and combined again at the end. The remaining PSF mismatch at wavelengths far away from the Balmer lines does affect the recovered slope in 
the stellar continuum for spaxels close to the QSO position. However, our attention is focused on the emission lines of the ionized gas 
within $\sim$200\AA\ from the Balmer lines in this paper for which such a PSF mismatch is not significant.

\subsection{Results and quality check}
\begin{figure}
\includegraphics[width=84mm]{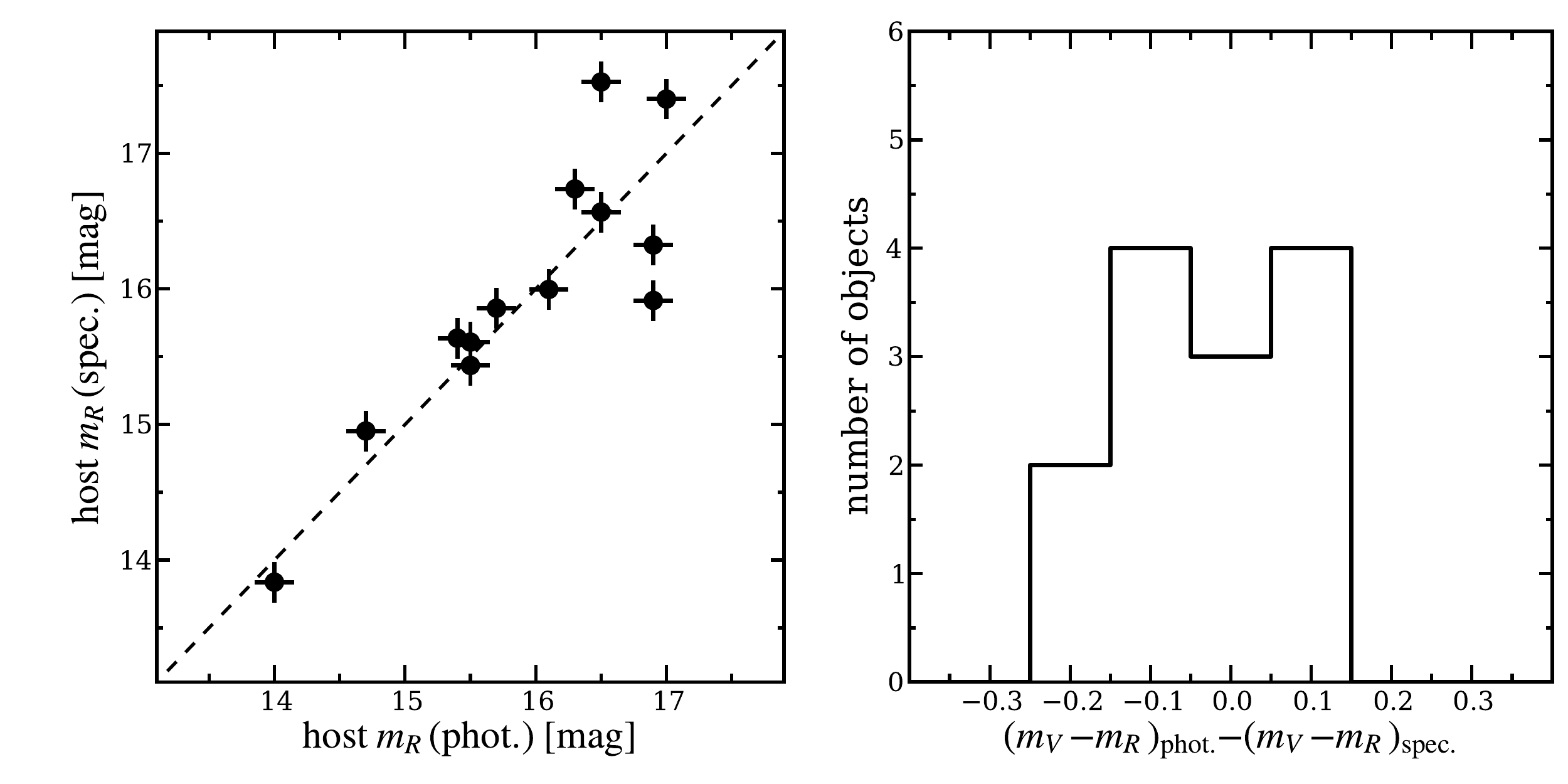}
\caption{Photometric comparison of our host galaxy spectra with the archival broad-band photometry of \citet{Jahnke:2004b}. \emph{Upper panel:} Synthetic $R$ band magnitudes of our deblended host galaxy spectra against the photometric ones.
$3\sigma$ error bars are shown as inferred from dedicated Monte Carlo simulation presented in the appendix. \emph{Lower panel:} Histogram of the $V-R$ colour difference between our 
spectroscopic and broad-band photometric data.  }
\label{fig:compare_phot}
\end{figure}

The decomposed spectra of all are shown in Fig.~\ref{fig:decomp_all} in Appendix~\ref{sect:decomp} to show the results of the spectral deblending 
process. Because of the limited sensitivity of our observations   we recovered the stellar continuum emission of the host galaxy for all 
objects except HE~1201$-$1201, HE~335$-$0847, HE~1416$-$1256 and HE~1434$-$1600. For the majority of objects, however, we recovered various 
prominent stellar absorption lines in the continuum. No remaining residuals of the broad emission lines are seen in the host galaxy spectra 
as a clear signature for the reliable separation of host and QSO light using \QDeb.

To roughly test the quality of the host galaxy spectra, we compare the broad-band photometry with our spectroscopy data in 
Fig.~\ref{fig:compare_phot} for a consistency check. We find that the $V$ and $R$ broad-band magnitudes computed from our host galaxy 
spectra are in agreement with the photometric host magnitudes of the multi-colour images \citep{Jahnke:2004b} within their 3$\sigma$ uncertainties. 
We determined the error from dedicated Monte Carlo simulations as described in the appendix~\ref{sect:simulations}. Here, we assume that the errors are similar for
both data sets since the uncertainties are dominated by the systematics of the QSO-host deblending process. Additional uncertainties from
the complex VIMOS flat-fielding are not included in the simulation. We expect that this effect is small because we find very little systematic offsets. 

Significant scatter in the absolute photometry occurs mainly for the faintest host galaxies with $m_R\ga16.5$ where the S/N of the IFU becomes critical 
for the QSO-host deblending process, but it is still consistent within the errors. Considering the $V-R$ colours, we find a very good agreement with respect to the multi-colour imaging data 
with a colour difference of typically $\pm$0.1\,mag except two cases. This is a more quantitative indication that the shape of the underlying stellar continuum 
is not significantly contaminated by emission from the QSO nucleus.

 \begin{figure}
 \includegraphics[width=84mm,clip]{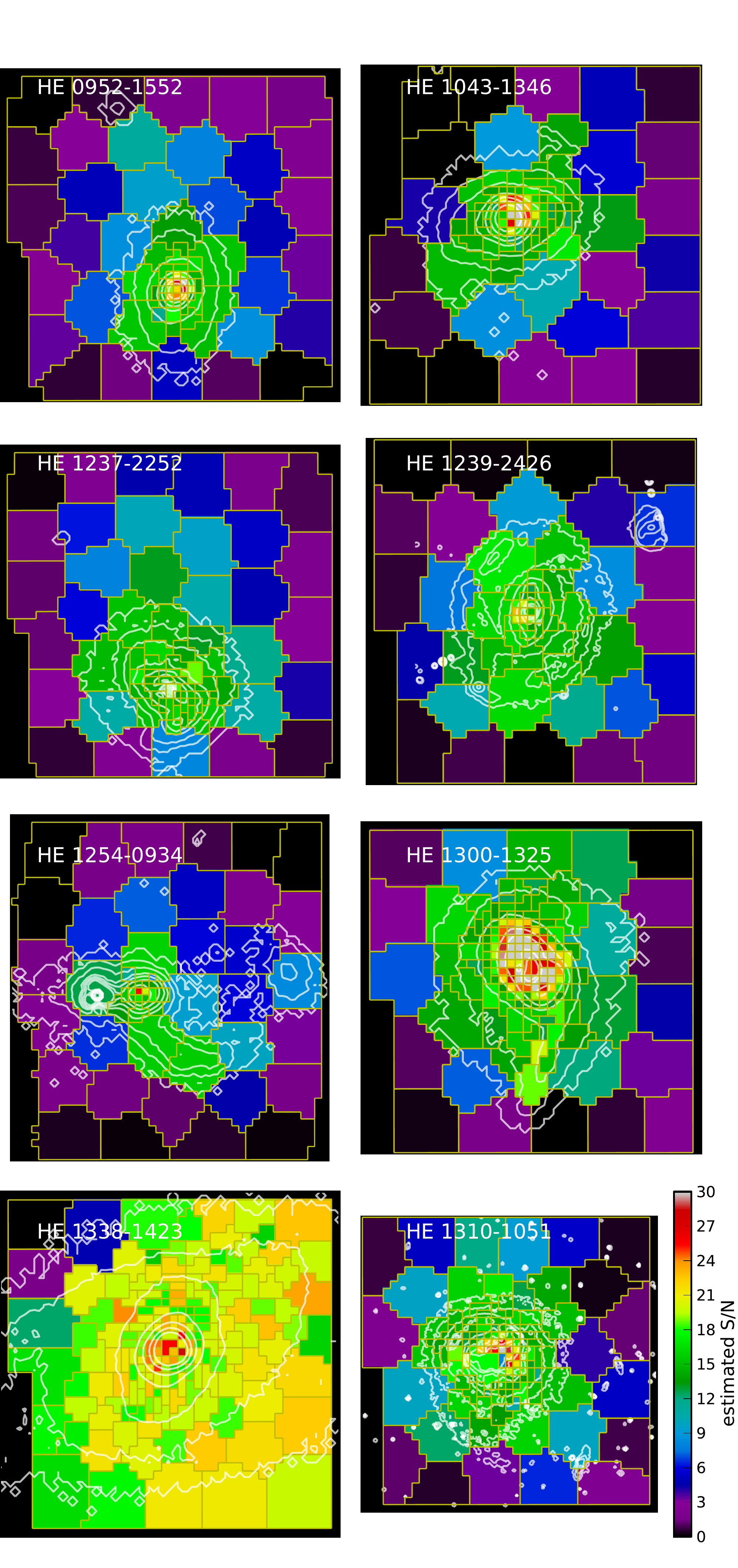}
 \caption{Results of the Voronoi binning for all host galaxies with sufficient extended continuum S/N. The colour maps represent the 
 estimated S/N of the host spectrum after binning of spaxels. The broad-band surface brightness distribution of the host galaxies is indicated by white contours.}
 \label{fig:voronoi}
 \end{figure}

\subsection{Estimation of measurement and systematic uncertainties}
The deblending process is not free from random and systematic uncertainties. Certain wavelength regions covering either the QSO broad emission 
lines or their adjacent continuum were \emph{manually} selected such that no residual broad line was apparently visible in the host galaxy 
spectrum. Small changes in those selected wavelength regions may have a significant impact on the result. We incorporated this effect in the 
error analysis via a Monte-Carlo approach. 

We generated 500 Monte-Carlo realizations for each observed datacube for which we changed the flux in each pixel within the error distribution 
as inferred from the variance cube. The deblending process was then applied to each realization exactly in the same way as the observed data, 
except that the boundaries of the manually selected wavelength regions were randomly varied within $\pm10\AA$ assuming a flat distribution. 
An additional constraint was that the new wavelength regions have a width of at least 2\AA, otherwise new boundaries were randomly chosen to 
match this criterion.

Total host and QSO spectra were extracted and stored from the 500 deblended realizations, which we used to estimate the measurement errors 
of the integrated spectra. The first 50 datacubes of each Monte Carlo run were fully stored for the error analysis of the 
\emph{spatially resolved} QSO host properties.

Apart from the measurement errors, the QSO-host deblending process is also subject to systematic uncertainties. We perform
extensive simulation to study the systematic effects of the process on the main quantities we obtain from our observation. We create 
create realistic IFU data at the redshifts of our QSOs based on existing IFU data of very nearby galaxies from the CALIFA survey \citep{Sanchez:2012a}. Details of the simulations
and their results are presented in Appendix~\ref{sect:simulations}.

\section{Emission-line diagnostics of the ionized gas}\label{sect:emlines}
\subsection{Estimating the stellar continuum}\label{sect:vimos_cont}
Before analysing the emission lines of the ionized gas, the stellar continuum had to be subtracted because the Balmer emission lines 
can be heavily blended with the corresponding stellar absorption lines. One possibility to estimate an appropriate stellar continuum 
spectrum is to find the best linear combination of a set of stellar template spectra. We used the spectral synthesis code \textsc{starlight} 
\citep{CidFernandes:2005} for this purpose. \textsc{starlight} searches for the best linear combination of template spectra taking into account 
the effect of reddening ($A_V$), smooths the templates to the optimal stellar velocity dispersion ($\sigma_*$), and shifts them to the 
systematic velocity ($v_*$) matching the observed spectrum ($O_\lambda$) in terms of kinematics and overall shape. Comparing the model 
spectra ($M_\lambda$) with the observed spectrum ($O_\lambda$), \textsc{starlight} searches for the minimum of 
\begin{equation}
 \chi^2 = \chi^2(\vec a, A_V,v_*,\sigma_*)=\sum_\lambda\left(\left(O_\lambda-M_\lambda\right)w_\lambda\right)^2\qquad .
\end{equation}
The weights $w_\lambda$ are the inverse errors of the observed spectrum. We specifically doubled the weights for the spectral regions 
around the G-band and MgI\,$\lambda5166$ absorption lines to improve the kinematic measurements and set the weights to zero for every 
spectral region containing prominent emission lines.

The high spectral resolution single stellar population (SSP) library generated by \citet{Gonzalez-Delgado:2005} from a library of 
synthetic stellar spectra \citep{Martins:2005} served as our template spectra. We selected a set of 39 different SSPs with ages 
of 1, 3, 5, 10, 25, 40, 100, 300, 600, 900, 2000, 5000 and 10000\,Myr and metallicities $2\mathrm{Z}_{\sun}$, $1\mathrm{Z}_{\sun}$, 
and $0.2\mathrm{Z}_{\sun}$.  Additionally, we used a set of 15 stellar spectra ranging in spectral type from O to K from the Indo-U.S. 
library \citep{Valdes:2004} at a spectral resolution of $\sim$1\AA\ to check the reliability of the results and to estimate systematic 
uncertainties related to the choice of the template library.

The spectrum of the best-fitting combination of SSP spectra for the integrated QSO host galaxy spectra is shown in the bottom panels of 
Fig.~\ref{fig:decomp_all} in the Appendix. Although we performed an integration over the entire host galaxy, the continuum is mainly 
dominated by the bright central region of the galaxy. In most cases, we find a good match between the observed and modelled continuum 
in most cases, considering that the bright QSO emission had to be removed before.

For the present article, we solely focus on measuring emission line fluxes from the ionized gas for diagnostic purposes. Although this 
requires a good representation for the continuum spectra, an accurate determination of the stellar population parameters and star
formation histories is not required. However, well known degeneracies of stellar populations parameters, like the age-metallicity 
degeneracy \citep{Worthey:1994}, reduce the ability to accurately recover the age of the stellar population \citep[e.g.][]{CidFernandes:2004}. 
We tested whether these degeneracies significantly affect the robustness of our emission line measurements by  comparing the continuum 
subtracted \Hb\ fluxes of all our host spectra based on two synthetic stellar library spectra.  We found that the scatter in the H$\beta$ 
flux is only  $\simeq$0.05\,dex, so that our emission-line measurements are robust and independent of the chosen stellar library for our 
observations to first order.

The modelling of the stellar continuum could often be extended to a spatially resolved scheme for the bright disc-dominated QSO host galaxies 
in our sample. Because the S/N of the individual spaxels are in most cases below 10, we employed weighted Voronoi tessellations 
\citep[][]{Diehl:2006} to adaptively bin adjacent spaxels. This is a generalization of the Voronoi binning algorithm of \citet{Cappellari:2003}, 
which has been widely used in the IFU community \citep[e.g.][]{Emsellem:2004,Sarzi:2006,McDermid:2006, Gerssen:2006,Dumas:2007,Stoklasova:2009, 
CidFernandes:2013a}. We estimated a S/N map directly from the data by measuring the mean and standard deviation in the rest-frame spectral 
window of $5500<\lambda<5700$. This is a conservative S/N estimate due to the presence of some weak absorption lines in that wavelength range. 
We chose a target S/N of 20 per bin for the brightest galaxy HE1338$-$1423 and a target S/N of 15 for the rest of galaxies. We limited the 
maximum number of spaxels per bin to 50 for the weighted Voronoi algorithms. The resulting S/N 
maps after binning are shown in Fig.~\ref{fig:voronoi} for all galaxies with sufficiently extended continuum signal.  When less than 5 bins had a S/N in 
the continuum greater than 15, we decided to use only the model of the integrated spectrum as a global continuum template spectrum for all spaxels.
For a few objects, mainly those observed with 0\farcs33 spatial resolution, no significant stellar continuum light could be recovered for 
subtraction. Table~\ref{tab:continuum_scheme} indicates which of those three schemes were applied to a specific object.

After binning, we modelled the co-added stellar continuum of each bin with \textsc{starlight} again as described above for the integrated spectra.
Because the seeing often changed between observations of different instrumental setups, the absolute continuum flux level for small bins 
(e.g. a single spaxel) may not necessarily be the same for the consecutive spectra. We therefore normalized the spectra of the two instrumental 
setups to match in the overlapping spectral region before combination.

Pure emission-line datacubes were subsequently created by subtracting the best-fitting continuum model, either spatially resolved or based on 
the model of the integrated spectrum.  The corresponding continuum model was scaled in absolute flux before subtraction to match that of each 
individual spaxel. We repeated the process for our 50 Monte Carlo realizations, previously constructed during 
the QSO-host deblending process, to generate 50 pure emission-line datacubes that we used to estimate the systematic uncertainties of 
the continuum subtraction on the emission line measurements.

\begin{table}
\caption{Stellar continuum subtraction schemes}
\input{tab3.tex}
\label{tab:continuum_scheme}
\end{table}

\subsection{Narrow-band images}
After removing the QSO emission and the underlying stellar continuum, we extracted 20\AA\ wide narrow-band images  (Fig.~\ref{fig:data_overview},
left panel) from the datacubes centred on the strongest emission lines, \Ha\ (or \Hb\ in the case of HE~1335$-$0847) and \Ox\,$\lambda5007$ 
(hereafter \Ox), to characterize the spatial distribution of the ionized gas.  These images reveal the presence of extended ionized gas in
almost all QSO host galaxies.

\begin{figure*}
 \includegraphics[width=0.97\textwidth,clip]{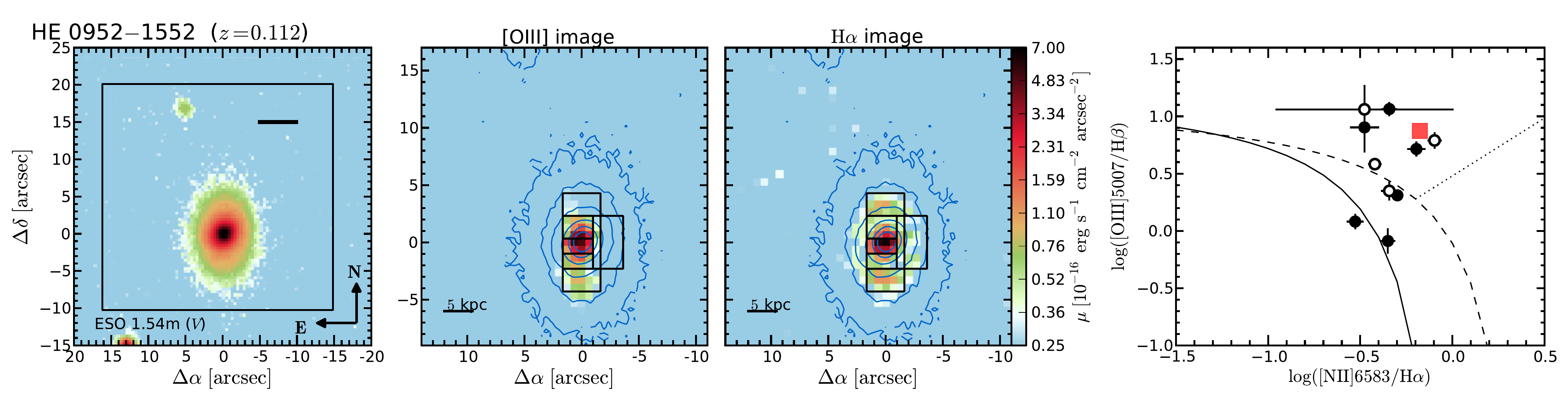}
  \vspace*{-2mm}\\
 \includegraphics[width=0.97\textwidth,clip]{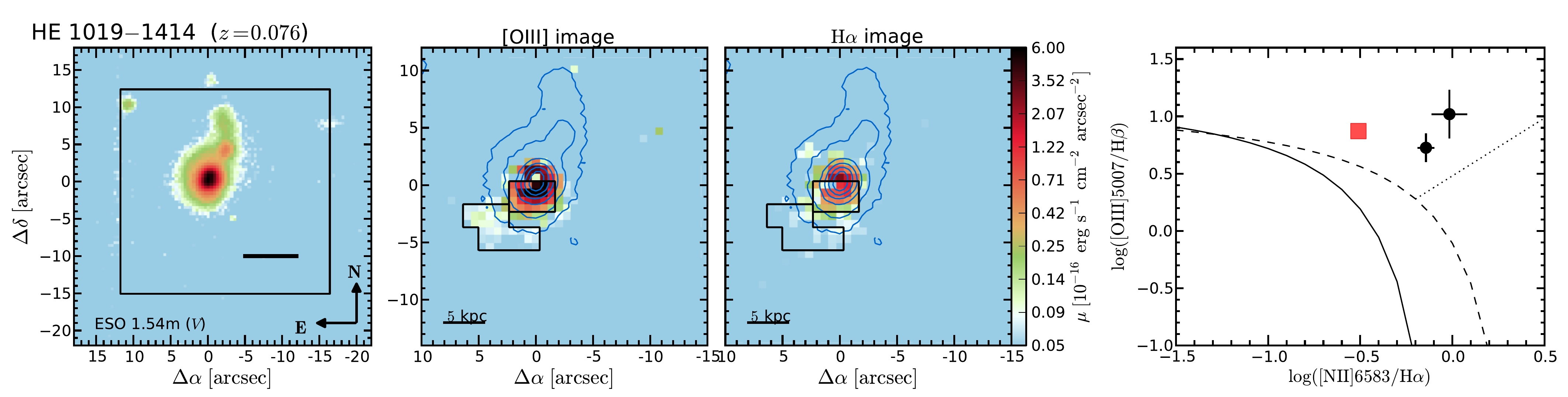}
 \vspace*{-2mm}\\
 \includegraphics[width=0.97\textwidth,clip]{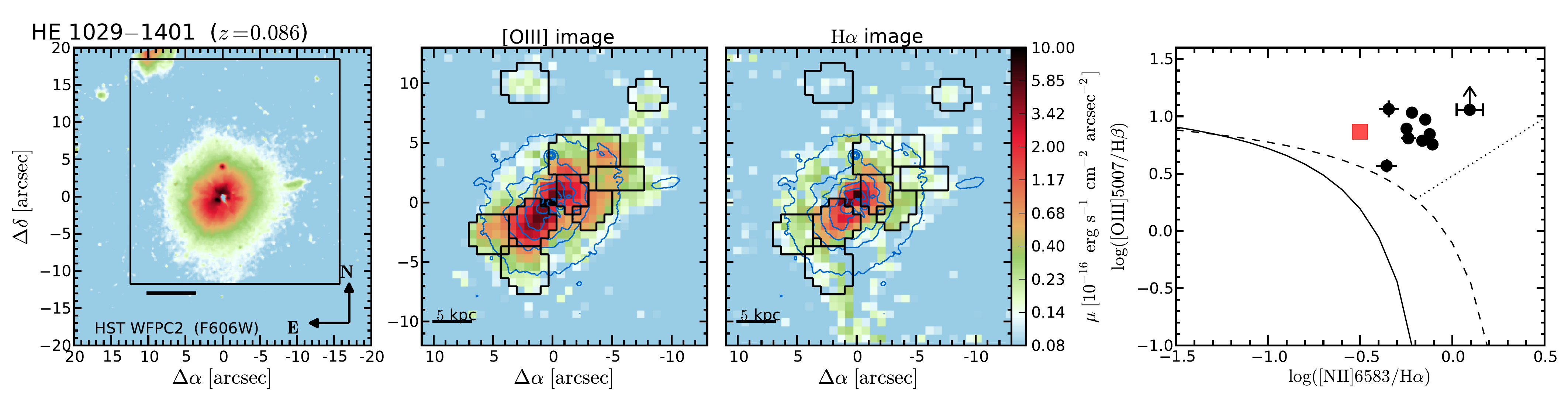}
 \vspace*{-2mm}\\
 \includegraphics[width=0.97\textwidth,clip]{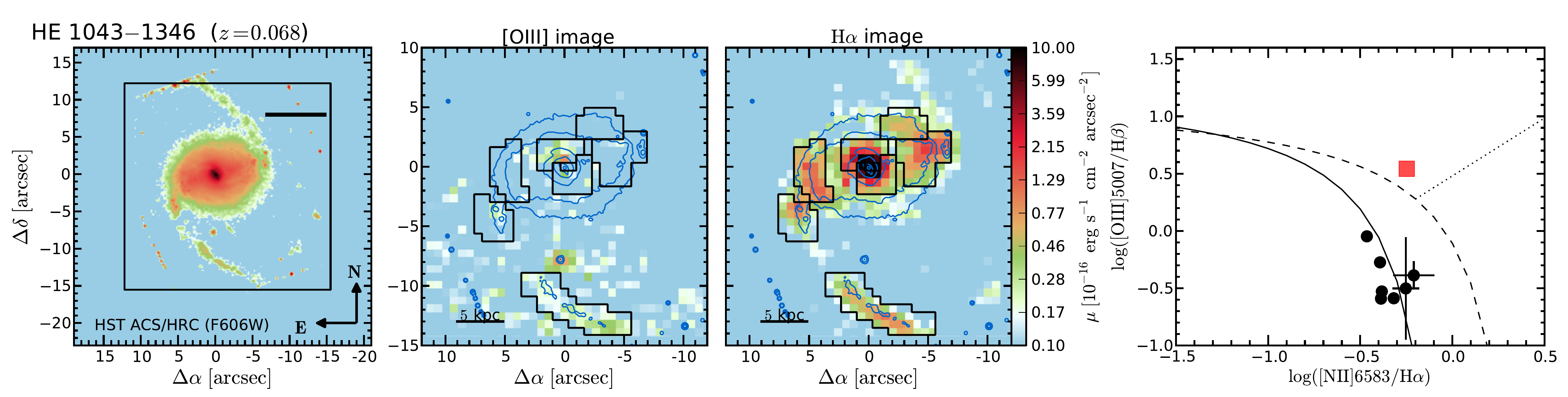}
 \vspace*{-2mm}\\
 \includegraphics[width=0.97\textwidth,clip]{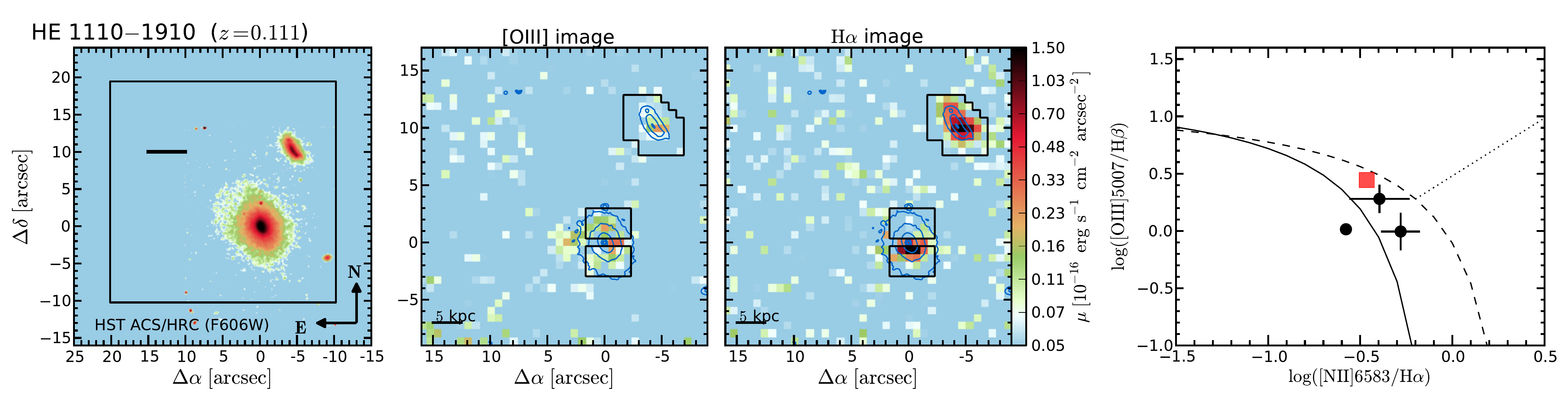}
\caption{Sample overview of the emission-line characteristics after QSO subtraction. 
\emph{Left panel:} QSO subtracted broad-band images of the host galaxies with the VIMOS FoV indicated by the rectangle.  
The black scale bar corresponds to 10\,kpc at the redshift of the QSO. \emph{Mid panel:} QSO and continuum subtracted \Ox\ and \Ha\ 
arrow-band images  in a logarithmic scaling. Apertures of specific regions are indicated by the black border lines and labelled alphabetically. 
Contours of the broad-band continuum surface brightness are over-plotted in blue to aid direct comparison. \emph{Right panel:} Standard BPT 
diagram for the individual regions indicated on the broad-band images. A second kinematic system of  emission-lines in those regions are indicated by open 
symbols if present. The red squared symbol corresponds to the line
ratios of the narrow lines above the broad lines in the QSO spectrum as extracted from the brightest spaxel at the QSO position.  The 
solid line is the demarcation curve for star-forming galaxies of \citet{Kauffmann:2003}, the dashed line is the theoretical maximum starburst line of \citet{Kewley:2001} and the dotted line represents 
the boundary between Seyfert and LINER-like emission proposed by \citet{CidFernandes:2010}.}
\label{fig:data_overview}
\end{figure*}

\begin{figure*}
\includegraphics[width=0.97\textwidth,clip]{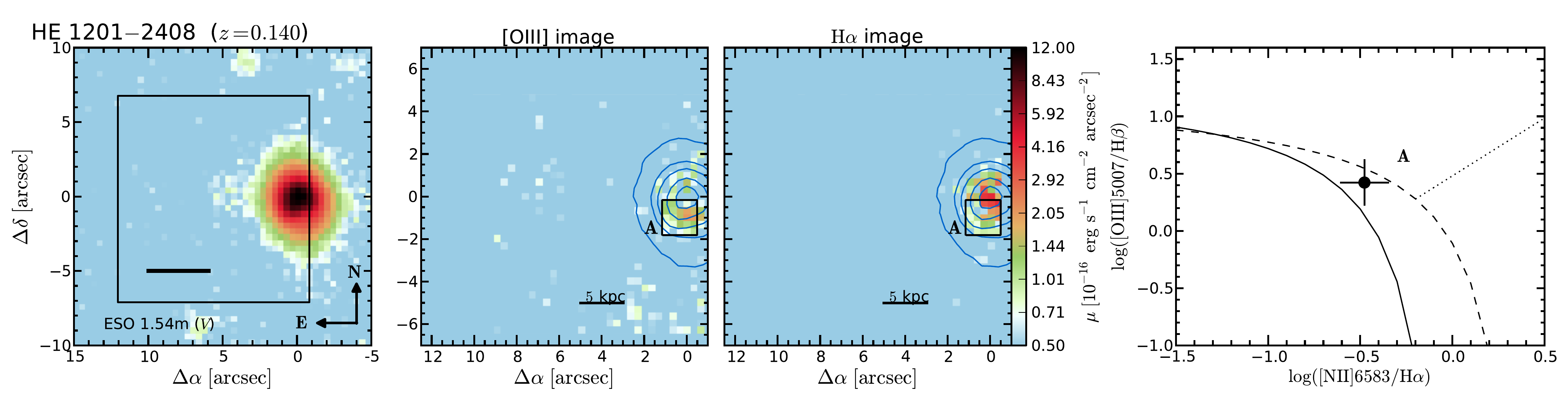}\\
\includegraphics[width=0.97\textwidth,clip]{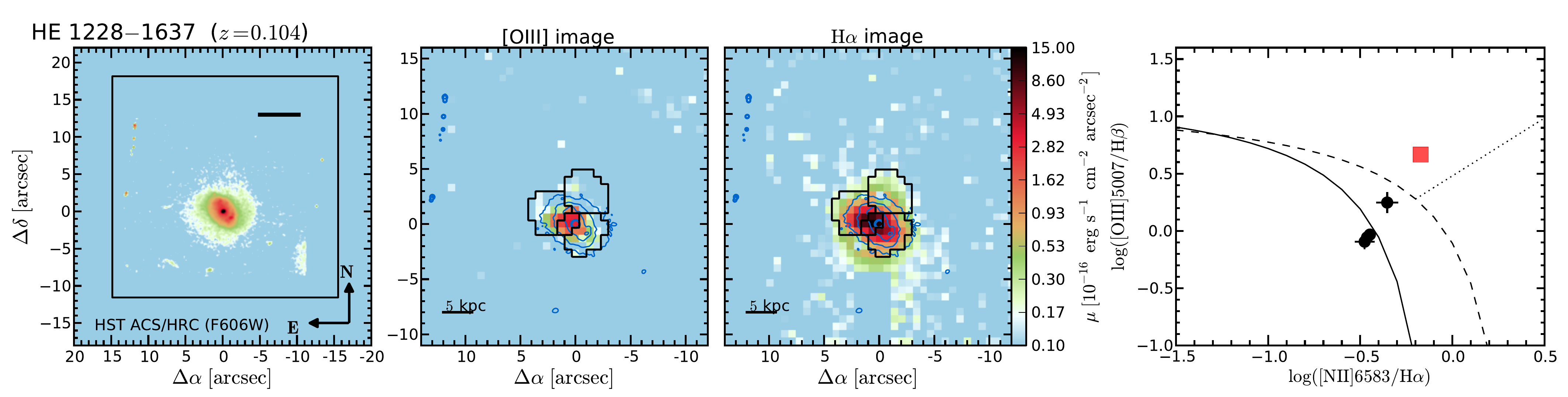}\\
\includegraphics[width=0.97\textwidth,clip]{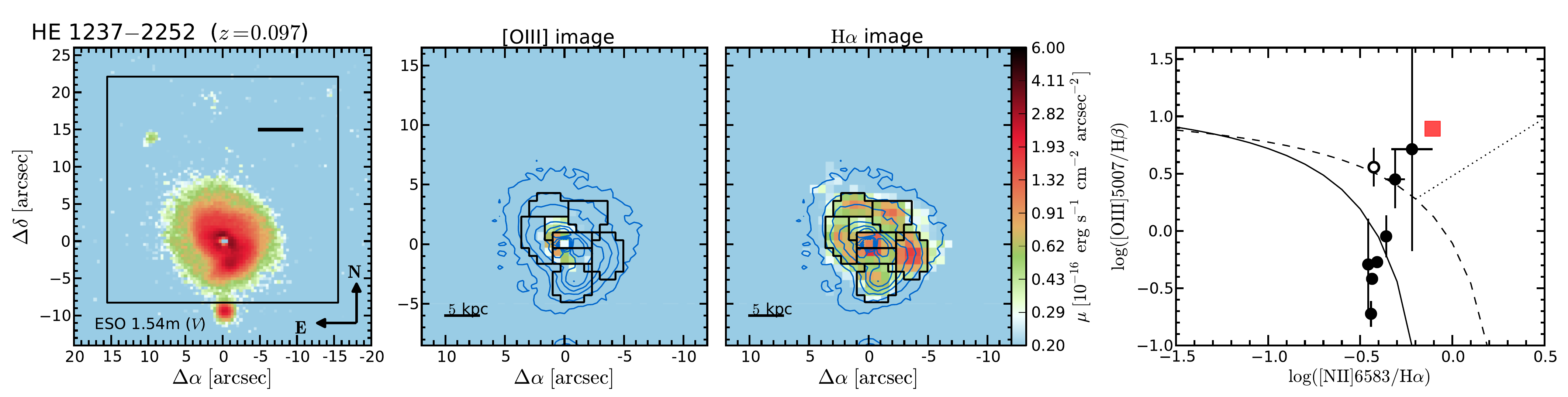}\\
\includegraphics[width=0.97\textwidth,clip]{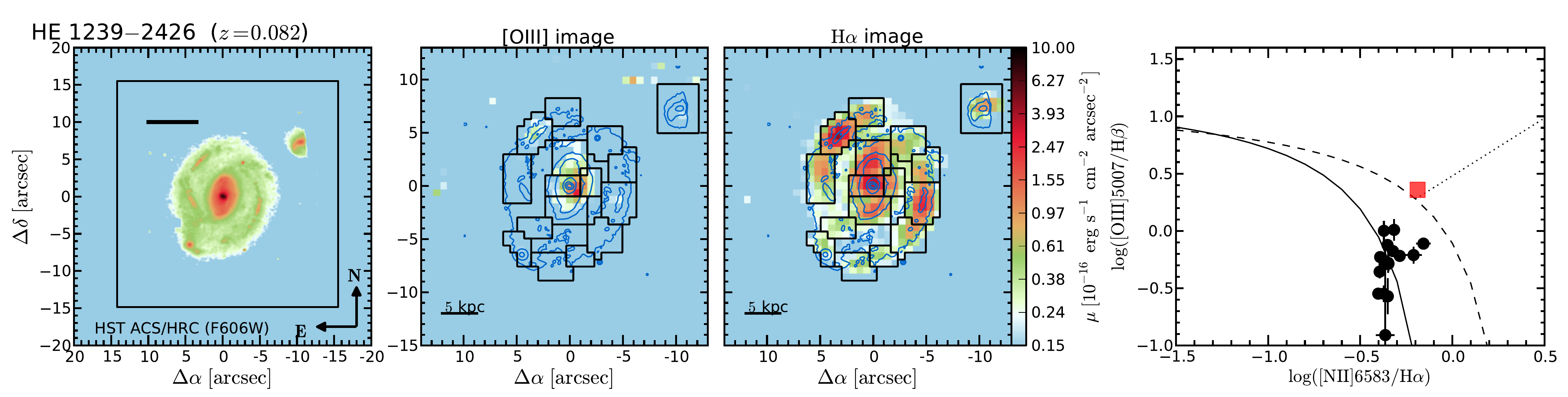}\\
\includegraphics[width=0.97\textwidth,clip]{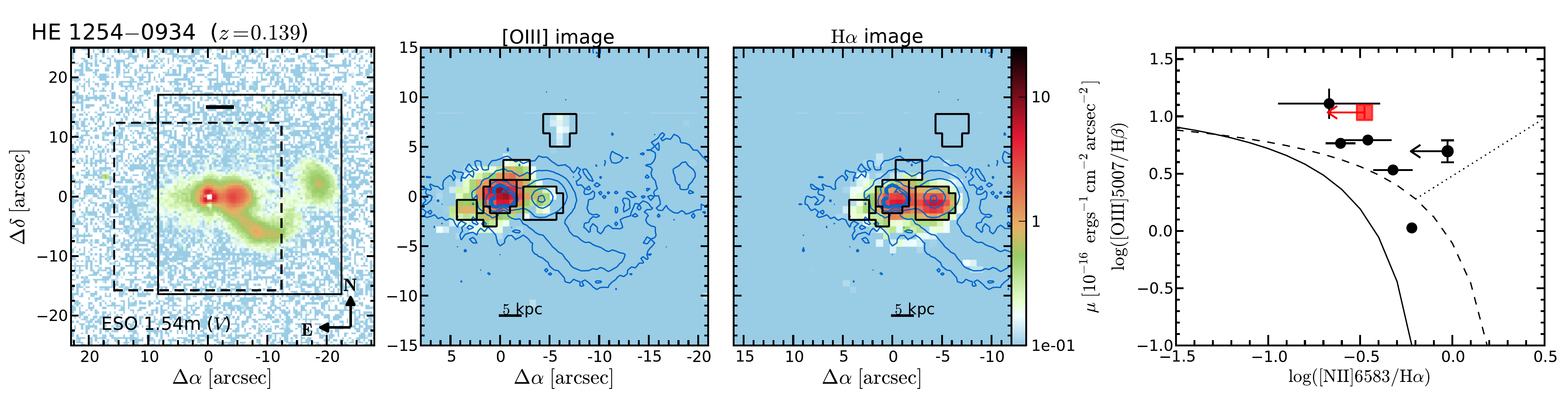}\bigskip\\
\contcaption{The line ratios of the narrow lines in the QSO spectrum of HE 1201$-$2408 could not be robustly measured because the narrow [NII] lines could not be resolved above the broad 
H$\alpha$ line. 
The HR orange (solid line) and the HR red (dashed line) observations of HE 1254$-$0934 have slightly different FoVs with a substantial overlap.}
\end{figure*}

\begin{figure*}
\includegraphics[width=0.97\textwidth,clip]{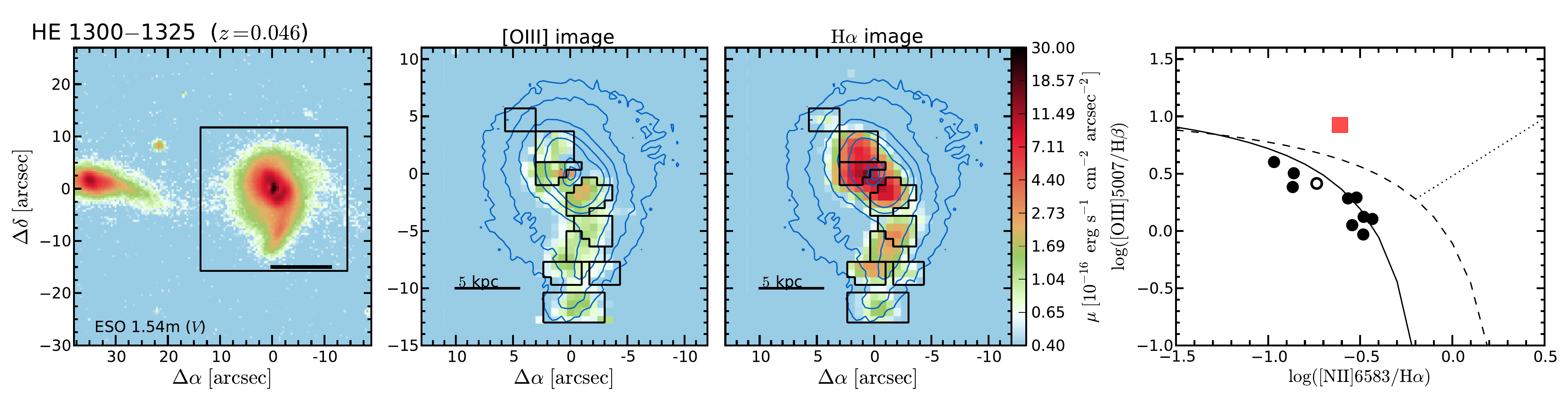}\\
 \includegraphics[width=0.97\textwidth,clip]{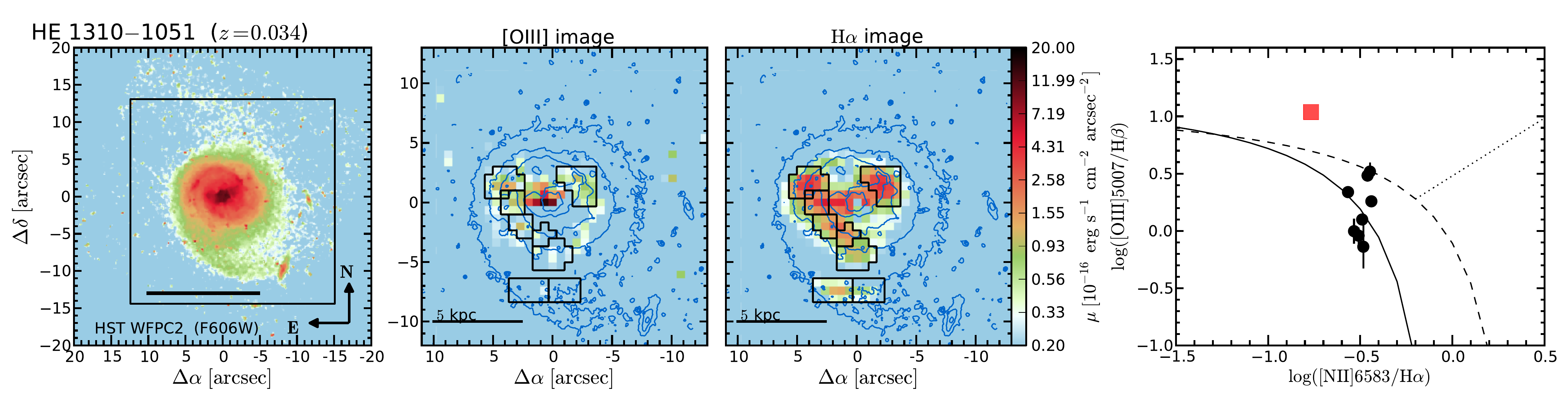}\\
 \includegraphics[width=0.97\textwidth,clip]{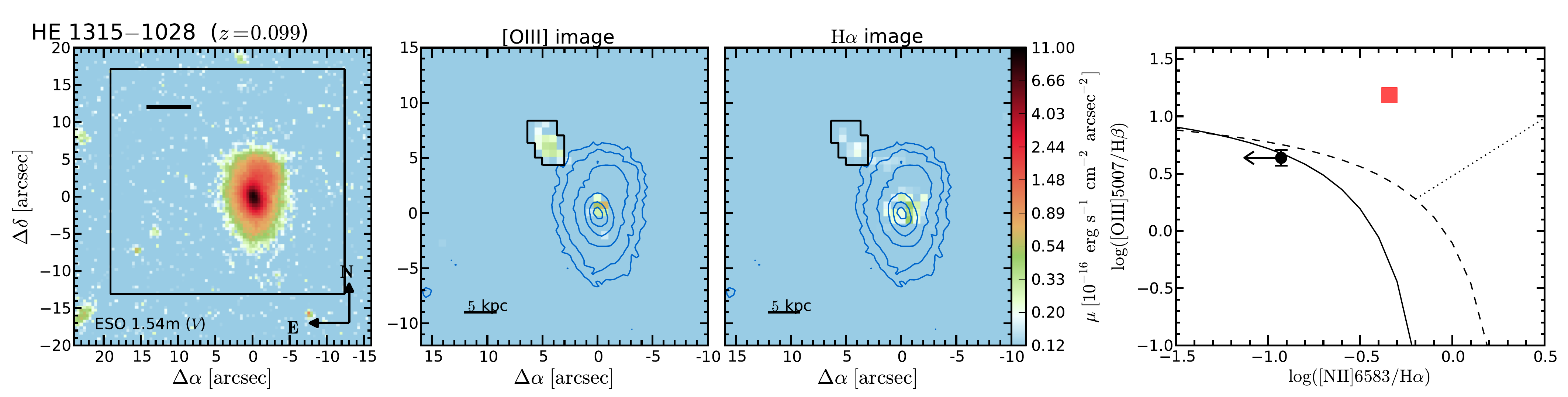}\\
 \includegraphics[width=0.97\textwidth,clip]{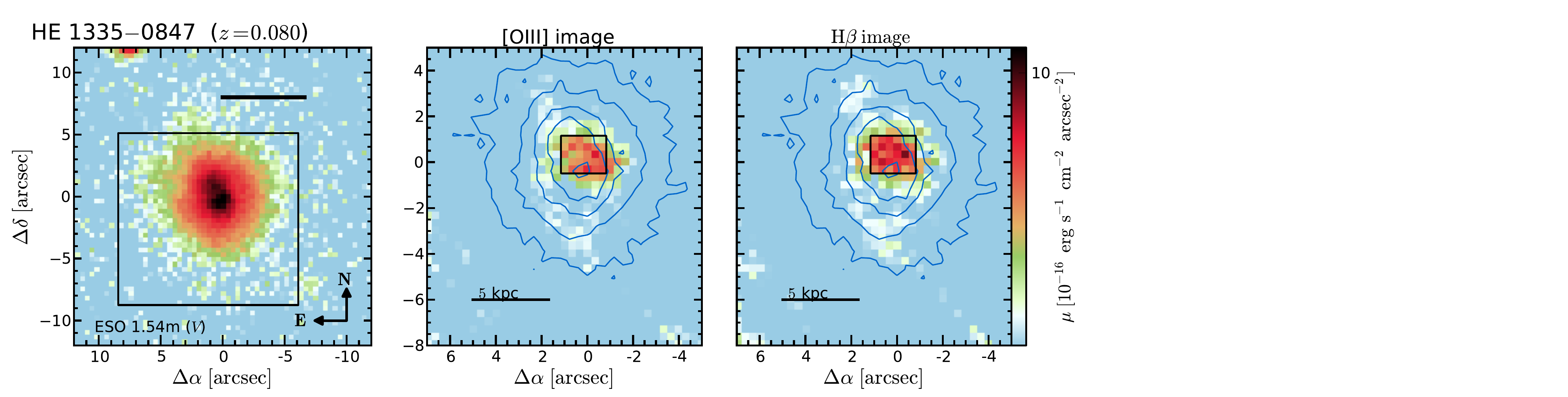}\\
\includegraphics[width=0.97\textwidth,clip]{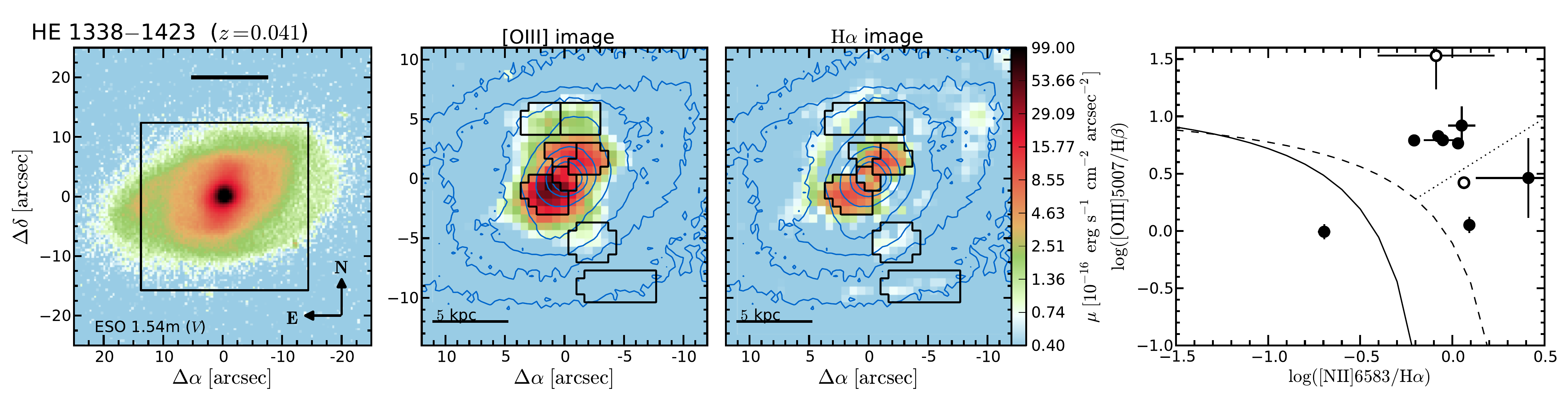}\\
\contcaption{The line ratios of the narrow lines in the QSO spectrum of HE 1338$-$1423 could not be robustly measured because the narrow Balmer lines could not be well resolved above the broad 
Balmer lines in this case.}
\end{figure*}

\begin{figure*}
\includegraphics[width=0.97\textwidth,clip]{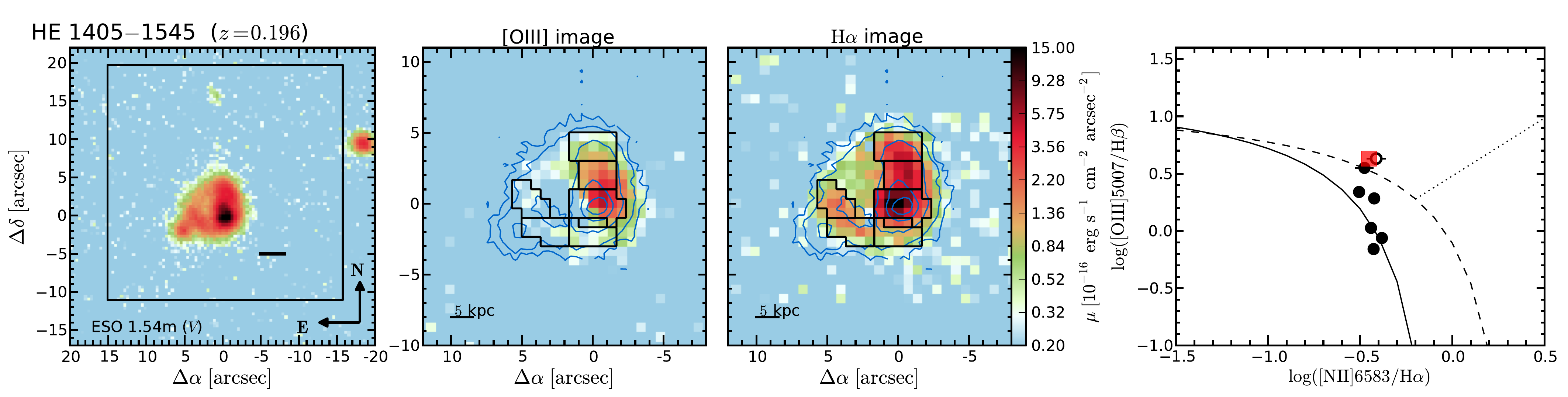}\\
 \includegraphics[width=0.97\textwidth,clip]{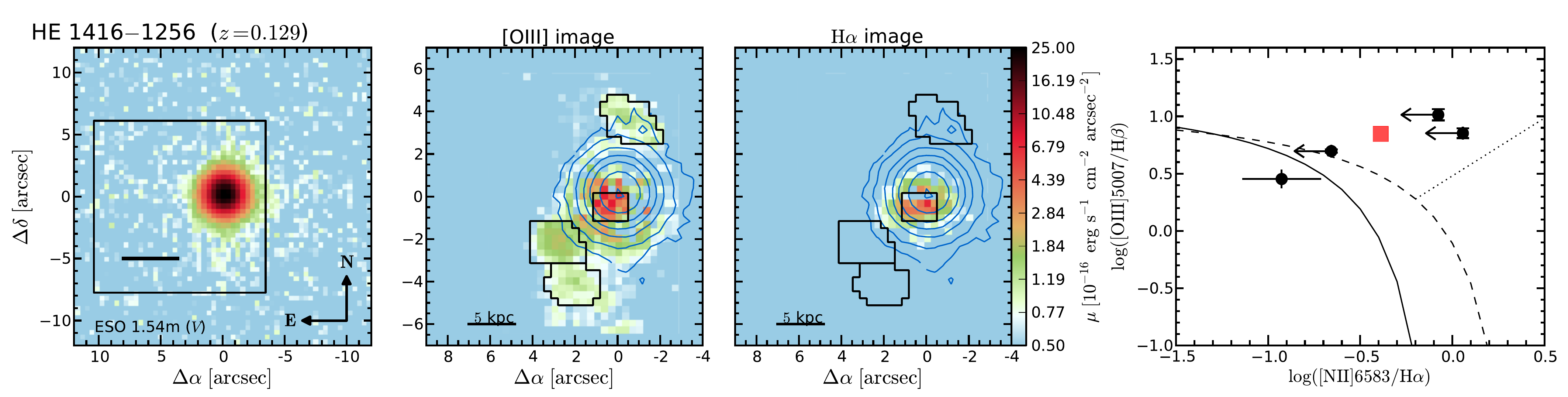}\\
 \includegraphics[width=0.97\textwidth,clip]{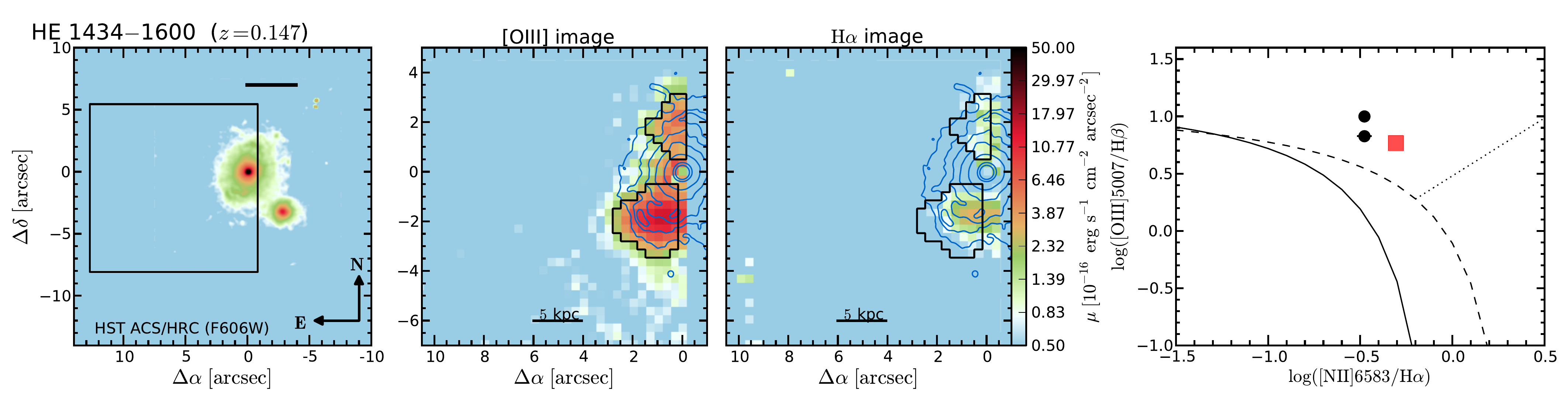}
\contcaption{}
\end{figure*}

In comparison with the broad-band continuum images, particularly the high-resolution \textit{HST} images, we find that the \Ha\ emission 
in most cases tracks very nicely the substructures in the morphology of the host galaxies. For example, the high surface brightness knots 
in the \textit{HST} ACS F606W broad-band images of HE~1043$-$1346 and HE~1239$-$2426 seen along the spiral arm and close to the nucleus 
coincide with the observed pattern of strong \Ha\ emission. The \Ha\ image of HE 1338$-$1423 shows strong residuals close to the nucleus. 
The QSO-host deblending is particularly difficult for the HR orange observation of this object. The broad H$\alpha$ emission line of the 
QSO is rather narrow ($\sim1600~\mathrm{km}~\mathrm{s}^{-1}$ FWHM) and the extended \Ni\,$\lambda\lambda6548,6583$ and \Ha\ emission lines
are bright so that it was impossible to define broad line spectral windows that are not contaminated by extended emission across the entire field.

The distribution in the ionized gas traced by the \Ha\ and \Ox\ lines is often consistent with each other, suggesting that a common 
ionization mechanism dominates throughout a galaxy. In a few galaxies, the \Ox\ emission is rather weak or even undetected compared 
to \Ha\ (e.g. HE~1239$-$2426) which is typical for high metallicity \HII-like regions.  HE~0952$-$1552 and HE~1254$-$0934, on the other hand, 
display quite different light distributions in \Ha\ and \Ox\ which presumably point to significant changes in the properties of the ISM across 
these galaxies.

\subsection{Emission-line fluxes of specific regions}
To investigate the physical conditions of the ISM in more detail, the fluxes of weaker emission lines, such as \Hb, \Ni\,$\lambda\lambda 
6548,6583$ and \Su\,$\lambda\lambda 6716,6731$, need to be accurately measured.  Those lines are often too weak to be detected in individual
spaxels, which requires a spatial binning of specific regions to increase the S/N. We manually defined several apertures (marked as black 
regions in Fig.~\ref{fig:data_overview}), covering a physically connected region as judged from the \Ha\ or \Ox\ light distribution.

We then modelled the emission-lines in the co-added spectrum of each aperture with simple Gaussians. Line ratios of the \Ox\,$\lambda\lambda
4960,5007$ and \Ni\,$\lambda\lambda 6548,6583$ doublets were fixed to their theoretical values. All emission lines, \Hb,\ \Ox\,$\lambda\lambda
4960,5007$,\ \Ha,\ \Ni\,$\lambda\lambda 6548,6583$ and\ \Su\,$\lambda\lambda 6716,6731$ were modelled simultaneously, and their line dispersions 
and redshifts were coupled to common values. This approach strongly reduced the number of free parameters and increased the robustness of the 
best-fitting model. We used a downhill simplex algorithm to find the best-fitting parameters at the minimal $\chi^2$.

\begin{figure}
\resizebox{\hsize}{!}{\includegraphics[clip]{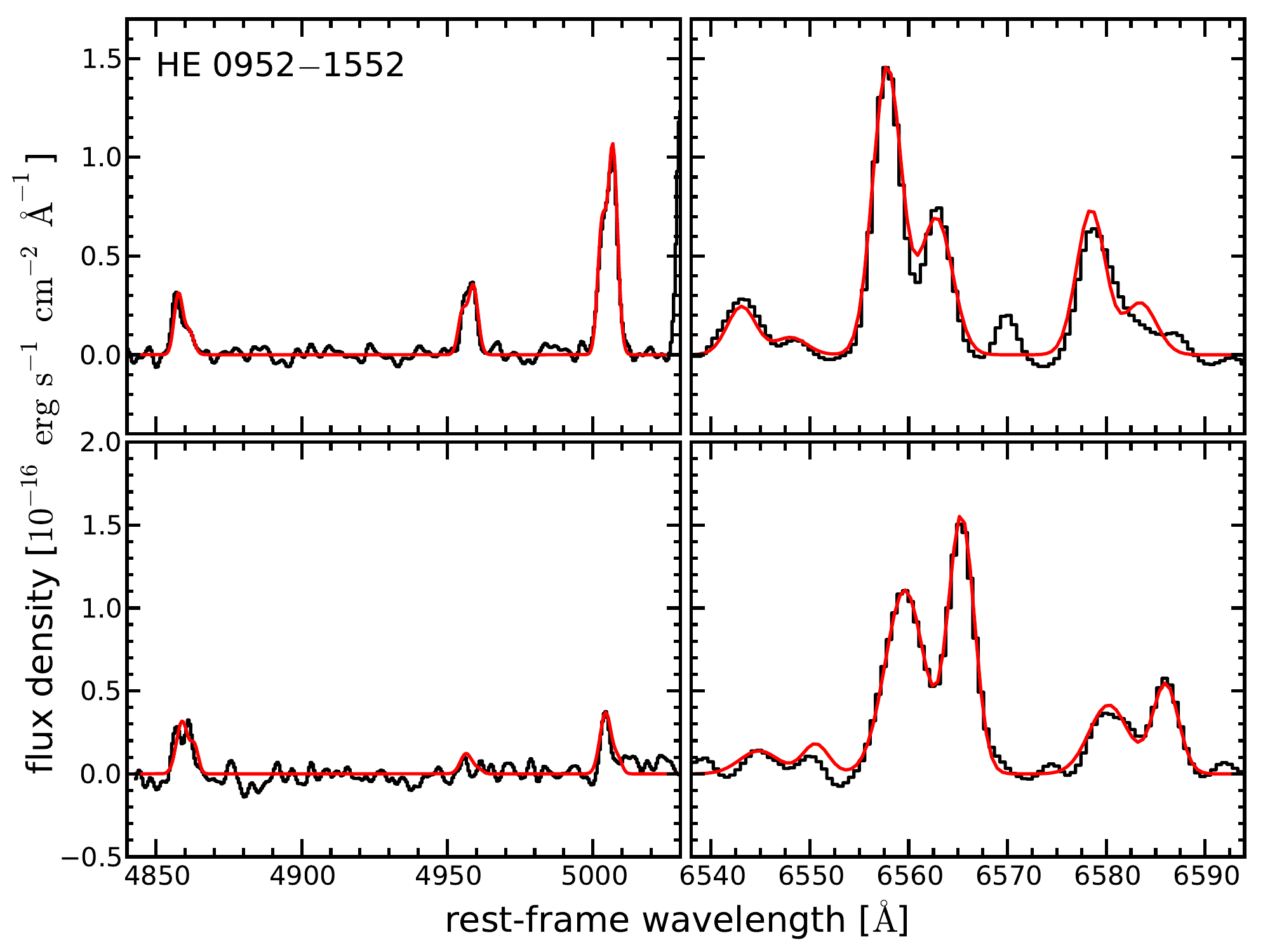}}
\caption{Examples of spectra requiring two kinematically distinct emission-line systems. A spectrum of HE~0952$-$1552 (region F) is presented in the top panels and a spectrum of HE~1237$-$2252 (region C) in the bottom panels. The red line represents the best-fitting model consisting of Gaussians for the various emission lines that are kinematically coupled to represent two independent emission line systems.}
\label{fig:line_modelling}
\end{figure}

With a spectral resolution of $\lambda/\Delta\lambda\sim2500$, corresponding to $\sim$100~$\mathrm{km}~\mathrm{s}^{-1}$ FWHM, we were able 
to resolve emission line systems with different kinematic components. This is important to disentangle the emission of companion galaxies 
along the same line-of-sight as the host galaxy, or for separating potential outflowing/inflowing gas associated with mergers, star formation, 
or AGN activity. We show in Fig.~\ref{fig:line_modelling} two examples of the spectral emission line modelling that required two different 
emission line systems. 

We derive the emission line errors on all parameter by analysing the 50 Monte-Carlo realizations of the datacubes 
and taking the standard deviation from the results. In this way we included the uncertainties of the QSO-host deblending and of the continuum subtraction 
into the error budget of the emission line measurements. We inferred $3\sigma$ upper limits for undetected \Hb, \Ox\ or \Ni\ lines based on
the noise from the adjacent continuum and the line width constrained by the detected lines.

\subsection{Distinguishing \HII-like and AGN ionization regions}
Emission-line diagnostic diagrams have been used to constrain the dominant ionizing source of the ISM in galaxies. The most frequently 
used one is the $\Ox\lambda5007/\Hb$ vs. $\Ni\lambda6582/\Ha$ diagram, often referred to as the BPT diagram \citep{Baldwin:1981}. It involves 
only emission-line ratios of the brightest lines in the rest-frame optical that are least sensitive to reddening. Several demarcation lines 
were proposed for the BPT diagram to discriminate between different underlying ionization mechanisms.

\citet{Kewley:2001} derived a boundary 
for the BPT up to which it could theoretically be produced by a massive starburst. An empirical boundary for star-forming galaxies was 
drawn by \citet{Kauffmann:2003} to include galaxies of the apparent star-forming branch in the SDSS data. These two demarcation curves 
have often been invoked to discriminate between star-forming or \HII-like regions, gas predominantly ionized by an AGN, and an intermediate 
region where both processes may significantly contribute to the ionization. However, those demarcation curves cannot be hard boundaries and
there would be a homogeneous sequence crossing those boundaries due to different levels of mixing. This implies that line ratios in the
intermediate region could be well produced entirely by a starburst, and line ratios in the AGN branch might still be substantially 
contaminated by star-forming regions as discussed by \citet{CidFernandes:2010}. Furthermore, shocks generated by jet-cloud interactions
and their precursors can be another source of ionization in AGN host galaxies \citep[e.g.][]{Morganti:1997,Clark:1998,Villar-Martin:1999,
Moy:2002,Rosario:2010} for which the BPT line ratios significantly overlap with the other ionization mechanisms.

We present the BPT diagrams for emission-line regions across the host galaxy after the QSO-host deblending in Fig.~\ref{fig:data_overview} (right panels) of all the objects in our sample. 
The HR orange observation, covering the \Ha\ and \Ni\ lines, is missing for HE~1335$-$0847, so that we could not construct a BPT diagram for this object. Nevertheless, 
we obtain a low $\Ox/\Hb$ line ratio of 0.3 for its prominent emission-line region which is consistent with an \HII-like region. In the case of HE~1315$-$1028 and HE~1416$-$1256,  
the \Ni\ lines are undetected and we can provide only upper limits for the \Ni/\Ha\ line ratio. In addition, we show the measured line ratios of the narrow lines above the broad lines in 
the QSO spectrum extracted from the brightest spaxel at the QSO position in Fig.~\ref{fig:data_overview} (right panels) as red squared symbols. 

The BPT diagrams reveal that the ionization state of the ISM is diverse and sometimes changes even across the host galaxy. 
In the following we consider BPT line ratios below or close to \citet{Kauffmann:2003} line to be dominated by \HII-like regions 
indicative of ongoing star formation, and line ratios above the \citet{Kewley:2001} to be dominated by ionization through AGN photoionization
or AGN driven shocks. 

The narrow emission-line ratios in the QSO spectrum closets to the nucleus are always in the region of AGN photoionization as expected 
also from other samples of luminous nearby AGN \citep[e.g.][]{Winter:2010}. How far AGN photoionization is dominating the gas ionization 
inside the host galaxy depends primarily on the AGN luminosity \citep[e.g.][]{Bennert:2002} and other ionization mechanisms such as star formation 
will take over at some point. Despite the obvious radial dependence on the mixing between star formation and AGN photoionization in the emission lines,
there is also a connection with the spectral properties of the QSO independent of luminosity \citep{Husemann:2008}. QSOs with strong \Fe\ emission
and broad lines with FWHM$<$4000\,$\mathrm{km}\,\mathrm{s}^{-1}$ seem to be deficit of extended AGN photoionization for still unknown reasons, which complicates the simple picture.

On kpc scales across the host galaxies, we identified clear evidence for ongoing star formation in 10 out of 18 host galaxies, showing \HII-like emission. 
In combination with the morphological classification, these numbers split up into 4/8 (50 per cent) for bulge-dominated QSO hosts, 5/8 (62.5 per 
cent) for disc-dominated QSO hosts and 1/2 (50 per cent) for major-merger systems. If we assume that the emission-line ratios in the intermediate 
region of the BPT have indeed contributions by \HII-like regions, the number of QSO hosts with evidence for current star formation increase 
by 3 objects including the second major-merger system. On the other hand, we find evidence for AGN photoionization on kpc scales also in 10 out of 
the 18 cases, which occurs in at least 3 QSO hosts together with ongoing star formation at different locations.

The fact that we find only very few regions with intermediate line ratios suggests that the spatial resolution helps 
to reduce mixing effects compared to single fibre spectroscopy, so that regions of a given ionization mechanism are spatially separated. 
Also, LINER-like emission \citep{Heckman:1980} for which the dominant ionization mechanism is strongly debated does not play a role for our study as all regions but one are above the 
corresponding demarcation line presented by \citet{CidFernandes:2010}.

\subsection{Extended narrow-line region sizes and QSO luminosities}\label{sect:ENLR_measurements}
The \Ox\ line is the brightest optical narrow emission line in case of AGN photoionization and can extent up to several kpc from the AGN, 
the so-called extended narrow-line region \citep[ENLR,][]{Unger:1987}. We constructed  \Ox\ narrow-band images from the QSO-subtracted 
IFU datacubes to measure the characteristic sizes of their ENLRs. Here, we estimated the sizes in the same manner as in our previous study 
on the ENLR around luminous low-redshift QSOs \citep{Husemann:2013a}. All pixels below the $3\sigma$ background noise level of the narrow-band 
image were masked out as well as all previously confirmed \HII-like regions, but no region with intermediate line ratios. We then measured 
the effective ENLR radius ($r_\mathrm{e}$) as the luminosity-weighted radius of all unmasked pixels. Here, we do not use the maximum ENLR size, 
because it would be more likely affected by the low surface brightness features that are very sensitive to the depth and seeing of the 
observations.

\begin{figure}
 \resizebox{\hsize}{!}{\includegraphics[clip]{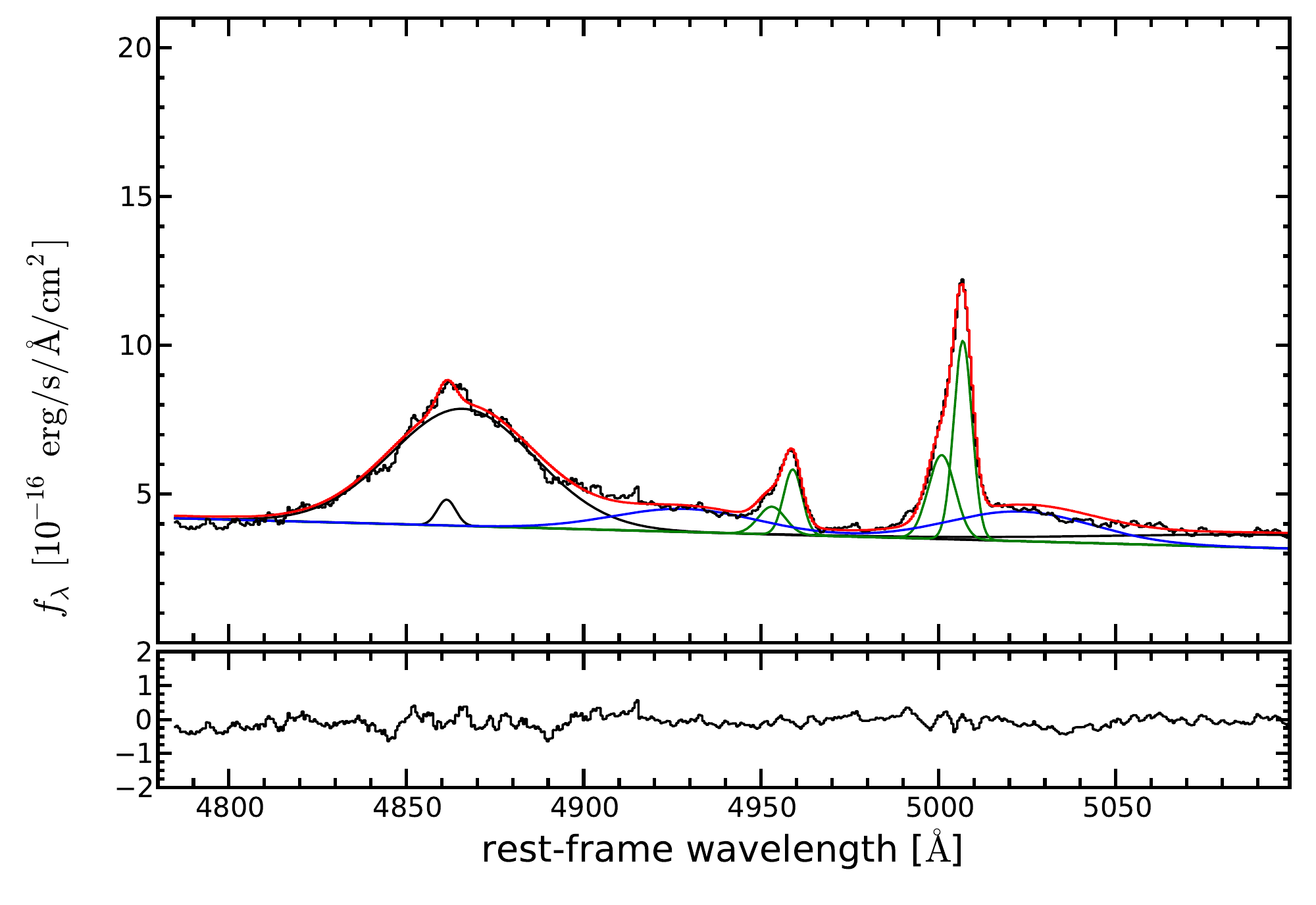}}
  \caption{Multi-component fit to the QSO spectrum of HE~1043$-$1346 in the H$\beta$-\Ox\ spectral region. The best-fitting model is indicated by the red solid line and the individual Gaussian components are plotted above the local linear continuum with the following colour coding: black\,--\,H$\beta$ line, green\,--\,\Ox\ doublet lines, blue\,--\,\Fe\ doublet lines. The residuals of the model are shown in the lower panel  with a refined scaling.}
 \label{fig:qso_example}
 \end{figure}

In several cases,  the extended emission lines are solely dominated by \HII-like emission even close to the QSO at our spatial resolution. 
We provide upper limits on the ENLR size in these cases based on the smallest distance to a detected \HII-like region. The objects
HE~1201$-$2409 and HE~1315$-$1028 were excluded from this analysis because the ionization source of the ISM could not be unambiguously 
constrained. The ENLR of HE~1434$-$1600 is only partially covered by our observations and a robust size could also not be estimated.

The deblended datacubes also provide high S/N spectra of the QSOs that are uncontaminated from host galaxy emission. For the purpose of this 
paper, we only inferred the QSO continuum luminosity at 5100\AA\ ($L_{5100}$) and the \Ox\ line luminosity ($L_{\mathrm{[\textsc{oiii}]}}$) 
from the spectra. We modelled the spectral region around the broad \Hb\ line with a set of Gaussians for the narrow and broad emission lines 
plus an underlying local continuum as described in \citet{Husemann:2013a}. Two Gaussians were usually required to account for the asymmetry of 
the \Ox\ line profile. An example of the QSO spectral modelling is shown in Fig.~\ref{fig:qso_example} for illustration. The measured QSO 
continuum luminosities, integrated \Ox\ luminosities of the QSO and ENLR  are reported in Table~\ref{tab:EELR_measurments} together with the 
effective ENLR sizes.

 \begin{table}
 \begin{footnotesize}
 \caption{QSO and ENLR properties}
 \input{tab3.tex}
 \label{tab:EELR_measurments}
 \end{footnotesize}
 \end{table}

\subsection{Gas-phase oxygen abundances}
\subsubsection{\HII-like regions}
Oxygen abundances of \HII\ regions can be estimated from various empirically calibrated emission-line ratios of strong lines, the so-called 
strong-line method. Commonly used strong-line calibrations are the $R_{23}$ index \citep[e.g.][]{Pagel:1979,Edmunds:1984,Dopita:1986,
Pilyugin:2001,Pilyugin:2005}, the N2 index \citep{Storchi-Bergmann:1994,Denicolo:2002,Pettini:2004} and the O3N2 index 
\citep{Alloin:1979,Pettini:2004}. Since the [\textsc{oii}]\,$\lambda3727$ is not covered within the wavelength range of our observations,
we could not compute the $R_{23}$ index. The N2 index is based only on the \Ni\ and \Ha\ lines, but is not very sensitive at high metallicities. 
Therefore, we used the O3N2 index and the linear calibration derived by \citet{Pettini:2004} (hereafter PP04) as the most suited oxygen 
abundance calibrator for our available set of emission lines. The systematic uncertainty of $\sim$0.15\,dex for this calibration is 
significantly larger than other methods because the ionization parameter cannot be taken into account. On the other hand, it has the 
advantage of being single-valued in contrast to the $R_{23}$ index.

All estimated oxygen abundances need to be used with care as every method has its strengths and weaknesses, which are still heavily debated in 
the literature \citep[e.g.][]{Pagel:1980,Kennicutt:1996,Kewley:2002,Perez-Montero:2005,Kewley:2008,Lopez-Sanchez:2012}.
Comparing oxygen abundances obtained with different calibrations  reveals large systematic offsets up to 0.5\,dex in
$\log(\mathrm{O}/\mathrm{H})$ \citep{Liang:2006,Kewley:2008}. Transformations between the different calibrations, in particular with the
O3N2 index, were determined by \citet{Kewley:2008}.  We scaled our inferred PP04 oxygen abundances to the calibration adopted by 
\citet{Tremonti:2004} (hereafter, T04) as a reference using the correction function from \citet{Kewley:2008}. This allows us to 
compare the oxygen abundances of our QSO host galaxies with respect to the stellar mass-metallicity relation as presented by T04 for star 
forming SDSS galaxies. Our approach is valid here, because we are only interested in a relative comparison between the oxygen abundances 
rather than in its absolute value.

\subsubsection{AGN-ionized regions}
Gas-phase element abundance calibrations for the AGN photoionized ENLR are largely unexplored. Currently, only photoionization models can 
be used to quantitatively estimate the element abundance of the gas in that case. \citet{Storchi-Bergmann:1998} computed the oxygen abundances 
for an artificial grid of line ratios with the photoionization code \texttt{CLOUDY} using an empirical AGN spectrum for the ionizing source. 
They estimated a calibration based on the $\Ox\lambda\lambda\,4960,5007/\Hb$ and $\Ni\lambda\lambda\,6548,6583/\Ha$ line ratios by fitting a 
two-dimensional polynomial of second-order to the resulting grid of oxygen abundances with an additional dependence on the electron density
($n_\mathrm{e}$).

For a rough estimate of $n_\mathrm{e}$, we employed the density-sensitive \Su\,$\lambda\lambda6716,6731$ doublet line ratio 
\citep[e.g.][]{Osterbrock:2006}. The inferred electron densities in the ENLR range between $100$--$300\,\mathrm{cm}^{-3}$, so that we 
adopted an electron density of $n_\mathrm{e}=200\,\mathrm{cm}^{-3}$ for the oxygen abundance calibration of \citet{Storchi-Bergmann:1998}. 
We emphasize that a change of $n_\mathrm{e}$ by a factor of 2  would alter the oxygen abundance by 0.03\,dex only.

Since the calibration of \citet{Storchi-Bergmann:1998} was  based on \texttt{CLOUDY} photoionization models, the oxygen abundances are 
comparable to the ones based on the T04 calibration for star-forming galaxies. One remaining systematic uncertainty is introduced by the 
choice of the adopted AGN ionizing spectrum. \citet{Storchi-Bergmann:1998} found that oxygen abundances are up to 0.5\,dex larger when the 
AGN spectrum is approximated by a pure power-law function. Here, we empirically tied the ENLR oxygen abundances with those of \HII-like regions 
in the host galaxies of HE~0952$-$1552 and HE~1405$-$1545 in which both could be determined at a similar distance from the galaxy centre. We 
found a systematic difference of $\approx$0.2\,dex between the estimate oxygen abundances of the ENLR and the \HII-like regions and corrected 
then ENLR oxygen abundances accordingly. 

\section{Results and Discussion}\label{sect:results}
\subsection{The size\,--\,luminosity relation for the ENLR}
\begin{figure*}
\includegraphics[width=0.48\textwidth,clip]{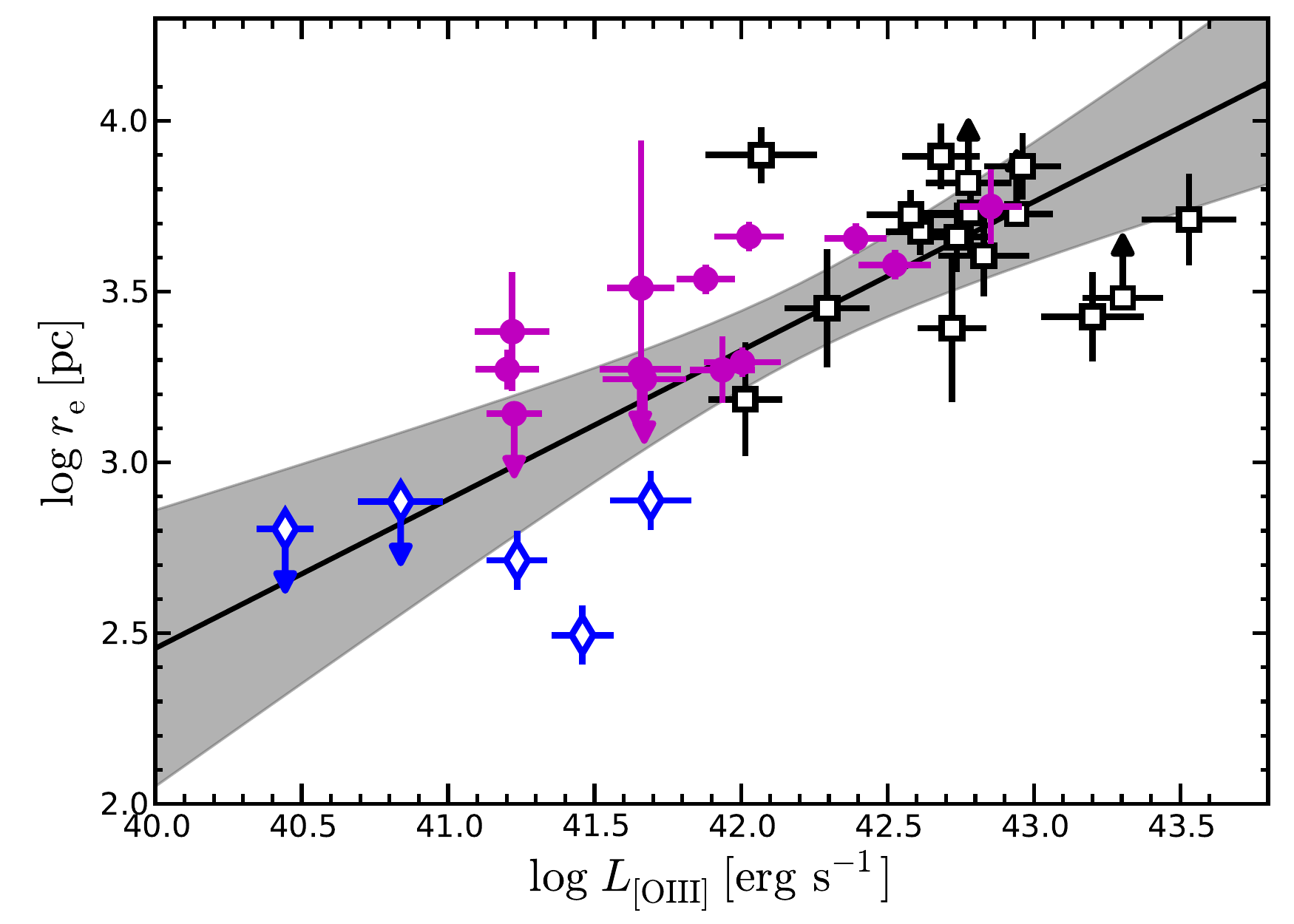}
\includegraphics[width=0.48\textwidth,clip]{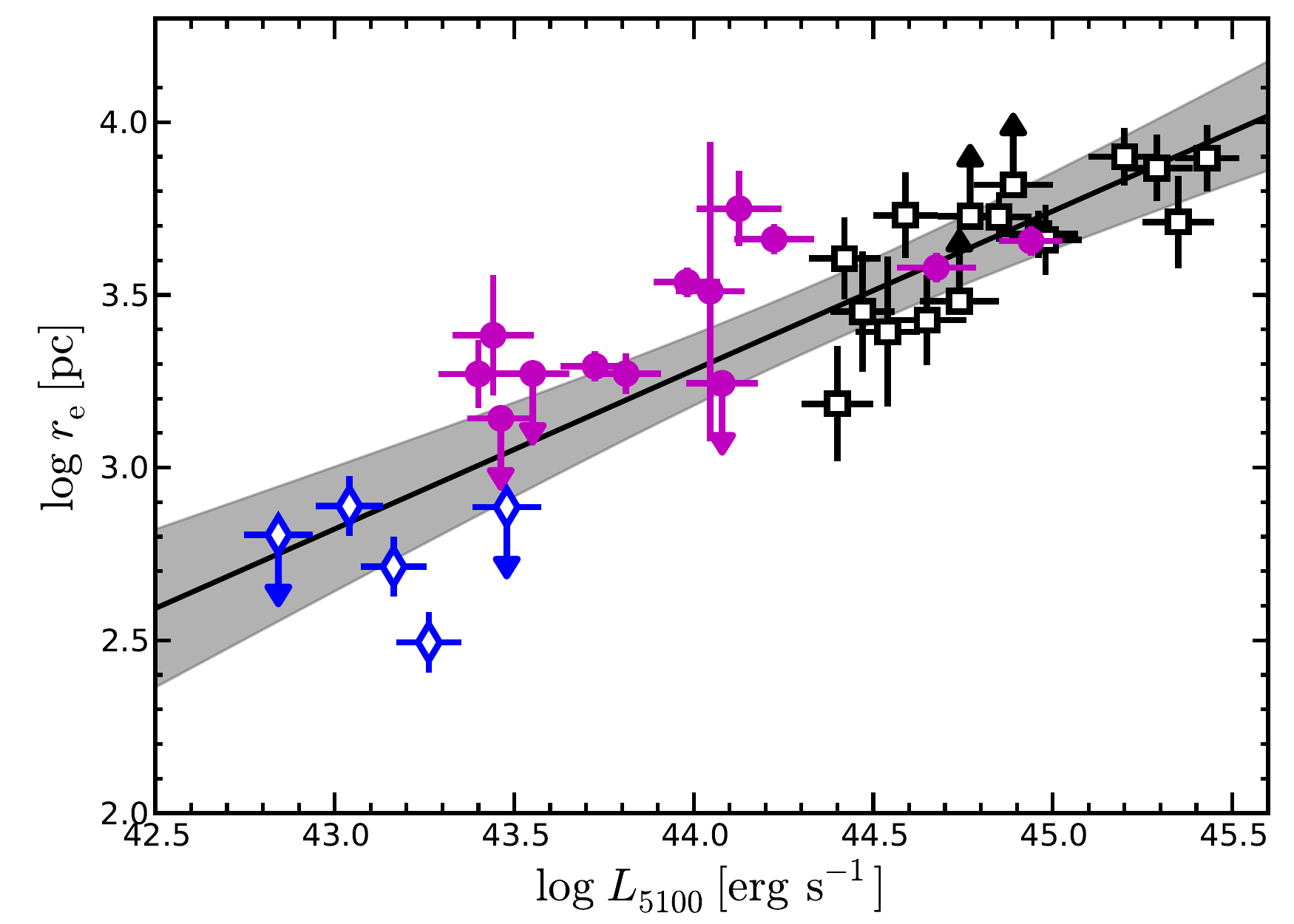}
\caption{ENLR size as a function of integrated \Ox\ luminosity (left panel) or AGN continuum luminosity at 5100\AA\ (right panel). 
The data from this sample are shown as filled circles in magenta colour. Additional measurements for luminous type-1 QSOs \citep{Husemann:2013a} 
and from 5 Seyfert 1 galaxies (Kupko et al. in prep.) are shown as black open squares and blue open diamonds, respectively. 
Arrows indicate upper limits on the ENLR size. The solid line represent the linear relations inferred from the Buckely-James method.
The shaded areas correspond to the 1$\sigma$ error of the relation which we derived with a bootstrap approach.}
\label{fig:EELR_sizes}
\end{figure*}

In \citet{Husemann:2013a}, we studied the size of the ENLR for a sample of luminous QSOs. The covered luminosity range of that sample alone
was too narrow to reliably constrain the slope of a presumed  ENLR size-luminosity relation. With the sample of type 1 QSOs presented in this work,
we significantly extend the range towards lower luminosities allowing us to investigate this relation in more detail. Additionally, VIMOS IFU 
data for five Seyfert 1 galaxies (Fairall~265, Fairall~51, Mrk~915, NGC~3783, NGC~4593) are available (Kupko et al. in prep.), which further 
complement our sample. In Fig.~\ref{fig:EELR_sizes}, we compare the ENLR sizes with the integrated \Ox\ luminosities (left panel) and the 
QSO continuum luminosities at 5100\AA\ (right panel), which are both frequently used proxies for the intrinsic bolometric luminosities of AGN.

A clear correlation between the ENLR sizes and the corresponding luminosities is evident in both cases. For a proper statistical analysis of 
these correlations we took the censored data points into account. The Astronomy Survival Analysis  package  \citep[ASURV,][]{LaValley:1992} was 
specifically developed for this task and incorporates several statistical methods to deal with bi-variate censored data \citep{Isobe:1986}. 
We computed the generalized Spearman's Rho correlation coefficient and found a significant correlation between the ENLR size and the AGN continuum 
luminosity with a coefficient of 0.87 at a confidence level greater than 99.99 per cent. A somewhat smaller correlation coefficient of 0.76 was 
computed for the correlation with the integrated \Ox\ luminosity at a confidence level of 99.99 per cent.

The ENLR size-luminosity relation has usually been approximated by a power law function, $\log(R_\mathrm{ENLR}) = \alpha\log(L_\mathrm{AGN})+R_0$. 
Here, we use the effective ENLR radius ($r_e$) as described in Sec.~\ref{sect:ENLR_measurements} for the ENLR size ($R_\mathrm{ENLR}$) and the 
integrated \Ox\ luminosity ($L_{\mathrm{\Ox}}$) and the AGN continuum luminosity at 5100\AA ($L_{5100}$) as two independent estimates for the 
intrinsic AGN luminosity ($L_\mathrm{AGN}$).  A linear regression analysis taking into account censored data points is the Buckley-James method 
\citep{Buckley:1979} which yields the following parameters
\begin{eqnarray}
\log r_\mathrm{e} &=& (0.44\pm0.06)\times \log L_{[\mathrm{OIII}]} - (14.98\pm2.47)\ ,\\
\log r_\mathrm{e} &=& (0.46\pm0.04)\times  \log L_{5100} - (16.95\pm1.61)\ .	
\end{eqnarray}
We estimate the errors on the parameters with a bootstrap approach. We repeat the regression analysis for 200 samples 
containing only 80\% of the objects which we randomly selected from the parent sample. In addition, we vary the data points within the
corresponding normal distribution given by their 1$\sigma$ uncertainties.

The correlation between the ENLR size and the AGN continuum luminosity is tighter than the correlation with \Ox\ luminosity as 
reported in \citet{Husemann:2013a} and confirmed for this expanded sample. Our IFU dataset now contains more than 40 type 1 AGN and covers 3 
orders of magnitude in luminosity, which shows that the relation is consistent with a slope of $\alpha\sim0.5$ within the errors. It was 
already discussed in \citet{Greene:2011} and \citet{Husemann:2013a} that the absolute zero-point $R_0$ of this relation is difficult to compare 
among the previous studies when different depth-dependent definitions for the ENLR size are used. In the following,  we restrict the discussion 
to the slope $\alpha$ of the relation.

A slope of $\alpha\sim0.5$ for the ENLR size-luminosity relation was initially reported by \citet{Bennert:2002} and can be easiest explained 
by a constant ionization parameter for gas clouds across the ENLR. This has also been the favoured scenario for the BLR size-luminosity relation
\citep[e.g.][]{Bentz:2009a}. Because the scales of the ENLR are orders of magnitude larger than the BLR, it is still unclear whether a constant 
ionization parameter can be a realistic scenario on host galaxy scales.  A different slope of the relation would indicate that other physical 
conditions and processes play a key role. A flatter slope of $\alpha = 0.33\pm0.04$ was inferred by \citet{Schmitt:2003b} from a sample of 60 
nearby Seyfert galaxies observed with \textit{HST}. However, the \textit{HST} narrow-band images used by \citet{Bennert:2002} and 
\citet{Schmitt:2003b} were much shallower compared to current ground-based observations. Moreover, it was not possible to distinguish between
\Ox\ emission from the ENLR and \HII-like regions.

An even flatter slope was recently presented by \citet{Greene:2011} and \citet{Liu:2013} of $\alpha=0.22\pm0.04$ and $\alpha=0.25\pm0.02$, 
respectively. They used deep ground-based optical long-slit and IFU spectroscopy to map the ENLR of luminous type 2 QSO at redshift $0.1<z<0.5$ 
combined with literature data of low-redshift Seyfert 2 galaxies. \citet{Greene:2011} proposed a model where the ENLR is matter-bounded, i. e. 
the emission is limited by the gas density rather than the density of ionizing AGN photons. However, their sample is quite heterogeneous and 
collected from different samples in the literature that used different observing techniques. Recently, \citet{Hainline:2013} added 8
type 2 AGN to the sample and confirmed the flat relation between the ENLR size and [OIII] luminosity. Interestingly, they find a slope of $\alpha\sim0.5$ 
instead by using the 8$\mu$m luminosity as an alternative AGN luminosity indicator, which is consistent again with the slope we report here.

From our homogeneous sample of genuine type 1 QSOs
we infer robust upper limits for ENLR sizes for the lower luminosity QSOs where the ionization is dominated by \HII-like regions close to the QSO 
at our resolution limit. Those limits are consistent with detected ENLRs at similar luminosities and significantly smaller than assumed for the
type 2 Seyfert galaxies inferred from the  \citeauthor{Fraquelli:2003} sample. On the other hand, it still remains open how undetected ENLRs 
around some luminous type 1 QSOs, presented in \citet{Husemann:2008} and \citet{Husemann:2013a}, fit into a common ENLR size-luminosity relation 
for all AGN and what role radio jets have on the ENLR properties \citep[e.g.][]{Wu:2009,Husemann:2013a, Mullaney:2013}.

\begin{table*}
\begin{minipage}{150mm}

\caption{Integrated properties of \HII-like regions}
\input{tab4.tex}
\label{tab:SFR_mean}
\end{minipage}
\end{table*}

\subsection{Ongoing star formation in QSO host galaxies}
\subsubsection{\Ha-based star formation rates from the IFU data}
We used the integrated \Ha\ luminosity ($L_{\mathrm{H}\alpha}$) of \HII-like regions in our QSO host galaxies to estimate the current SFR, 
adopting the calibration of \citet{Kennicutt:1998},
\begin{equation}
 \frac{\mathrm{SFR}}{\mathrm{M}_{\sun}\mathrm{yr}^{-1}} = \frac{L_{\mathrm{H}\alpha}}{1.26\times10^{41}\,\mathrm{erg\,s}^{-1}}\ .\label{eq:vimos_SFR}
\end{equation}

The effect of reddening was estimated for the individual \HII-like regions in the host galaxies from the observed H$\alpha$/H$\beta$ Balmer 
decrement in comparison to the theoretically expected value of 2.86 at $T_\mathrm{e}=10\,000$K assuming case B recombination 
\citep{Osterbrock:2006}. We computed the $V$ band attenuation ($A_V$) from the Balmer decrement adopting a standard Milky way attenuation law 
\citep{Cardelli:1989} and  $R_V=3.1$. The variance weighted mean attenuation of all individual regions of the host galaxy is included in 
Table~\ref{tab:SFR_mean} to allow a global comparison of the dust attenuation between the different host galaxies and companions. We find that 
the attenuation is typically modest with $A_{\mathrm{V}}<1.7$\,mag except for a very high attenuation of $A_\mathrm{V}$\,$\sim$\,$4.1$\,mag in 
the close companion/second nucleus of the ongoing major-merger system HE~1254$-$0934. A significant difference exists between the 
bulge and disc-dominated systems in general. The mean attenuation for the bulge-dominated hosts ($\langle A_\mathrm{V}\rangle=0.9$\,mag) is on 
average only 2/3 that for the disc-dominated hosts ($\langle A_\mathrm{V}\rangle=1.4$\,mag) in our sample. 

We measure the total H$\alpha$ fluxes by summing up all the H$\alpha$ flux from \HII-like regions (including the intermediate ones) for each 
host galaxy. From the attenuation-corrected H$\alpha$ luminosities, we estimate the SFRs with Eq.~\ref{eq:vimos_SFR} and compute the specific 
SFRs ($\mathrm{SSFR}\equiv\mathrm{SFR}~{M_*}^{-1}$) with the SED-based estimates of $M_*$ for our host galaxies. All values are listed in 
Table~\ref{tab:SFR_mean} including the systematic uncertainties and offsets of the QSO host-deblending process as determined in Appendix~\ref{sect:simulations} 
from extensive simulations. For HE~1335$-$0847 we obtain the SFR from the \Hb\ luminosity, adopting the mean reddening of $A_V=0.9$\,mag for 
bulge-dominated systems because the spectral region containing the \Ha\ line was not observed for this object. Similarly, we also adopt the 
mean reddening for HE~1110$-$1910, given that the Balmer decrement suffered from an unacceptably large uncertainty in \Hb\ flux. For  all the 
other QSOs that are exclusively dominated by emission from the ENLR, we constrain robust upper limits on the SFR by integrating the entire \Ha\ 
emission which represents the maximum contribution of \HII-like emission to the ENLR. Exceptions are HE~1201$-$2409 and HE~1434$-$1600 which are 
not fully covered within the instrument FoV. 

\begin{figure}
\resizebox{\hsize}{!}{\includegraphics{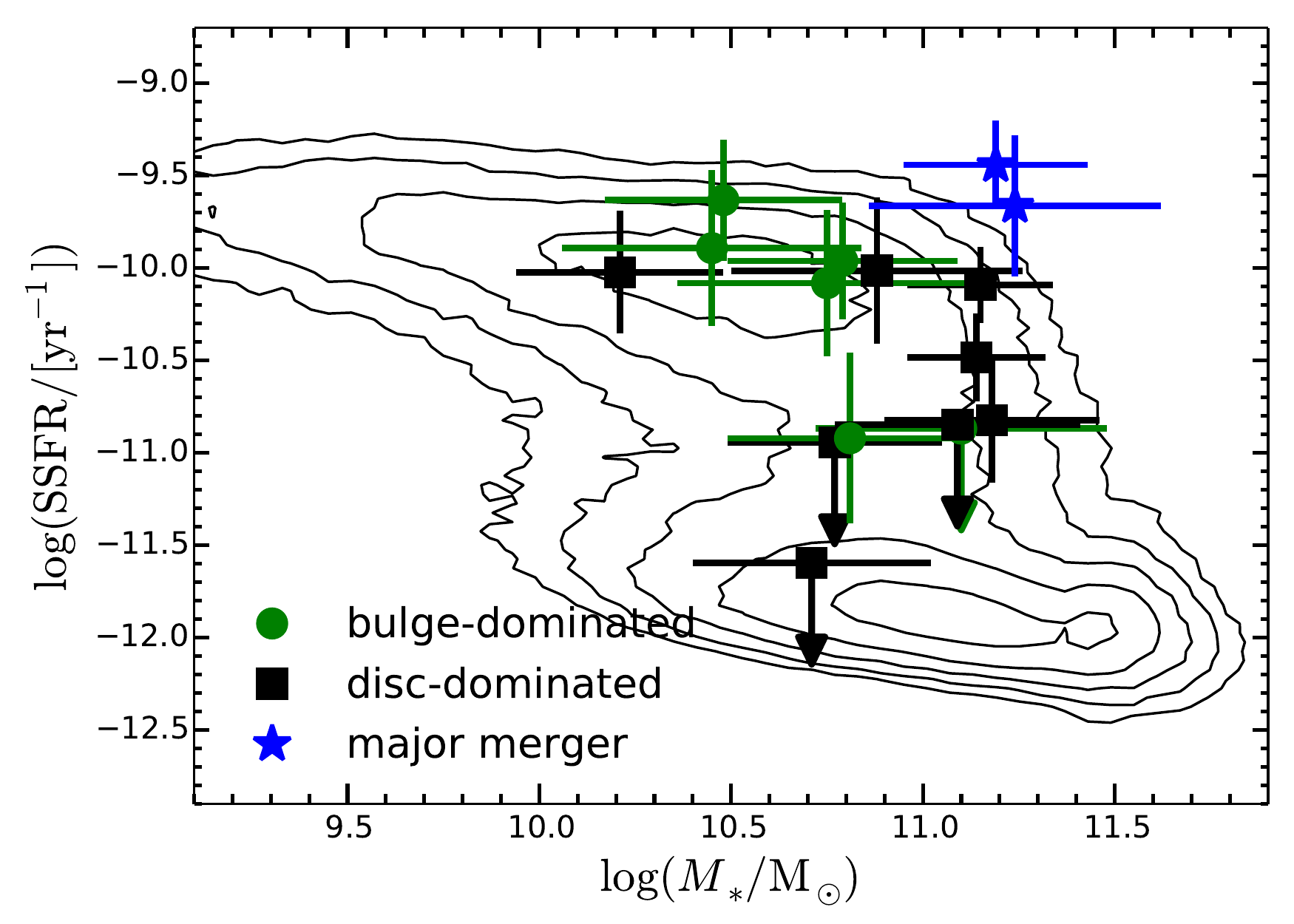}}
\caption{Specific SFR against the total stellar mass of the QSO host galaxies. The SFRs are estimated from the integrated attenuation-corrected H$\alpha$ luminosity of \HII-like region 
 (filled coloured symbols) and the 60+100$\mu$m FIR luminosity (open grey symbols) using the calibrations of \citet{Kennicutt:1998} and \citet{Bell:2003b}, respectively. Upper limits for 
the H$\alpha$ luminosity are determined for cases where the ENLR dominates the emission across the host galaxy. They are based on the integrated \Ha\ luminosity as the maximum possible 
contribution from any underlying \HII-like regions. Despite the numerous upper limits, the IR luminosity appears to be systematically contaminated by AGN emission and the H$\alpha$ 
luminosity provides a more reliable SFR indicator for these QSOs. For comparison, the distribution in SSFR of galaxies in the SDSS MPA/JHU galaxy catalogue \citep{Brinchmann:2004} is shown as 
contours in logarithmic scaling.}
\label{fig:mass_SSFR}
\end{figure}

\subsubsection{Comparison with IR and radio-based star formation rates}
The H$\alpha$-based SFRs are affected by dust attenuation and systematic uncertainties of the QSO-host deblending at our seeing-limited spatial resolution. Thus, we compare them with SFRs 
inferred from alternative IR and radio tracer. 
It was shown that the extinction-corrected H$\alpha$ luminosity agrees very well with the IR \citep{Kewley:2002b,DominguezSanchez:2012} and radio luminosity \cite[e.g.][]{Bell:2003b} in normal 
star-forming galaxies and therefore leads to consistent SFRs. For luminous AGN this is not correct, because jets can contribute to the radio luminosity and the
AGN-heated dust contribute to the 60$\mu$m and 100$\mu$m accessible with IRAS. The IR and radio luminosities can therefore only provide
upper limits on the SFR and need to be considered with care.

Here we adopt the calibrations of \citet{Bell:2003b} to estimate the SFR from the IR and radio luminosities as follows:\\
For $L_\mathrm{1.4GHz}>6.4\times10^{21}\,\mathrm{W\,Hz}^{-1}$
\begin{equation}
 \mathrm{SFR}_\mathrm{radio} = 5.52\times10^{-22}L_\mathrm{1.4GHz}
\end{equation}
and for $L_\mathrm{1.4GHz}\leq6.4\times10^{21}\,\mathrm{W\,Hz}^{-1}$:
\begin{equation}
 \mathrm{SFR}_\mathrm{radio} = \frac{5.52\times10^{-22}}{0.1+0.9(L_\mathrm{1.4GHz}/6.4\times10^{21}\,\mathrm{W\,Hz}^{-1})}L_\mathrm{1.4GHz}
\end{equation}
Similarly, for $L_\mathrm{IR}>10^{11}L_\odot$
\begin{equation}
 \mathrm{SFR}_\mathrm{IR} = 1.57\times10^{-10}L_\mathrm{IR}\left(1+\sqrt{\frac{10^9}{L_\mathrm{IR}}}\right)                
\end{equation}
and for $L_\mathrm{IR}\leq10^{11}L_\odot$
\begin{equation}
 \mathrm{SFR}_\mathrm{IR} = 1.17\times10^{-10}L_\mathrm{IR}\left(1+\sqrt{\frac{10^9}{L_\mathrm{IR}}}\right)  \quad .              
\end{equation}
$L_\mathrm{IR}$ is the total IR luminosity between $8$--$1000\mu$m. We assume that the total IR flux is a factor of $\sim1.75$ higher 
\citep{Calzetti:2000} than the FIR flux between $40$--$120\mu$m  which we compute from the IRAS fluxes given in Janskys 
as $\mathrm{FIR} = 1.26\times10^{-14}(2.58\mathrm{F}_{60\mu\mathrm{m}}+\mathrm{F}_{100\mu\mathrm{m}})\,\mathrm{W\,m}^{-2}$ \citep{Helou:1988}.

The IRAS IR luminosity or upper limits and NVSS radio luminosity often correspond to SFRs for our AGN that are several factors higher than the ones derived from the 
H$\alpha$ luminosity (see Fig.~\ref{fig:mass_SSFR}). The radio-based SFRs are also often lower than the IR-based ones which indicates that the AGN contribution to the FIR is significant 
despite the fact that the radio flux itself is
contaminated by additional flux from jets. Thus, the spatially resolved extinction-corrected H$\alpha$ luminosity appears to be the most reliable SFR tracer for our AGN host galaxies. It also offers
the highest spatial resolution compared to IR and radio surveys.

\subsubsection{Discussion of specific star formation rates}
Although the attenuation correction significantly increases the uncertainty for the estimated SFRs, the SFRs of our AGN host galaxies 
typically range between 1\,$\mathrm{M}_{\sun}\,\mathrm{yr}^{-1}$ and 10\,$\mathrm{M}_{\sun}\,\mathrm{yr}^{-1}$. However, the absolute SFR is 
well correlated with stellar mass, so that the SSFR is nearly constant for the bulk population of disc-dominated galaxies. We compare the 
SSFR as a function of stellar mass for our QSO host galaxies in Fig.~\ref{fig:mass_SSFR} with the overall distribution of star forming galaxies
within SDSS presented by \citet{Brinchmann:2004}. We find that the majority of our disc-dominated system shows signatures of ongoing star
formation and  are consistent with the main sequence of star-forming galaxies. Three of them have upper limits placing them below the main 
sequence of which HE~1315$-$1028 is an extreme case. This galaxy does not show any detectable emission of ionized gas except of an ENLR at 
its outskirts and therefore is very likely a gas poor system compared to the rest.  The bulge-dominated QSO hosts appear to have diverse 
properties. Two bulge-dominated QSO hosts, HE~1228$-$1637 and HE~1300$-$1325 are located on the main sequence, whereas HE~1335$-$0847 and 
HE~1110$-$1910 display roughly an order of magnitude lower SSFRs compared to the main sequence. Still they are certainly not consistent 
with the SSFR expected for bulge-dominated galaxies on the red sequence, which is unclear for HE~1029$-$1401 as its bright and extended 
ENLR makes it impossible to infer the contribution from star formation to the \Ha\ luminosity. On the other hand, the SSFR is significantly enhanced
for the two ongoing major mergers in our sample. To summarize, we find that 2 out of 16 QSO hosts (13 per cent) exhibit 
enhanced star formation, 8 of 16 QSO hosts (50 per cent) are consistent with the main sequence of star-forming galaxies, and 6 of 16 QSO 
hosts (38 per cent) show at least intermediate levels of star formation between the main sequence and the quiescent red-sequence or even lower.

Previous attempts to characterize the SFR of luminous QSOs used the [\textsc{oii}] line in the integrated spectrum as a SFR indicator.
For example, \citet{Ho:2005} and \citet{Kim:2006} found that most of the [\textsc{oii}] emission in a sample PG and SDSS type-1 QSO 
originates mainly from the NLR with a low SFR and therefore conclude that star formation is generally suppressed in QSO host galaxies. 
On the contrary, \citet{Silverman:2009} used the same technique to infer the SFR for a large X-ray selected sample of AGN and found
that their host galaxies exhibit on average a higher SFR compared to normal galaxies at a given stellar mass. Such enhanced levels 
of star formation are also found in studies that used the mid- and far-infrared wavelength to disentangle the relative contribution 
of AGN and star formation \citep[e.g.][]{Schweitzer:2006,Netzer:2007,Lacy:2007}. They additionally report that the SFR is correlated 
with the QSO luminosity.  

Only the ongoing major mergers in our sample exhibit enhanced SSFRs, but for a large fraction of the undisturbed QSO host galaxies 
the SSFR is consistent with those of inactive galaxies. This pictures is in agreement with recent results from far-infrared studies
of AGN host galaxies based on Herschel data \citep{Santini:2012,Rosario:2012}. On the other hand, we also find that six QSO hosts have 
intermediate or very low SSFRs based on the H$\alpha$ luminosity independent of their morphological type.  Either the star formation 
efficiency in the galaxy are directly reduced by the presence of the AGN or the intrinsic gas content is lower compared to the other
galaxies on galaxy-wide scales. This is something we cannot address with this dataset alone as the information on the molecular gas content
is missing. HE~1310$-$1051 and HE~1338$-$1423 were observed with the IRAM 30m telescope in CO emission so far \citep{Bertram:2007}, but only 
upper limits on the molecular gas masses of around $<0.5\times 10^9\,M_\odot$ could be inferred. The reasons for the non-detection of clear
\HII-like regions therefore remain open. However, AGN feedback generally cannot be present, or must be significantly delayed after the onset 
of the QSO phase, considering the large fraction of galaxies with normal or enhanced levels of star formation. The construction of control 
samples of inactive galaxies with matching galaxy properties is therefore crucial to infer the immediate impact of AGN.

\begin{figure*}\centering
\includegraphics[width=\textwidth]{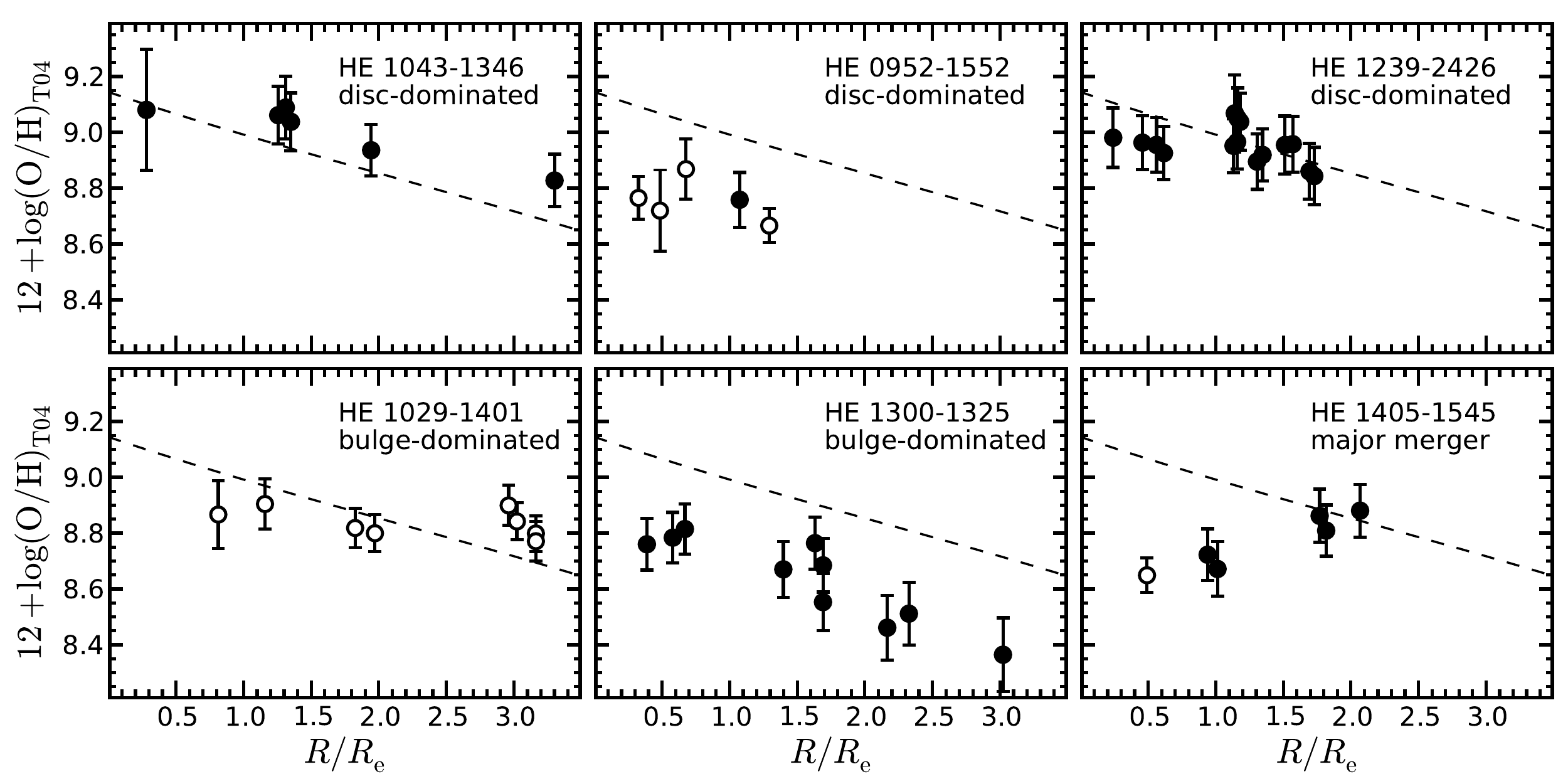}

\caption{Oxygen abundance as a function of effective radius for the six QSO host galaxies with sufficient radial coverage. Filled symbols denote abundance estimates based on \HII-like regions using the O3N2 index from \citet{Pettini:2004}, whereas open symbols refer to abundances based on the NLR adopting the calibration of \citet{Storchi-Bergmann:1998}. The radial oxygen abundance gradient derived from 38 face-on disc-dominated galaxies \citep{Sanchez:2012b} is shown as the dashed line for comparison. All abundances are rescaled to match the \citeauthor{Tremonti:2004} calibration applying the correction derived by \citet{Kewley:2008}.}
\label{fig:metal_grad}
\end{figure*}

\subsection{Oxygen abundances distribution in QSO host galaxies: Signature for internal processes or external gas supply?}
We construct the gas-phase metallicity gradients for six of our QSO host galaxies with sufficient radial coverage (Fig.~\ref{fig:metal_grad}).
The galactocentric distances are normalized by the effective radius ($R_\mathrm{e}$) of the corresponding host galaxies to put them on a comparable scale.
For undisturbed inactive late-type galaxies, negative gas-phase metallicity gradients were recovered in numerous spectroscopic 
studies of the \HII\ regions \citep[e.g.][]{McCall:1985,Vila-Costas:1992,Zaritsky:1994,Henry:1999,Kennicutt:2003,Magrini:2007,
Viironen:2007, Sanchez:2012b}. A reference for a statistically robust oxygen abundance gradient for disc-dominated galaxies was recently provided
by \citet{Sanchez:2012b} based on data from the PINGS \citep{Rosales-Ortega:2012} and the CALIFA survey \citep{Sanchez:2012a}, which we indicated
as a dashed line in Fig.~\ref{fig:metal_grad} for comparison.

These few QSO host galaxies with recovered gas-phase metallicity gradients do not represent a statistically meaningful sample. Nevertheless, we
note that the late-stage merger HE~1405$-$1545 exhibits a strong positive gradient. This is in contrast to the disc-dominated galaxies which are statistically
consistent with the expected negative gradient despite a global metallicity offset. The results are less clear for the two bulge-dominated QSO host galaxies.
Although HE~1300$-$1325 appears to have at least a similar slope than the disc-dominated ones, the gas-phase metallicity remains constant up to 1.5 $R_\mathrm{e}$ and exhibits 
a pronounced drop in the oxygen abundance at this location which mimics a global negative gradient. The galaxy is in a very early interaction phase with its major companion 
and  the gradient is a superposition of an inner rotating gas disc and a kinematically detached tidal arm that
extends out to $3\,R_\mathrm{e}$ radii.  On the other hand, the abundance gradient  in the bulge-dominated galaxy HE~1029$-$1401 appears to be flat as already discussed in \citet{Husemann:2010}. A 
caveat could be that it is solely based on the oxygen abundance measurements from the ENLR which might be prone to other systematic effects, i. e. QSO luminosity, as 
pointed out by \citet{Stern:2013}.

\begin{table}
\caption{Gas-phase oxygen abundances of QSO host galaxies}
\input{tab5.tex}
\label{tab:OH_re}
\end{table}

To take better advantage of the sample size, we measure the oxygen abundance at $\sim$1$R_\mathrm{e}$ 
(Table~\ref{tab:OH_re}) as the average of abundances between $0.75R_\mathrm{e}<R<1.25R_\mathrm{e}$.  This measurement can be done for the 
majority of our objects. In  the case of HE~0952$-$1552, HE~1029$-$1401, HE~1043$-$1346, HE~1239$-$2426, and HE~1300$-$1325 the 
radial distribution is sufficient to estimate a linear relation, which we evaluate at  $1R_\mathrm{e}$ to provide a consistent measurement 
with respect to the other objects. Assuming the slope of the radius-metallicity relation from \citet{Sanchez:2012b}, we extrapolated the 
oxygen abundance at $1R_\mathrm{e}$ towards the centre of the host galaxies.

\begin{figure}
\resizebox{\hsize}{!}{\includegraphics{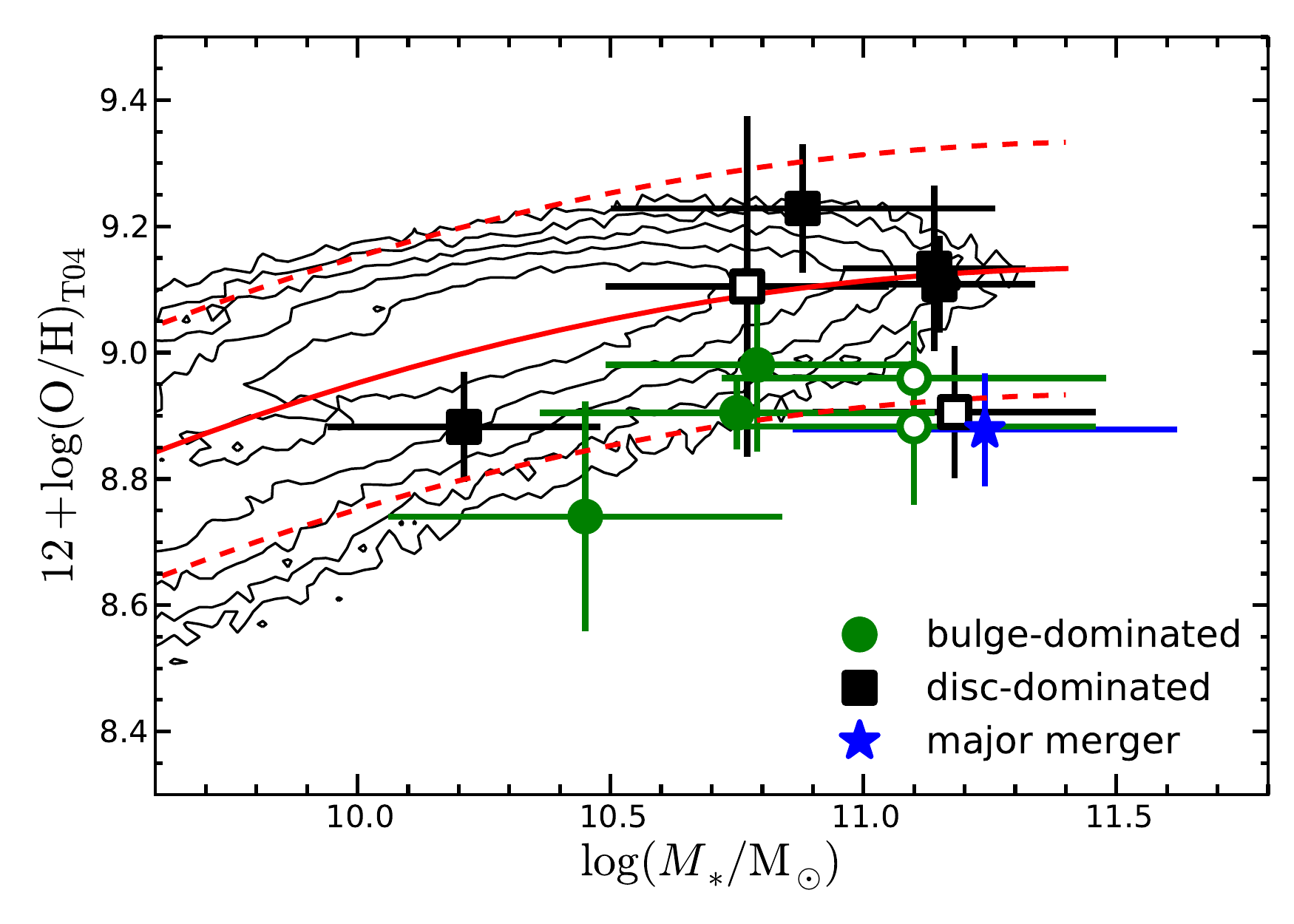}}
\caption{Oxygen abundance against the total stellar mass of QSO host galaxies. The oxygen abundances are extrapolated from 1$R_\mathrm{e}$ towards the galaxy centre adopting the radius-metallicity 
slope of disc-dominated inactive galaxies presented by \citet{Sanchez:2012b}. Open and filled symbols correspond to oxygen abundance measurements based on the strong line NLR calibration of 
\citet{Storchi-Bergmann:1998} and the O3N2 calibration of \citet{Pettini:2004}, respectively. The symbols denote different galaxy morphologies as defined in Fig.~\ref{fig:mass_SSFR}.  The distribution 
in oxygen abundances of star forming galaxies in the SDSS MPA/JHU galaxy catalogue \citep{Tremonti:2004} is shown as contours in a logarithmic scaling for comparison. The red solid line represents the 
best-fitting line to the SDSS data  with the dashed lines indicating roughly its $3\sigma$ uncertainty.}
\label{fig:mass_metal}
\end{figure}

We compare the extrapolated central oxygen abundances against the stellar masses determined from the broad-band SED of the galaxies in
Fig.~\ref{fig:mass_metal}. Our disc-dominated QSO host galaxies follow the mass-metallicity relation of galaxies \citep{Tremonti:2004} 
whereas the bulge-dominated ones have systematically lower oxygen abundances at a given stellar mass. Interestingly, the bulge-dominated 
host HE~1416$-$1256 and the major merger HE~1405$-$1545 exhibit particularly low oxygen abundances that are well below the mass-metallicity 
relation. The significance of our result might be reduced by the systematic uncertainties of 0.15\,dex for the strong-line oxygen abundance 
calibrators. However, such uncertainties must \emph{randomly} affect the overall sample and not selectively affect only a certain morphological 
type, except when there is a physical reason, e. g. when the ionization parameter for the \HII-like regions depends on the host morphology. 

Given that the general mass-metallicity relation is almost flat in the high-mas regime, it is ruled out that the lower abundance can be explained 
only with a strong offset in stellar mass. It would require the stellar masses to be overestimated by an order of magnitude to be on the 
relation, which is beyond the uncertainties of our measurements and cannot be explained by the doubling of mass during a 1:1 merger. However, 
it is likely that a dilution in oxygen abundance and the immediate mass increase of the merger act together in this respect. The measurement
of the effective radius can also be affected by systematic uncertainties. In particular, for bulge-dominated systems the effective radius can 
be systematically underestimated during the QSO-host deblending caused by flux transfer from the QSO to the host galaxy profile. This would 
reduce the distance of our \HII-like regions with respect to the effective radius and lower the extrapolated central oxygen abundance even 
further compared to the disc-dominated system in favour of our conclusions.

One difficulty to interpret our results is that the gas-phase metallicities of inactive elliptical and bulge-dominated galaxies are not well 
explored. It is known that more than 60 per cent of the elliptical galaxies display detectable emission lines of warm ionized gas 
\citep[e.g.][]{Trinchieri:1991,Goudfrooij:1994,Macchetto:1996,Sarzi:2006}. The gas is often concentrated at the nucleus with LINER-like
emission which appears to be predominately ionized by post-AGB stars or a low-luminosity AGN rather than ongoing star formation. 
Metallicities of elliptical galaxies were mainly obtained from their hot gas phase and their stellar population \citep[e.g.][]{Trager:2000a}.  
Only a few recent studies tried to infer the metallicity of the warm-ionized gas,  concentrating on the galaxy centre. \citet{Annibali:2010} 
found that the gas-phase oxygen abundances in a sample of 65 inactive bulge-dominated galaxies were systematically lower than the corresponding 
stellar abundances. They attributed this discrepancy either to a systematic mismatch between the gas-phase and stellar abundance calibrations or 
to the accretion of external gas, likely via minor mergers. Additional evidence for the external origin of the gas in a large fraction of  
bulge-dominated galaxies comes from the misaligned kinematics of ionized gas \citep[e.g.][]{Caon:2000, Sarzi:2006} and molecular gas 
\citep{Davis:2011} with respect to the stars. Contrarily, \citet{Athey:2009} argued that the solar metallicity gas in elliptical galaxies
is unlikely to be accreted from the environment, and they favoured mass loss from AGB stars as the origin of the gas.

It is therefore intriguing to find that the bulge-dominated QSO host galaxies and the systems with ongoing interactions are offset 
from the mass-metallicity relation. While the assumed radius-metallicity slope introduces an additional uncertainty, the reported 
trend can only be removed if the slope of the metallicity gradient are much steeper. However, the opposite seems to be the case for 
interacting galaxies \citep{Kewley:2010,Rupke:2010,Rich:2012}. Therefore, we interpret the lower oxygen abundances as an indication of 
metal dilution through gas inflow by external gas accretion or tidal forces during strong galaxy interactions. 

\citet{Letawe:2007} also inferred oxygen abundances for their sample of luminous QSOs and found typically low abundances 
($12+\log(\mathrm{O/H})<8.6$) for their host galaxies. They attributed the lowest abundances to those host galaxies with  
distorted morphologies, and indeed observational and theoretical evidence is mounting that interactions can be responsible for the
significant dilution of metals. For example, the central oxygen abundance is statistically lower by 0.05-0.1\,dex in galaxies with close companions 
compared to isolated galaxies \citep{Kewley:2006,Ellison:2008}. Furthermore, \citet{Peeples:2009} and \citet{Alonso:2010} found 
that the majority of extreme low-metallicity outliers from the mass-metallicity relation are distorted galaxies indicative of recent or 
ongoing interactions. The infall of gas is also an interpretation to explain the fundamental relation between mass, metallicity and SFR in galaxies \citep[e.g.][]{Mannucci:2010},
in which galaxies with higher SFR at a given stellar mass have lower oxygen abundance. \citet{Montuori:2010} showed with numerical simulations of galaxy mergers that a dilution in the central 
metallicity is expected to occur at various stages of galaxy interactions, even during fly-bys. However, the efficiency with which triggered star formation 
enriches the gas with metals again needs to be taken into account \citep{Torrey:2012}. Finally, \citet{Yates:2012} combined observations and
semi-analytic models to show that massive galaxies with lower oxygen abundance had experienced a major merger in the past exhausting all the gas, 
but that subsequent accretion of low-metallicity gas leads to the gradual decline in abundances at the present time. 

The systematically lower oxygen abundance of our bulge-dominated QSO host galaxies with respect to disc-dominated counterparts (at a given 
stellar mass) might be a common feature that coincides with ongoing star formation as the likely cause for their exceptionally blue colours 
\citep{Jahnke:2004b}. The gas inflow of low-metallicity gas on galaxy wide scales towards the centre of bulge-dominated systems might also be
responsible for triggering the current phase of black hole accretion. Whether our bulge-dominated systems are currently in a late stage of a 
major merger after final coalescence, during a minor merger, or in a phase of gas accretion from the environment is something we cannot 
address with the current data. Contrary, the matching oxygen abundances and radial gradients for several disc-dominated QSO hosts with 
their inactive counterparts on galaxy wide scales suggest that their evolution is driven by internal galaxy processes that may also lead 
to the accretion on to the black hole.

\section{Summary and Conclusions}\label{sect:conclusions}
In this paper we presented a comprehensive spatially resolved spectroscopic analysis of a flux-limited sample of QSOs at low redshift ($z<0.2$). 
It is the first paper of a series investigating in detail the properties of these QSO host galaxies. Here, we focused on the characterization 
of the spatially resolved ionized gas properties across the galaxies. The main results can be summarized as follows:
\begin{itemize}
 \item[1.] All QSO host galaxies exhibit ionized gas with a variety of ionization mechanisms from classical \HII\ regions to  ENLRs extending 
 across the entire host galaxy.
\item[2.] The ENLR size, expressed as the \Ox\ luminosity-weighted ENLR radius ($r_\mathrm{e}$), strongly correlates  with the intrinsic AGN 
luminosity as traced either by the continuum luminosity at 5100\AA\ or by the integrated \Ox\ luminosity. The scatter in the relation based on
the \Ox\ luminosity is significantly larger compared to the continuum luminosity, which indicates that the \Ox\ luminosity does not solely
depend on the AGN luminosity.  We inferred a best-fitting relation of $r_\mathrm{e}\propto L_{5100}^{0.46\pm0.04}$. 
 \item[3.] In more than 50 per cent of the host galaxies, irrespective of their morphology, we find ionized gas from \HII-like regions 
 indicating ongoing star formation. The specific star formation rates (SSFRs) based on the dust-corrected H$\alpha$ luminosity are consistent 
 with those of the star forming main sequence for most of these QSO hosts. Comparison with the SSFRs based on the 60$\mu$m+100$\mu$m FIR luminosity
 suggests that the FIR luminosity is contaminated by AGN emission and H$\alpha$ is a more robust and sensitive tracer for the current SFR.
 Significantly enhanced star formation is thus rare and always associated with strong galaxy interactions. The upper limits on the SSFR 
 for the rest of our sample, including disc-dominated galaxies, place them clearly below the main sequence. Either their overall gas content is 
 systematically lower or the star formation is possibly suppressed as a consequence of AGN feedback.
\item[4.] For a subsample of six QSO hosts we construct radial oxygen abundance gradients for the first time. The major merger
exhibit a strong positive metallicity gradient in contrast to the negative gradients expected for undisturbed inactive disc-dominated QSO hosts.
While the negative gradients in three disc-dominated hosts are statistically consistent with the reference slope despite a global metallicity offset, 
the slopes for the two bulge-dominated host galaxies are inconclusive.
\item[5.] Bulge-dominated QSO host galaxies exhibit systematically lower gas-phase oxygen abundances compared to their disc-dominated 
counterparts, placing them below the general mass-metallicity relation of galaxies. We interpret this as evidence for recent minor or 
advanced major mergers stages, which are causing the current metal dilution in these system.
\end{itemize}

The properties of the ionized gas draw a quite diverse picture of the population of nearby QSO host galaxies. Ionization from \HII-like 
regions and AGN photoionization are found in host galaxies of different morphological types from disc- or bulge-dominated galaxies to 
strongly interacting systems. Several recent studies have suggested that there is no significant difference in the frequency of major-mergers 
between AGN and inactive galaxies \citep[e.g.][]{Cisternas:2011,Kocevski:2012,Boehm:2013}. On the contrary, a correlation between 
AGN fraction and various signatures of interaction, like close companions, ongoing mergers or post-merger systems has been reported 
\citep{Koss:2010,RamosAlmeida:2010,Ellison:2011,Bessiere:2012,Cotini:2013,Sabater:2013}. Deep AO-assisted NIR imaging of the luminous Palomar Green QSOs presented by
\citet{Guyon:2006} shows that 30\% of the galaxies are interacting systems and major mergers. One explanation for this apparent
discrepancy is that selection effects of the samples play a crucial role and interaction and major-mergers only play a dominant role
for the triggering the most luminous AGN \citep{Treister:2012}. Our measurements of the gas-phase oxygen abundances in QSO hosts 
show that disc-dominated ones have matching properties to their inactive counterparts. This indicates that the QSO activity in disc-dominated AGN host galaxies
is driven by internal galaxy processes. Interactions seem to play a crucial role not only for major-mergers, but also for the bulge-dominated QSO hosts
that may not have been fully accounted for in some imaging studies.

To understand whether and how quenching of star formation is related to AGN feedback, it will be crucial in the future to compare AGN 
with control samples of inactive galaxies that robustly match in most host galaxies properties. Our sample is too small in this respect 
given the diversity in the QSO host galaxies properties. However, the SSFRs in many of our disc-dominated QSO hosts match with those of 
inactive ones, while enhanced SSFRs are associated with mergers. It suggests that AGN  only have little effect on the global 
SFR in these systems, at least during the current QSO phase. This appears in contrast with several studies that find strong evidence
of AGN feedback. AGN-driven outflows have been identified in ULIRGs \citep[e.g.][]{Veilleux:2013,Cicone:2014} that can reach 
outflow velocities of 1000$\mathrm{km}\,{s}^{-1}$ \citep[e.g.][]{Rupke:2011}. Another indication for fast outflows are broad absorption line (BAL) QSOs
which display very broad absorption lines with line widths of several thousand $\mathrm{km}\,{s}^{-1}$ \citep[e.g.][]{Kool:2001,Chartas:2007}.

Since our sample is selected solely based on optical QSO brightness at low redshifts, it does not contain any ULIRGs. Those are ongoing major mergers with a very
high surface gas density leading to high SFR densities. The spatial scales of those outflows are often unconstrained and may
not reach beyond 1\,kpc in many cases. Despite the fact that our sample of QSOs is quite complementary to ULIRGs and BAL QSOs we are not able to resolve the central kpc 
around the QSOs. We are therefore blind to either circumnuclear starbursts or AGN-driven outflows
on kpc scales that could still be present in our systems. On the other hand, we are able to characterise the properties of the entire host galaxies
with 1-3\,kpc resolution. If AGN-driven outflows would be present in all host galaxies system, its effect need to be restricted to the central kpc 
considering that the majority of galaxies show ionized gas and ongoing star formation across the entire host galaxies.
 
Identifying signatures of interactions solely based on broad-band imaging data is often difficult given the low surface brightness of 
these features.  Our results show that the properties of the gas-phase oxygen abundances may provide additional constraints on the 
occurrence of ongoing or past galaxy interactions, even for galaxies hosting a luminous QSO. Given that the emission lines usually have
a higher surface brightness than the stellar continuum, it offers a high diagnostic value in particular with increasing redshifts. This 
will be important to construct robust AGN and inactive control samples separating the effects of internal galaxy processes and galaxy 
interactions when investigating the relation between AGN and their host galaxies. In particular for bulge-dominated galaxies, the oxygen abundance
could provide additional diagnostic power to identify ongoing or recent interactions compared to pure imaging studies.

\section*{Acknowledgements}
We thank the referee for valuable comments and specific instructions that substantially improved the clarity of the paper.
BH and LW acknowledge financial support by the DFG Priority Program 1177 ``Witnesses of Cosmic History: Formation and evolution of 
black holes, galaxies and their environment'', grant Wi 1369/22-1 and Wi 1369/22-2. Furthermore, BH, DK and LW gratefully acknowledge 
the support by the DFG via grant Wi 1369/29-1. SFS would like to thanks the 'Ministerio de Ciencia e Innovacion' project ICTS-2009-10, 
and the 'Junta de Andalucia' projects P08-FWM-04319 and FQM360. DN and KJ are funded through the DFG Emmy Noether-Program, grant JA 1114/3-1.
We like to thank J. Caruana and S. Dhawanfor checking the manuscript prior to submission.

\bibliographystyle{mn2e}
\bibliography{references}

\clearpage
\appendix
\section{Estimating systematic uncertainties of the QSO-host debelending process}\label{sect:simulations}
\subsection{Simulation of realistic IFU data}
Simulating realistic IFU data to estimate the systematic uncertainties of the QSO-host debelending process is a difficult task. Details of 
the 2D host galaxy surface brightness distribution, the 2D gas kinematics, the ionized gas distribution and brightness, 
the apparent host galaxy brightness, the nucleus-to-host ratio, the shape of QSO spectrum and the spatial resolution are all 
affecting the result of the deblending process to some extend. Since this study is the first one 
to characterize QSO host galaxies in such details, it is impossible to create mock data based on an analytic prescription of those components.

 \begin{figure}
 \resizebox{\hsize}{!}{\includegraphics[height=\hsize]{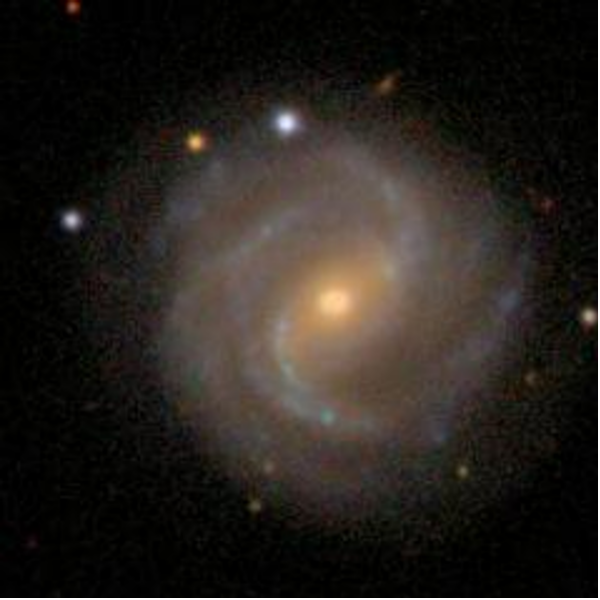} \includegraphics[height=\hsize]{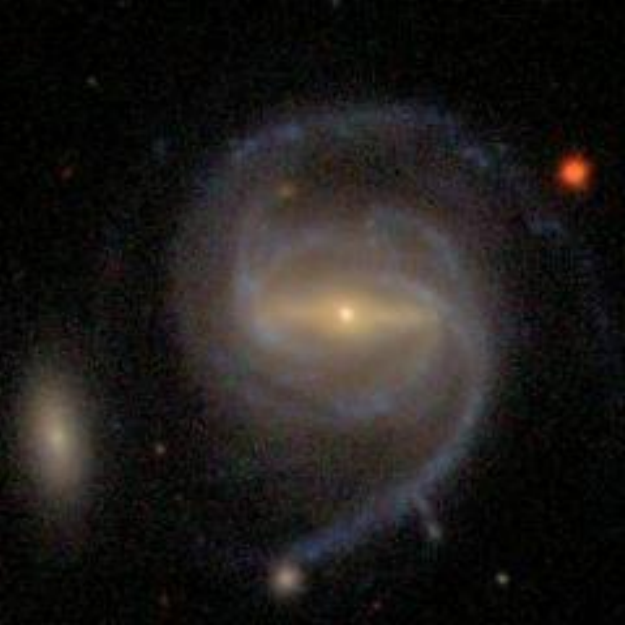}}
 \resizebox{\hsize}{!}{\includegraphics{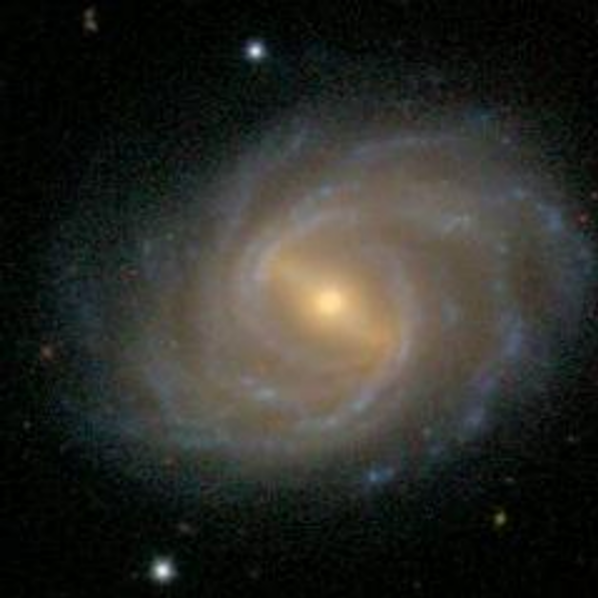} \includegraphics{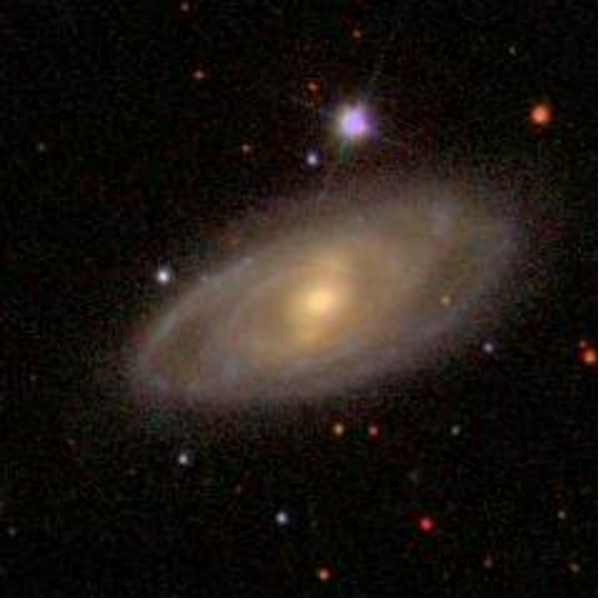}}
 \caption{SDSS colour images of CALIFA galaxies used for simulating QSO host galaxies at various redshift. The chosen galaxies are NGC~0776 (upper left), NGC~5000 (upper right),
 NGC~5406 (lower left), and NGC~6497 (lower right).}
 \label{fig:CALIFA_SDSS}
 \end{figure}

 \begin{figure}
 \resizebox{\hsize}{!}{\includegraphics{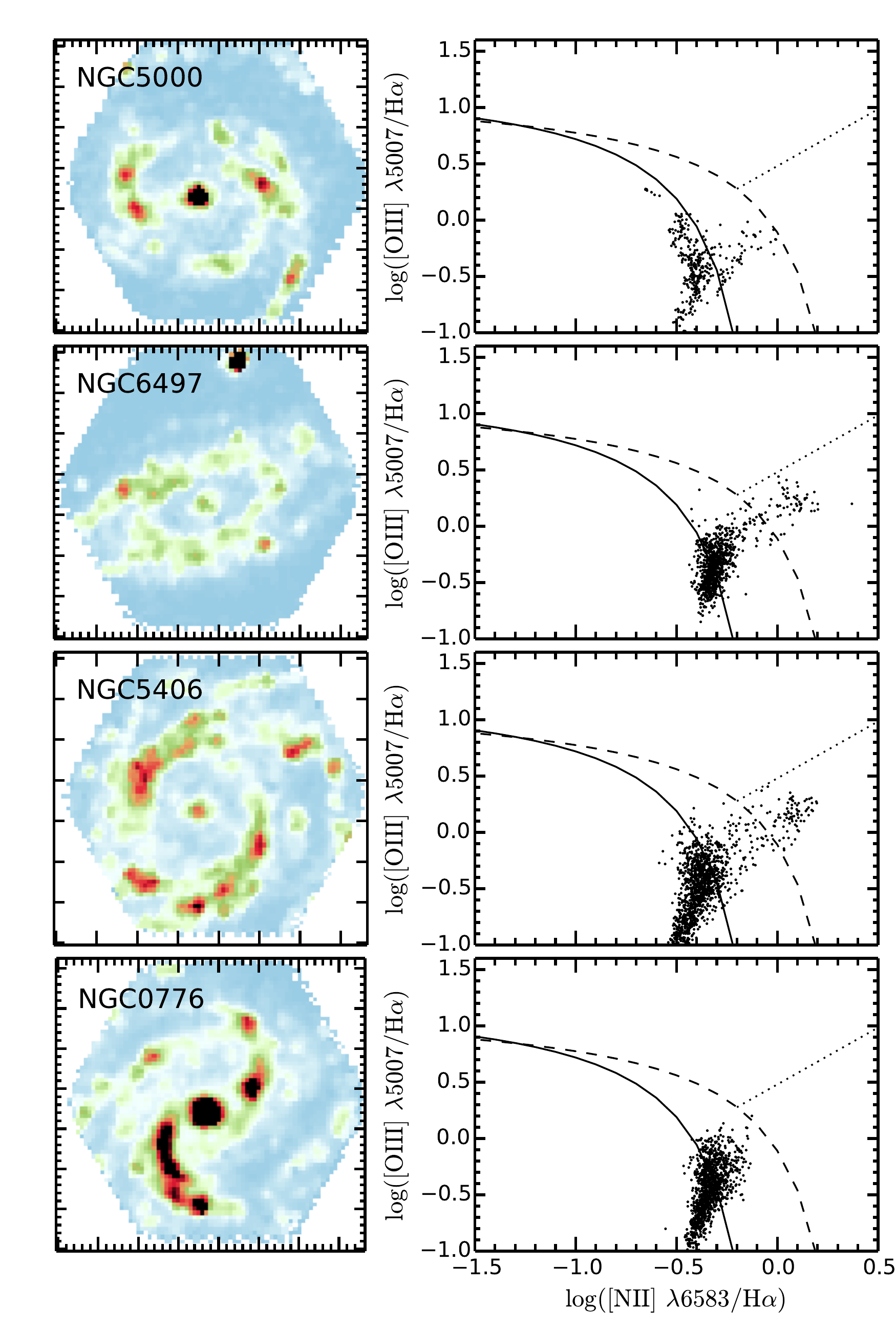}}
 \caption{\emph{Left panels:} Reconstructed H$\alpha$ narrow-band images of the CALIFA galaxies after stellar continuum subtraction. \emph{Right panels:} Standard BPT
 diagnostic diagram for all spaxels in the CALIFA cubes which have a S/N$>$3.}
 \label{fig:CALIFA_Ha_BPT}
 \end{figure}
 
Here, we took advantage of the CALIFA survey \citep{Sanchez:2012a} that already provided IFU data for 100 galaxies with its first data release 
\citep{Husemann:2013b}. We selected 4 different nearly face-on spiral galaxies (see Fig.~\ref{fig:CALIFA_SDSS}) with stellar masses 
between $2\times10^{10}M_\odot$ and $10^{11}M_\odot$ which all display significant star formation as seen through the strong H$\alpha$ emission. Those galaxies serve as our empirical
models for the QSO host galaxies simulations. Our template CALIFA galaxies are rather local galaxies with $z<0.02$ which provide much higher spatial
resolution as our VIMOS data at higher redshifts. We therefore binned the CALIFA data by 2-10 spaxels to simulate higher redshift galaxies. Each binning
corresponds to a certain redshift considering the ratio of the $1''$ CALIFA and $0.67''$ VIMOS sampling and the ratio of the physical half light radius between the CALIFA
galaxies and the targeted half light radius. We additionally smoothed the data by a 2D Gaussian so that the PSF
of the simulated data match with the median seeing of our VIMOS observation ($1.6''$ along right ascension and $1.2''$ along declination). In order
to scale the apparent host galaxy brightness independently from the ionized gas, we modelled the stellar continuum and ionized gas in the CALIFA 
data. Finally, we placed a high S/N QSO spectrum data at the centre 
of the simulated galaxy taken a certain nucleus-to-host ratio into account and added noise to the simulated cube matching with the depth of the 
VIMOS data.

Based on this scheme we ran a suite of simulations to produce QSO host galaxies based on all four CALIFA galaxies with the following parameters.
The apparent $V$ band host galaxy brightness was set to 15.0, 16.0, 16.5, and 17.0\,mag at any redshift. The H$\alpha$ luminosity was set to 0.25, 1
, 4 and 16 times the H$\alpha$ luminosity of the corresponding CALIFA galaxy. We used four different QSOs, namely HE~1019$-$1441, HE~1029$-$1401, 
HE~1239$-$2624, and HE~1338$-$1423, to add a QSO component to the simulations with nucleus-to-host ratios of 0.25, 1, 2, and 4 at the corresponding 
host galaxy brightness. All galaxies are simulated by assuming a half light radius of 5 and 10\,kpc to cover very roughly the physical sizes of the 
QSO host galaxies in our sample. Afterwards we ran the QSO-host debelending process on each simulated galaxy with \QDeb\ and recovered the
$V$ band host galaxy brightness. We further binned the simulated data by 3x3 spaxels to mimic our binning scheme before the stellar continuum 
and emission lines analysis. From that analysis we measured the total H$\alpha$ luminosity and the oxygen abundance at the half light radius 
which are the main observables from which we draw our main conclusions.

\subsection{Systematic uncertainties of the host galaxy brightness}
\begin{figure}
 \resizebox{\hsize}{!}{\includegraphics{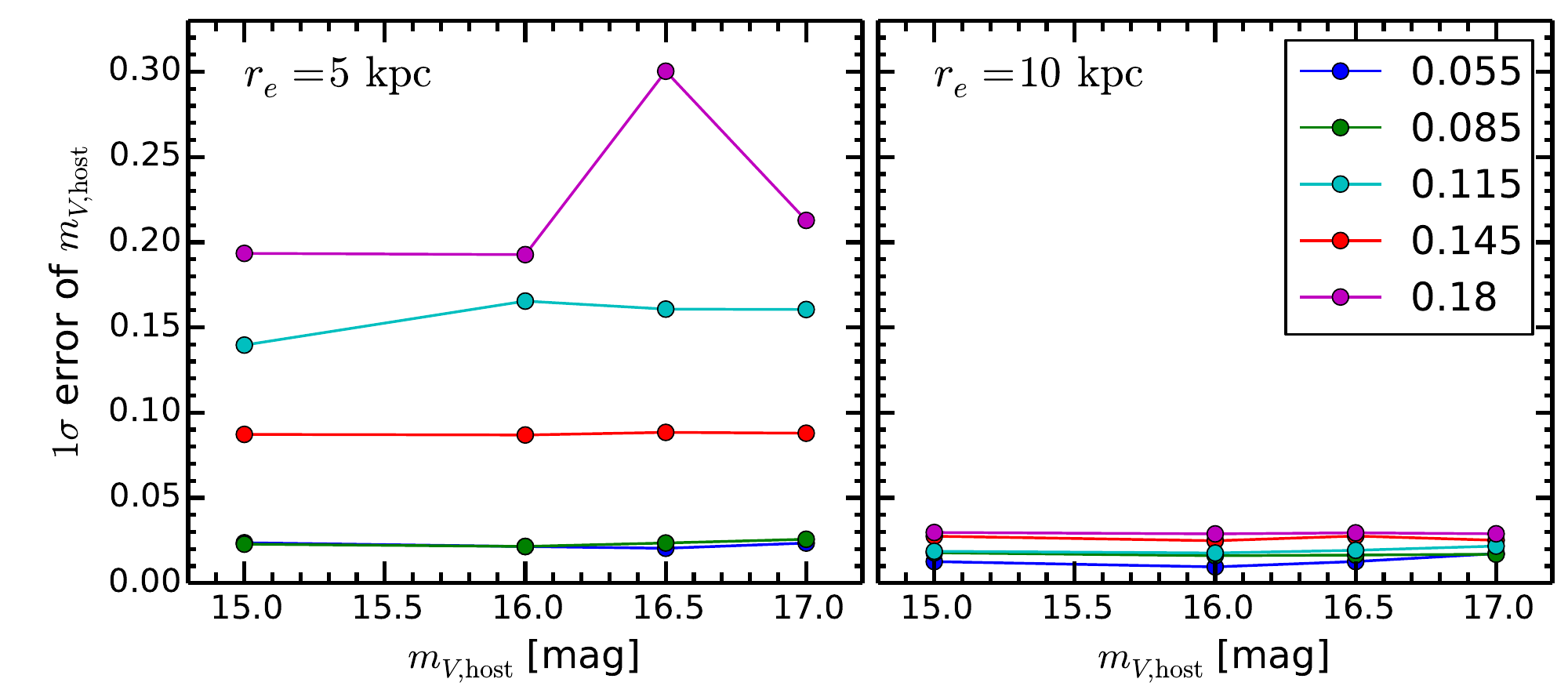}}
 \caption{Recovered $1\sigma$ uncertainty of the host galaxy brightness after the QSO-host deblending process. We present the results for galaxies with $r_e=5\,\mathrm{kpc}$ (left panel) 
 and $r_e=10\,\mathrm{kpc}$ (right panel).  The uncertainty of the recovered host galaxy brightness is plotted against the input host galaxy brightness for five different redshifts, 0.054, 0.085, 
 0.115, 0.145, 0.18, respectively.  We find no significant dependence on the nucleus-to-host ratio and therefore do not shown it in the plots.}
 \label{fig:simulations_br_result}
 \end{figure}

Although the recovery of the host galaxy brightness is not a primary quantity for the presented work, we still want to assess the uncertainty introduced by the QSO-host debelending technique. 
A similar assessment of the uncertainties for the broad-band imaging decomposition \citep{Jahnke:2004b} lead to an average error of 0.1\,mag. Instead of computing the average uncertainty we provide 
the results for the full parameter space covered by our simulation. We show the results in Fig.~\ref{fig:simulations_br_result} for the two different galaxy sizes with effective radii of 
$r_e=5\,\mathrm{kpc}$ (left panel)  and $r_e=10\,\mathrm{kpc}$ (right panel). We ignore the nucleus-to-host ratio within the range of our simulation, since we find no significant trend of the 
uncertainty with the nucleus-to-host ratio. 
 
The primary driver of the uncertainty in the host galaxy brightness is the redshift and the size of the galaxy. It is a natural consequence of the QSO-host deblending of seeing-limited 
observations, because  the QSO contamination covers an apparently larger part of the galaxy when the host galaxy become small or it is apparent size decreases with redshift. For galaxies with 
$r_\mathrm{e}=5$\,kpc at about a redshift of 0.11, we estimate a $V$-band uncertainty of 0.1\,mag in agreement with the results of the broad-band imaging studies. Those kind of galaxies represent the 
majority of our systems, but we also have larger galaxies at lower redshifts in the sample for which the systematic uncertainty in the galaxy brightness will be significantly lower.
 
In Fig.~\ref{fig:compare_phot} we provide the convolved error bars of the measurement and systematic errors as derived from the simulation at the matched galaxy properties. Since 
\citet{Jahnke:2004b} reported only  a fixed systematic uncertainty, we decided to use the uncertainties derived for the IFU data also for the broad-band imaging data to be conservative. We would 
expect that the systematic uncertainties in the IFU  QSO-host debelending scheme are equal or worse than those of the broad-band imaging study.
 
\subsection{Systematic uncertainties of the extinction-corrected SFR}
\begin{figure}
 \resizebox{\hsize}{!}{\includegraphics{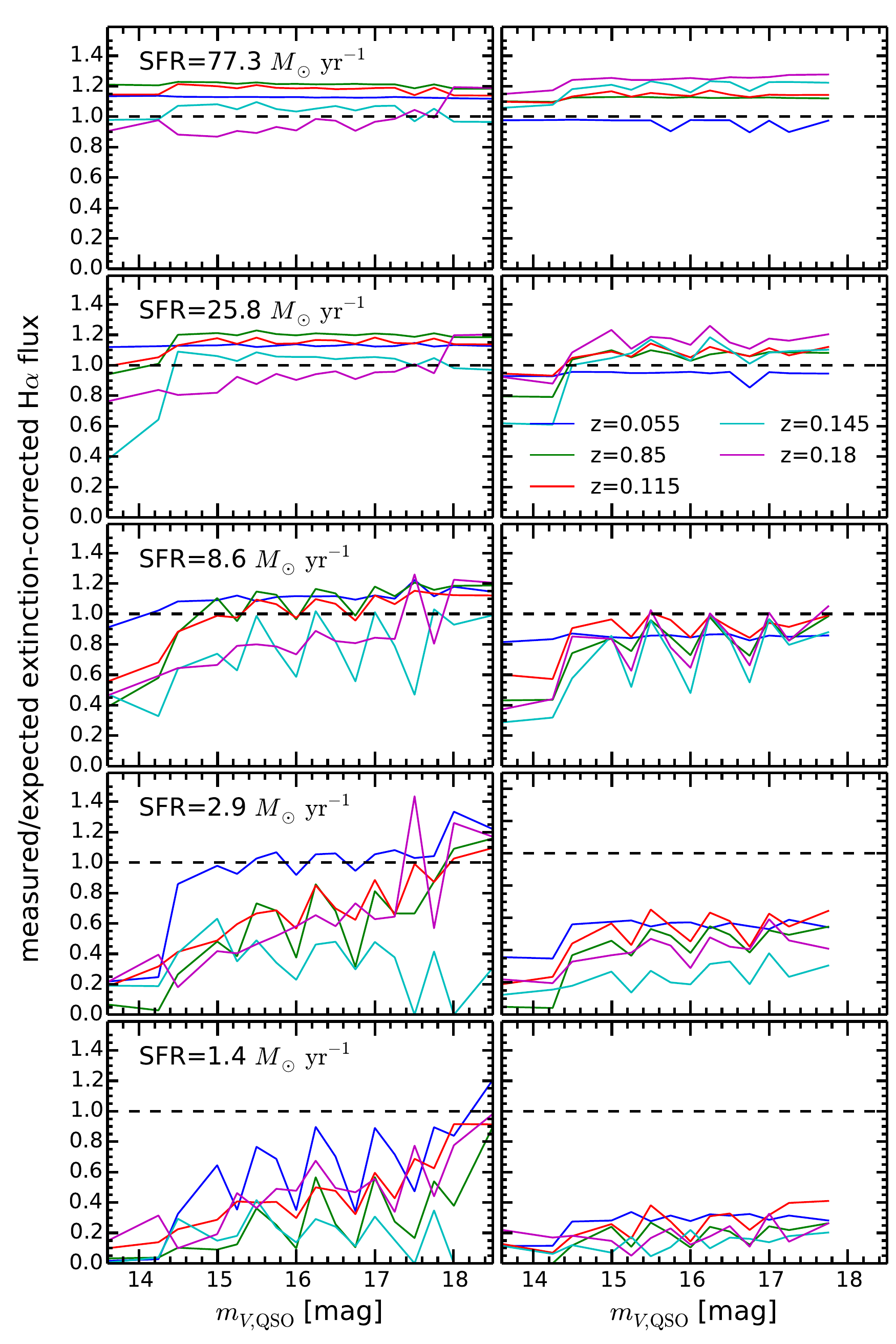}}
 \caption{Result for the recovery of the extinction-corrected H$\alpha$ luminosity from the simulation. The ratio between measured and expected extinction-corrected H$\alpha$ luminosity
 is shown as a function of QSO brightness. Different coloured lines correspond to redshifts of 0.055, 0.085, 0.115, 0.145 and 0.185 as indicated in the legend. The left panels correspond to 
 galaxies with an effective radius of $r_\mathrm{e}=5$\,kpc and the right panels to the ones with $r_\mathrm{e}=10$\,kpc. The dashed black lines indicate the best case of a 100\% recovery of the H$\alpha$
 luminosity.}
 \label{fig:simulations_Ha_result}
 \end{figure}
 
The systematic uncertainties of the extinction-corrected SFR depend on the recovered H$\alpha$ flux and the measurement accuracy of the H$\beta$/H$\alpha$  Balmer decrement to apply the 
extinction-correction. Both measurements may react differently on the QSO-host deblending. We are not trying to disentangle both effects, but quantify the combined systematics effects on the measured 
extinction-corrected H$\alpha$ luminosity.

In Fig.~\ref{fig:simulations_Ha_result} we show the ratio between the measured and expected extinction-corrected H$\alpha$ luminosity as a function of V-band QSO brightness for galaxies 
placed at redshifts 0.055, 0.085,0.115,0.145 and 0.18. We present the results for the two intrinsic galaxy sizes $r_e=5\,\mathrm{kpc}$ (left panels) and $r_e=10\,\mathrm{kpc}$ (right panels) as well 
as for five different input SFRs of 1.3$M_\odot\,\mathrm{yr}^{-1}$, 2.7$M_\odot\,\mathrm{yr}^{-1}$, 8.0$M_\odot\,\mathrm{yr}^{-1}$, 23.9$M_\odot\,\mathrm{yr}^{-1}$, and 
71.6$M_\odot\,\mathrm{yr}^{-1}$.
 
We find that the H$\alpha$ luminosity is severely underestimated for the lowest SFR at high redshifts as expected because part of the H$\alpha$ flux fall below the S/N limit of the observations. There 
is also a trend with QSO brightness in the sense that H$\alpha$ luminosity is better recovered for fainter QSOs, which  disappears with increasing SFR. It is easy to understand that the contrast 
between the QSO emission and the H$\alpha$ flux in the host galaxy determines how well the host galaxy emission can be recovered close to the QSO nucleus. In addition the size of the galaxy matters. 
The QSO will affect a smaller region of host galaxy with increasing size so that total H$\alpha$ flux of the host galaxy  can be better recovered.
 
Based on these simulation we take the systematics into account for each object individually by correcting the measured SFR according to the QSO brightness, galaxy size, redshift and level of 
SFR. Furthermore,  we include the systematic uncertainty of the correction into the error budget. 
\subsection{Systematic uncertainties of the oxygen abundance}
\begin{figure}
 \resizebox{\hsize}{!}{\includegraphics{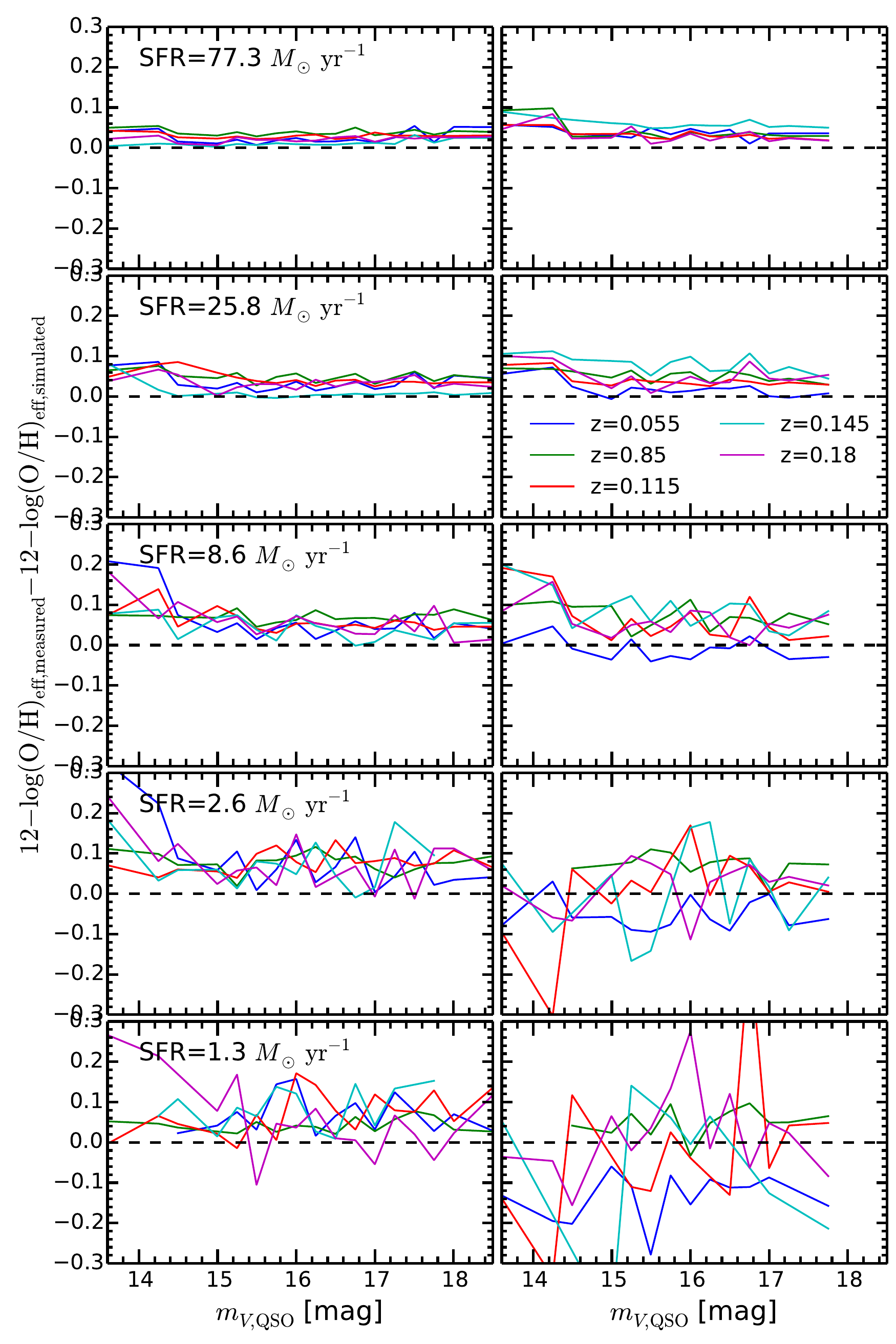}}
 \caption{Result for the recovery of the oxygen abundance $12+\log(\mathrm{O/H})$ from the simulation. The difference between measured and simulated  oxygen abundance is shown as a function 
of QSO brightness. Different coloured lines correspond to redshifts of 0.055, 0.085, 0.115, 0.145 and 0.185 as indicated in the legend. The left panels correspond to 
 galaxies with an effective radius of $r_\mathrm{e}=5$\,kpc and the right panels to the ones with $r_\mathrm{e}=10$\,kpc. The dashed black lines correspond to the case where the measured and simulated
 oxygen abundance matches one-to-one.}
 \label{fig:simulations_OH_eff}
 \end{figure}
 
The systematic effects on the measured oxygen abundance are presented in Fig.~\ref{fig:simulations_OH_eff}. We estimate the oxygen abundance at the effective radius $r_\mathrm{e}$. Depending
on the galaxy size and redshift, it is less affected by the QSO-host deblending process unless the effective radius becomes smaller than the seeing disc. Those resolution effects
were specifically studied by \citet{Yuan:2013} and \citet{Mast:2014} as function of redshift. They found that spatial resolution has a strong impact on the recovered metallicity gradients.

Our simulations show that the oxygen abundance at a fixed radius is more strongly affected with increasing redshift and at low S/N parametrized by the SFR. There is a trend to overestimate 
the oxygen abundance with increasing redshift, which is caused by the luminosity weighting of the beam smearing towards the galaxy centre with higher intrinsic oxygen abundance. S/N effects obviously set
in at very low SFRs at high redshifts. The effect of the QSO-host deblending seems to start playing a more significant role at a V-band QSO brightness of $<$15\,mag.

For galaxies with $z<0.1$ the measured and simulated oxygen abundance agree within the errors for almost all host galaxy parameters covered, which represent the majority of our galaxies. The 
systematic error on the measurements vary between 0.03 and 0.15 with redshift and host galaxy properties and we include those systematic errors into the uncertainties determined individually for each 
object.

\section{Results of the spectral QSO-host deblending}\label{sect:decomp}

\begin{figure*}
\centering
\includegraphics[width=0.95\textwidth]{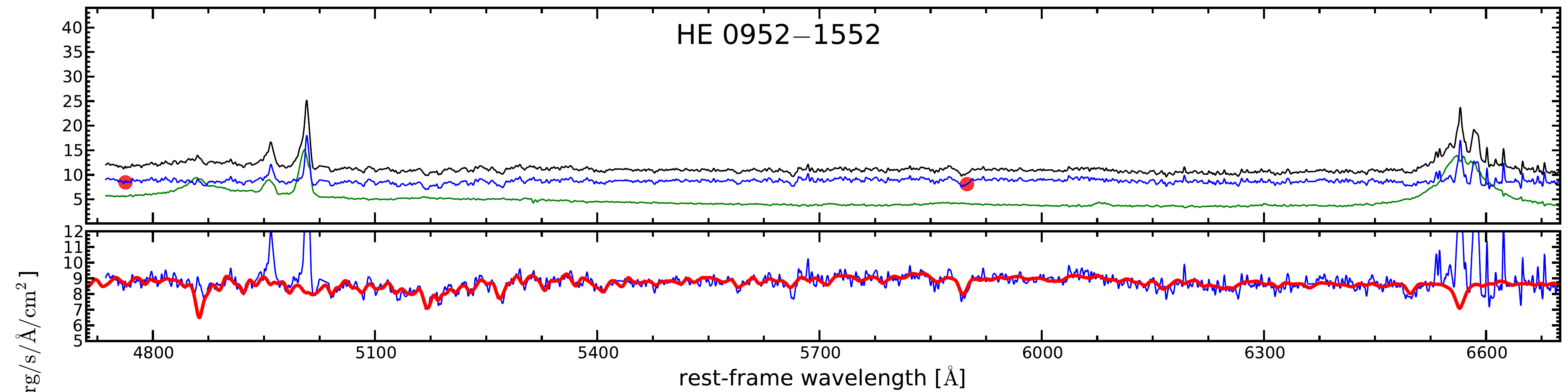}
\includegraphics[width=0.95\textwidth]{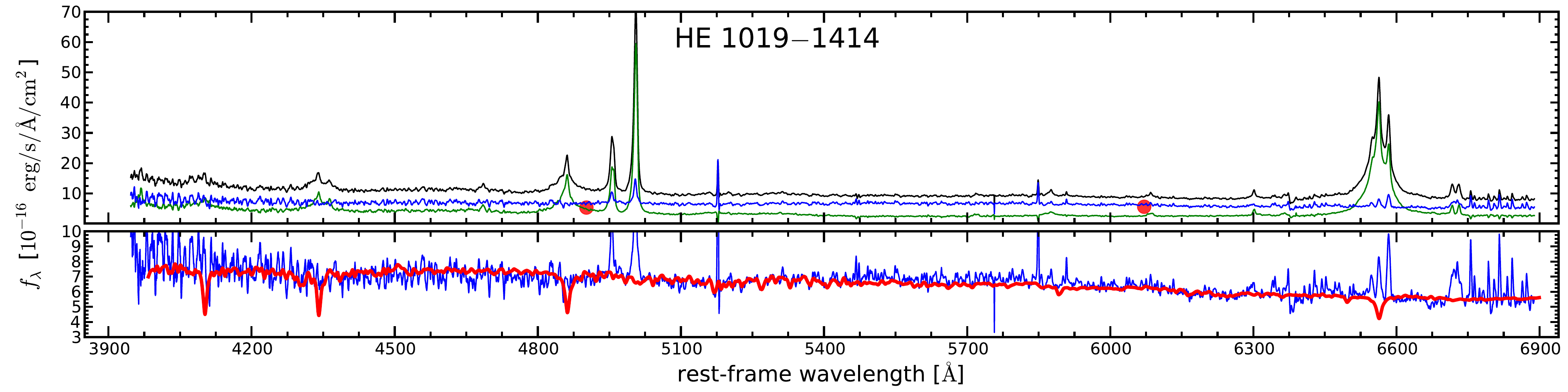}
\includegraphics[width=0.95\textwidth]{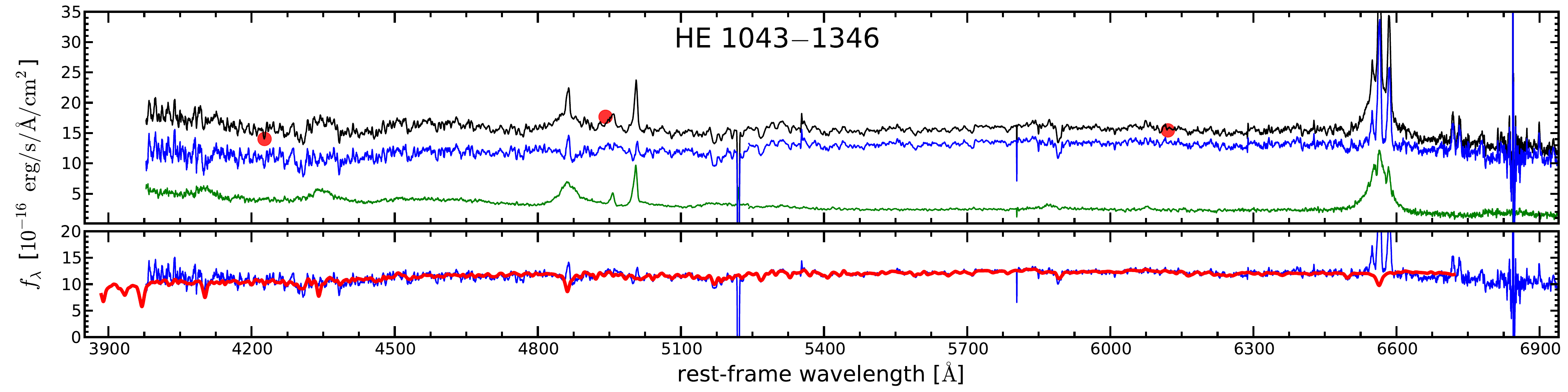}
\includegraphics[width=0.95\textwidth]{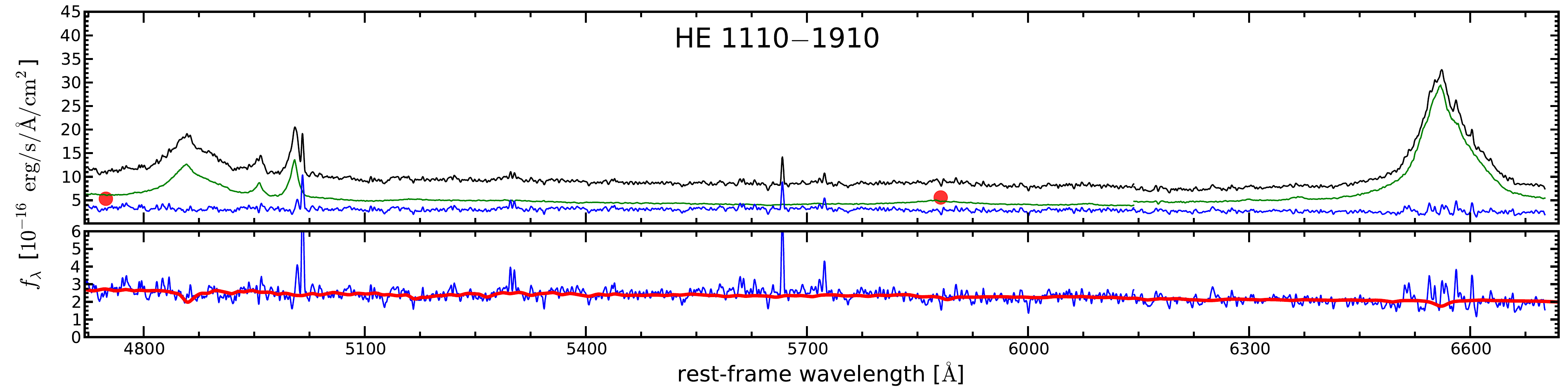}
\includegraphics[width=0.95\textwidth]{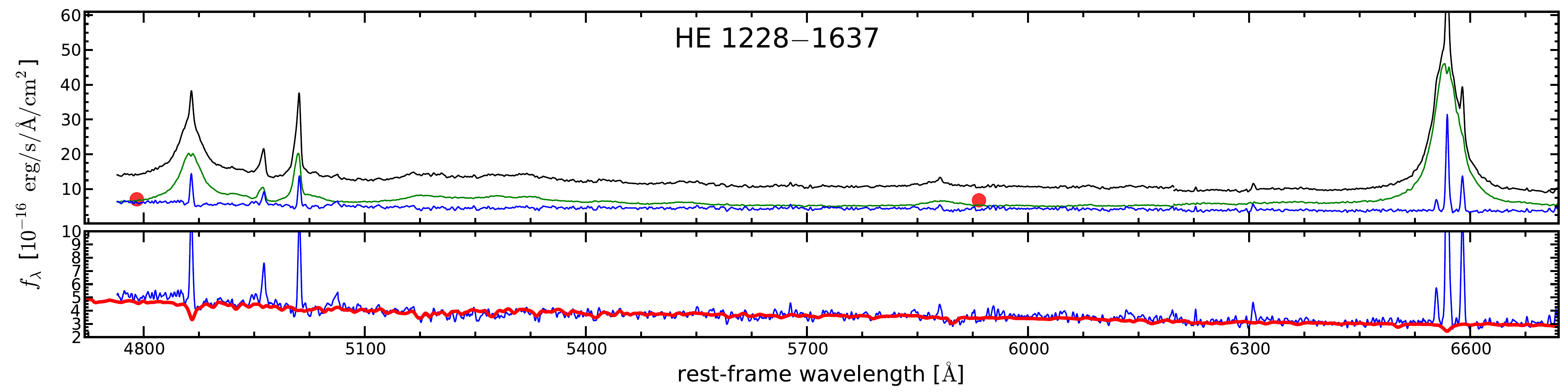}
\caption{Results of the spectral QSO-host deblending. The black line corresponds to the integrated spectrum, the green line to the QSO contribution and the blue line to the host galaxy contribution. The bottom panels show the host galaxy continuum emission with the best-fitting model spectrum (see Section~\ref{sect:vimos_cont} for details) overplotted as the red line.}
\label{fig:decomp_all}
\end{figure*}

\begin{figure*}
\centering
\includegraphics[width=0.95\textwidth]{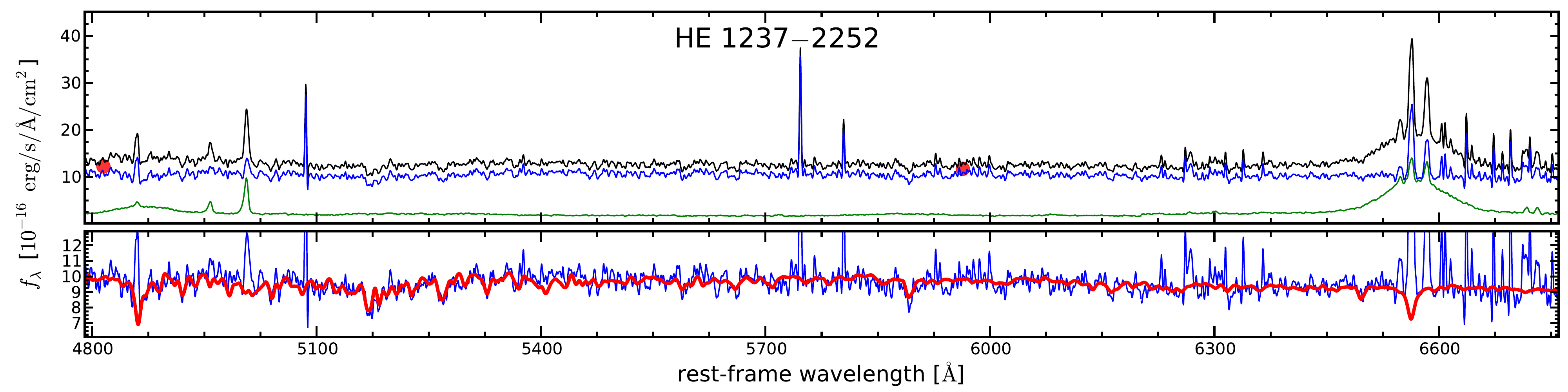}
\includegraphics[width=0.95\textwidth]{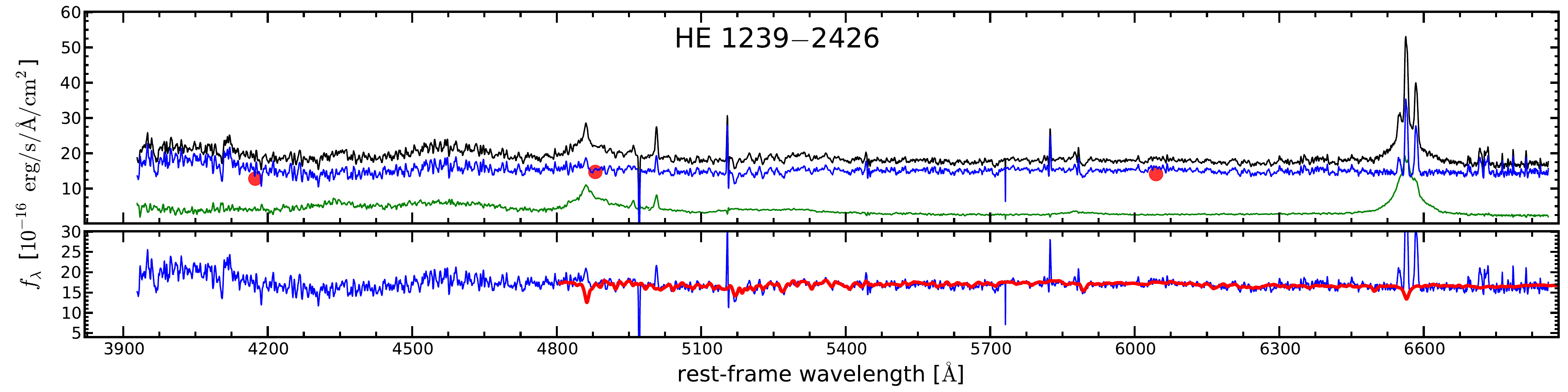}
\includegraphics[width=0.95\textwidth]{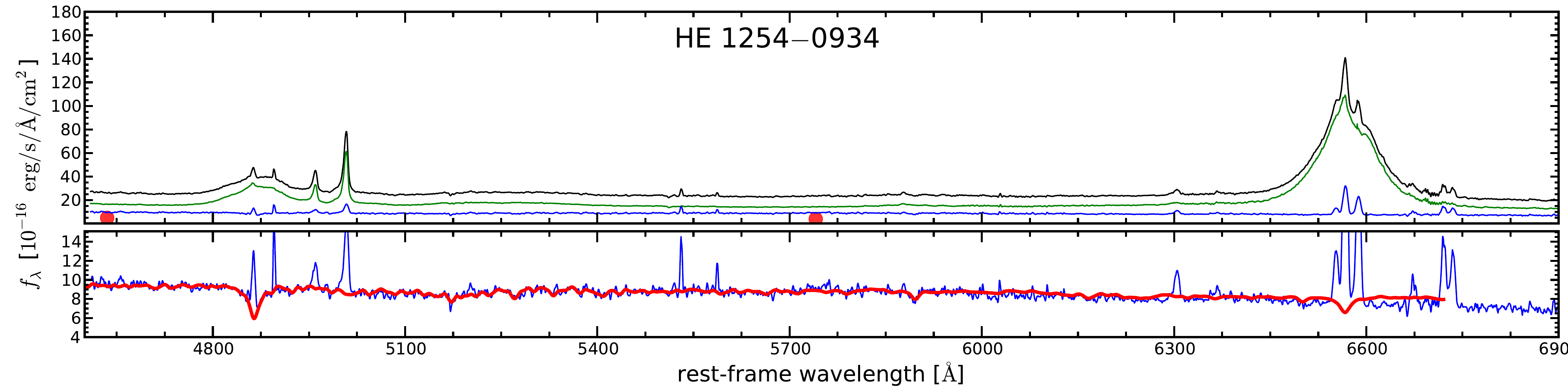}
\includegraphics[width=0.95\textwidth]{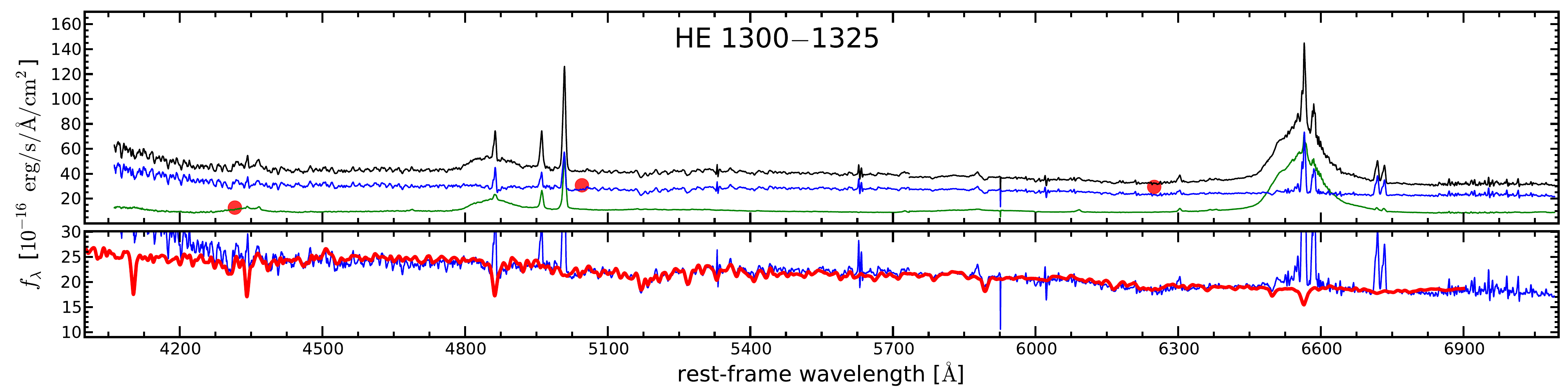}
\includegraphics[width=0.95\textwidth]{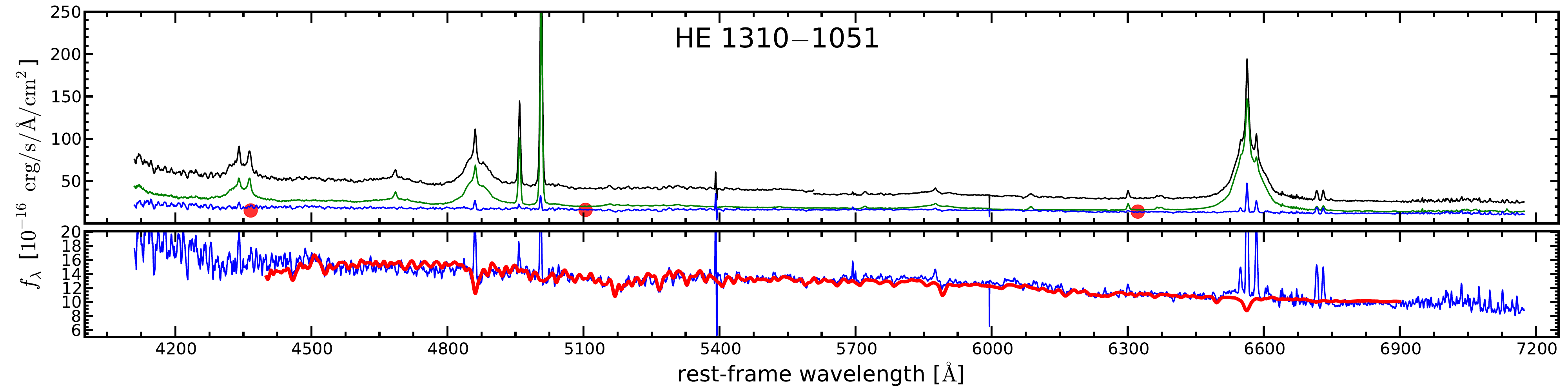}
\contcaption{}
\end{figure*}

\begin{figure*}
\centering
\includegraphics[width=0.95\textwidth]{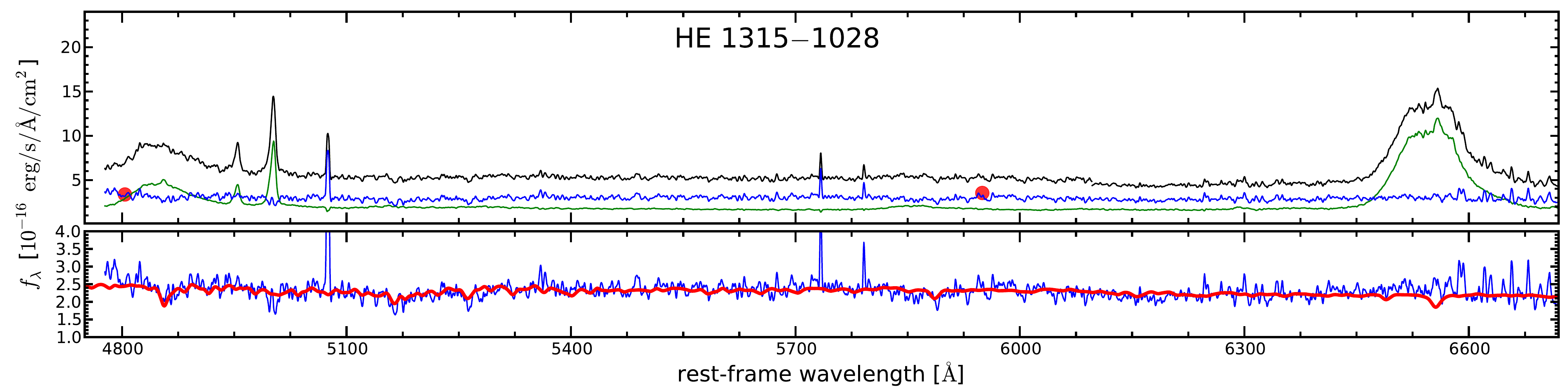}
\includegraphics[width=0.95\textwidth]{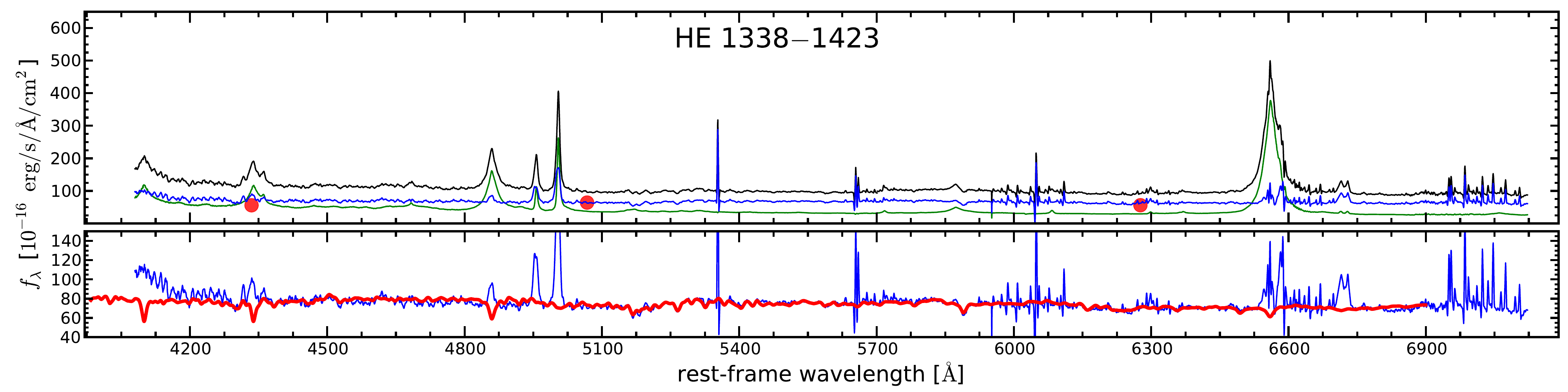}
\includegraphics[width=0.95\textwidth]{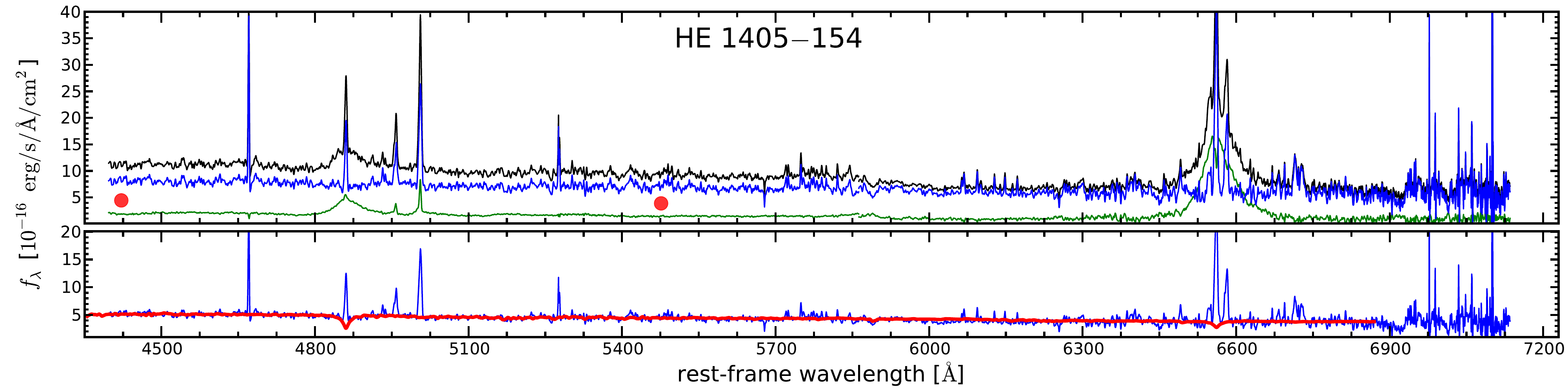}
\includegraphics[width=0.95\textwidth]{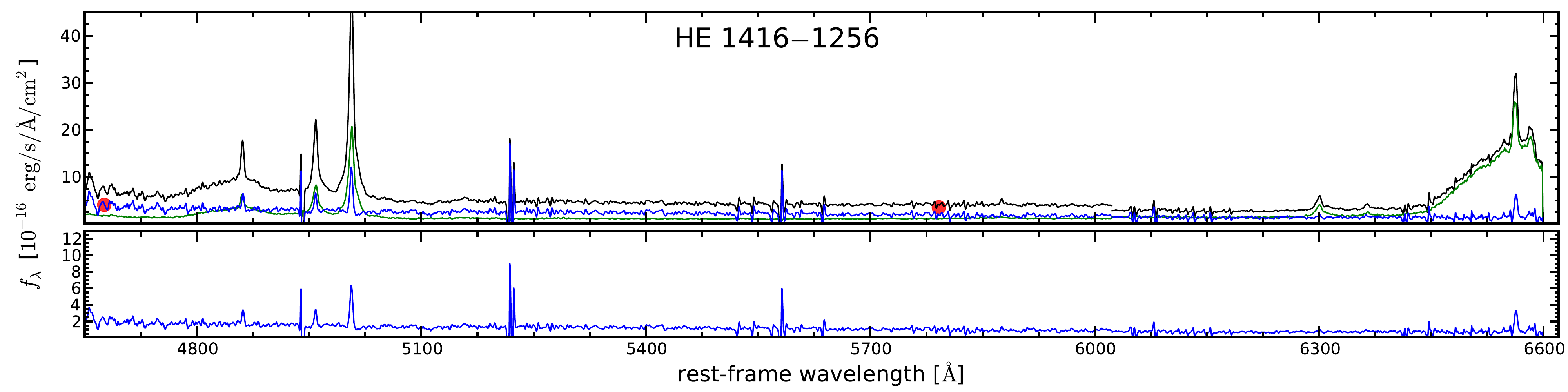}
\contcaption{}
\end{figure*}
\end{document}

%% file: tab1.tex
\begin{tabular}{lcccccccccc}\hline\hline
Object  &  $z$  &  $m_V^{a}$  & Morph.${}^{b}$ & $R_\mathrm{e}^{c}$ & $M_K^{d}$ &  $\log(M_*/\mathrm{M}_{\sun})^{e}$ & $R^{f}$ & $L_{1.4\,\mathrm{GHz}}^{g}$ & $L_{60\,\mu\mathrm{m}}^{h}$ & $L_{100\,\mu\mathrm{m}}^{h}$ \\ 
   &   &  &  & [kpc] &  &  &  &[W/Hz] & [Jy] & [Jy]  \\ \hline
HE  0952$-$1552 & 0.112 & 15.8 & D & 5.1 & -25.5 & $11.18\pm0.28$ & $3.4$ & $23.23$ & $<$0.2 & $<$1.0\\
HE  1019$-$1414 & 0.076 & 16.1 & D & 3.7 & -24.1 & $10.77\pm0.28$ & $2.6$ & $22.86$ & $<$0.2 & $<$1.0\\
HE  1020$-$1022${}^\dagger$ & 0.196 & 16.6 & B & 8.1 & -25.8 & $11.45\pm0.27$ & $779.7$ & $25.95$ & $<$0.2 & $<$1.0\\
HE  1029$-$1401${}^\dagger$ & 0.085 & 13.7 & B & 3.1 & -25.8 & $11.10\pm0.38$ & $0.6$ & $23.43$ & $<$0.2 & $<$1.0\\
HE  1043$-$1346${}^\dagger$ & 0.068 & 15.7 & D & 5.0 & -24.9 & $10.88\pm0.38$ & $<2.5$ & $<22.16$ & 0.36 & 1.51\\
HE  1110$-$1910${}^\dagger$ & 0.111 & 16.0 & B & 4.2 & -24.8 & $10.81\pm0.32$ & $<1.0$ & $<22.60$ & $<$0.2 & $<$1.0\\
HE  1201$-$2409 & 0.140 & 16.3 & B & 1.1 & -25.3 & $10.96\pm0.32$ & $<1.6$ & $<22.81$ & $<$0.2 & $<$1.0\\
HE  1228$-$1637${}^\dagger$ & 0.104 & 15.8 & B & 2.6 & -24.6 & $10.79\pm0.30$ & $<0.7$ & $<22.54$ & $<$0.2 & $<$1.0\\
HE  1237$-$2252 & 0.097 & 15.9 & D & 7.4 & -25.4 & $11.14\pm0.18$ & $<1.6$ & $<22.48$ & $<$0.2 & $<$1.0\\
HE  1239$-$2426${}^\dagger$ & 0.082 & 15.6 & D & 6.8 & -25.2 & $11.15\pm0.19$ & $1.9$ & $22.60$ & 0.37 & 1.12\\
HE  1254$-$0934 & 0.139 & 14.9 & M & 12.2 & -25.2 & $11.19\pm0.24$ & $2.0$ & $23.89$ & 0.92 & 1.09\\
HE  1300$-$1325 & 0.046 & 14.9 & B & 3.6 & -24.8 & $10.75\pm0.39$ & $<0.6$ & $<21.81$ & 0.47 & 1.24\\
HE  1310$-$1051${}^\dagger$ & 0.034 & 14.9 & D & 2.5 & -23.2 & $10.21\pm0.27$ & $<0.3$ & $<21.55$ & $<$0.2 & $<$1.0\\
HE  1315$-$1028 & 0.099 & 16.8 & D & 6.1 & -24.2 & $10.71\pm0.31$ & $<2.0$ & $<22.49$ & $<$0.2 & $<$1.0\\
HE  1335$-$0847 & 0.080 & 16.3 & B & 3.7 & -23.9 & $10.48\pm0.31$ & $<1.9$ & $<22.30$ & $<$0.2 & $<$1.0\\
HE  1338$-$1423 & 0.041 & 13.7 & D & 10.0 & -25.4 & $11.09\pm0.32$ & $0.5$ & $22.29$ & $<$0.2 & $<$1.0\\
HE  1405$-$1545 & 0.196 & 16.2 & M & 6.9 & -25.6 & $11.24\pm0.38$ & $<0.9$ & $<23.12$ & $<$0.2 & $<$1.0\\
HE  1416$-$1256${}^\dagger$ & 0.129 & 16.4 & B & 4.9 & -24.5 & $10.45\pm0.39$ & $5.9$ & $22.95$ & $<$0.2 & $<$1.0\\
HE  1434$-$1600${}^\dagger$ & 0.147 & 15.7 & B & 5.5 & -25.7 & $11.10\pm0.36$ & $417.5$ & $25.71$ & $<$0.2 & $<$1.0\\
\hline\end{tabular}

${}^{a}$Total apparent $V$ band magnitude.
${}^{b}$Morphological classification of the QSO hosts: D\,$-$\,disc-dominated/late-type galaxies, B\,$-$\,bulge-dominated/early-type galaxies, and M\,$-$\,ongoing major mergers.
${}^{c}$Effective (half-light) radius of the QSO hosts as reported by \citet{Jahnke:2004b} converted to our adopted cosmology.
${}^{d}k$-corrected absolute $K$-band magnitudes of the QSO hosts as reported by \citet{Jahnke:2004b}.
${}^{e}$Total stellar masses of the QSO hosts based on multi-color SED fits (Schramm et al. in prep.).
${}^{f}R$ parameter definied as the flux density ratio at 6\,cm over that at 4400\AA.
${}^{g}$Continuum radio luminosity at 1.4\,GHz.
${}^{h}$IRAS 60$\mu$m and 100$\mu$m fluxes and upper limits from the IRAS Faint Source Catalogue v2.0 \citep{Moshir:1990}.
${}^{\dagger}$Objects with available high resolution HST imaging data.

%% file: tab2.tex
\begin{tabular}{lccccccccc}\hline\hline
Object	                &   Date	&  Grating  &  Sampling  & $t_\mathrm{total}^{a}$   & $n_\mathrm{exp}^{b}$ &   Airmass  & Seeing${}^{c}$ & Resolution${}^{d}$ & Remarks${}^{e}$  \\\hline 
HE 0952$-$1552 &  2003--12--21 & HR Orange & 0\farcs67   &       2700        &     6     &     1.1    &  1\farcs0/1\farcs5 &  2.0/3.0\,kpc  & \\
HE 1019$-$1414 &  2009--05--18 & HR Blue   & 0\farcs67   &       2000        &     4     &     1.0    &  0\farcs9/1\farcs5 &  1.2/2.1\,kpc &  \\
			&  2009--05--18 & HR Orange & 0\farcs67   &       3000        &     4     &     1.1    &  1\farcs1/1\farcs6 &  1.5/2.2\,kpc &   \\
HE 1020$-$1022 &  2009--05--19 & HR Orange & 0\farcs33   &       2000        &     4     &     1.1    &  0\farcs9/1\farcs4 & 2.9/4.5\,kpc  & (i)  \\
                        &  2009--05--13 & HR Red    & 0\farcs33   &       3000        &     4     &     1.2    &  1\farcs2/1\farcs6 & 3.8/5.0\,kpc & (i)  \\
HE 1029$-$1401 &  2003--12--17 & HR Blue   & 0\farcs67   &       900         &     3     &     1.1    &  1\farcs0/1\farcs5 &   1.6/2.4\,kpc & \\
                        &  2003--12--23 & HR Orange & 0\farcs67   &       2700        &     6     &     1.2    &  0\farcs9/1\farcs5 & 1.4/2.4\,kpc &   \\
HE 1043$-$1346$^\dagger$ &  2003--12--22 & HR Orange & 0\farcs67   &       2700        &     6     &     1.2    &  1\farcs2/1\farcs6 &  1.6/2.1\,kpc &  \\
                        &  2009--05--19 & HR Blue   & 0\farcs67   &       2000        &     4     &     1.2    &  1\farcs1/1\farcs6 & 1.4/2.1\,kpc &   \\
HE 1110$-$1910 &  2003--12--20 & HR Orange & 0\farcs67   &       2700        &     6     &     1.1    &  1\farcs0/1\farcs5 &  2.0/3.0\,kpc &  \\
HE 1201$-$2408 &  2009--04--15 & HR Orange & 0\farcs33   &       1500        &     2     &     1.1    &    ...    &  ... &  (ii)      \\
                        &  2009--04--15 & HR Red    & 0\farcs33   &       1000        &     2     &     1.0    &    ...    &  ... &  (ii)      \\
HE 1228$-$1637 &  2003--12--30 & HR Orange & 0\farcs67   &       3150        &     7     &  1.2--1.6  &  1\farcs0/1\farcs6 & 1.9/3.0\,kpc & \\
HE 1237$-$2252 &  2003--12--31 & HR Orange & 0\farcs67   &       2700        &     6     &  1.6--2.3  &  1\farcs3/1\farcs8 & 2.3/2.9\,kpc & \\
HE 1239$-$2426 &  2003--12--28 & HR Blue   & 0\farcs67   &       900         &     3     &     1.5    &  1\farcs3/1\farcs6 & 1.9/2.4\,kpc & \\
			&  2003--12--31 & HR Orange & 0\farcs67   &       2700        &     6     &  1.2--1.5  &  1\farcs2/1\farcs4 & 2.8/2.1\,kpc &  \\
HE 1254$-$0934 &  2004--01--01 & HR Orange & 0\farcs67   &       2700        &     6     &  1.4--1.8  &  1\farcs2/1\farcs5 & 2.7/3.6\,kpc & \\
			&  2009--04--18 & HR Red    & 0\farcs67   &       2000        &     4     &     1.2    &  1\farcs2/1\farcs8 & 2.7/4.3\,kpc & \\
HE 1300$-$1325 &  2009--04--22 & HR Blue   & 0\farcs67   &       2000        &     4     &  1.2--1.5  &  1\farcs2/1\farcs9 & 1.1/1.7\,kpc & \\
			&  2009--04--25 & HR Orange & 0\farcs67   &       3000        &     4     &     1.2    &  0\farcs9/1\farcs7 & 0.8/1.5\,kpc & \\
HE 1310$-$1051 &  2009--04--22 & HR Blue   & 0\farcs67   &       2000        &     4     &     1.2    &  1\farcs3/1\farcs8 &  0.9/1.3\,kpc &\\
			&  2009--04--25 & HR Orange & 0\farcs67   &       3000        &     4     &  1.3--1.5  &  0\farcs9/1\farcs4 & 0.6/1.0\,kpc & \\
HE 1315$-$1028 &  2004--01--17 & HR Orange & 0\farcs67   &       3000        &     4     &  1.3--1.7  &  1\farcs3/1\farcs8 & 2.3/3.2\,kpc & \\
HE 1335$-$0847 &  2009--04--24 & HR Blue   & 0\farcs33   &       2000        &     4     &  1.3--1.6  & 0\farcs8/1\farcs6     &  1.2/2.4\,kpc &        \\
HE 1338$-$1423 &  2009--04--27 & HR Blue   & 0\farcs67   &       2000        &     4     &  	1.1    &  1\farcs1/1\farcs7 & 0.9/1.4\,kpc & \\
			&  2009--04--27 & HR Orange & 0\farcs67   &       3000        &     4     &  1.2--1.5  &  1\farcs/1\farcs7 & 0.9/1.4\,kpc & \\
HE 1405$-$1545 &  2004--01--22 & HR Orange & 0\farcs67   &       2700        &     6     &  1.2--1.5  &  2\farcs2/2\farcs6 & 7.0/8.3\,kpc & \\
			&  2004--01--27 & HR Red    & 0\farcs67   &        900        &     3     &     1.2    &  1\farcs2/1\farcs6 & 3.8/5.1\,kpc & \\
HE 1416$-$1256 		&  2009--04--18 & HR Orange & 0\farcs33   &       3000        &     4     &     1.1    &  0\farcs9/1\farcs4 & 2.1/3.2\,kpc & \\
HE 1434$-$1600 &  2009--04--17 & HR Orange & 0\farcs33   &       1500        &     2     &  1.2--1.5  &     ...   &  ... &(ii)       \\
			&  2009--04--18 & HR Orange & 0\farcs33   &       1000        &     2     &  	1.2    &     ...   & ...  &(ii)     \\\hline
\end{tabular}

${}^{a}$Total integration time of all exposures in seconds. 
${}^{b}$Number of exposures taken for a given object and instrumental setup. 
${}^{c}$Estimated seeing of the combined cubes for the minor and major axis of the asymmetric VIMOS point spread function at the wavelength of the broad Balmer lines. 
${}^{d}$Physical spatial resolution at the redshift of the object according to the seeing.
${}^{e}$Objects marked with (i) are rejected from the analysis because of exceptionally bad spectrophotometry as explained in the text. Objects marked with (ii) were positioned at the edge of the VIMOS FoV so that an absolute photometric calibration could not be performed.
$^\dagger$The HR Blue observation of HE 1043$-$1346 suffers from a bad spectrograph focus in the quadrant covering the QSO.

%% file: tab3.tex
\begin{tabular}{lccc}\hline\hline\noalign{\smallskip}
Object  &  $\log L_{5100}^{a}$  &  $\log L_{\Ox}^{b}$ &  $r_\mathrm{e}^{c}$   \\ 
   &  [\,erg/s]  &  [\,erg/s] &  [kpc]   \\ \hline
HE 0952$-$1552 & $43.98\pm0.09$ & $41.88\pm0.10$ & $3.4\pm0.1$ \\
HE 1019$-$1414 & $43.40\pm0.11$ & $41.94\pm0.11$ & $1.9\pm0.1$ \\
HE 1029$-$1401 & $44.94\pm0.09$ & $42.39\pm0.11$ & $4.5\pm0.1$ \\
HE 1110$-$1910 & $44.05\pm0.10$ & $41.66\pm0.11$ & $3.2\pm1.1$ \\
HE 1228$-$1637 & $44.08\pm0.10$ & $41.67\pm0.14$ & $<$1.8 \\
HE 1237$-$2252 & $43.44\pm0.11$ & $41.22\pm0.13$ & $2.4\pm0.3$ \\
HE 1239$-$2426 & $43.81\pm0.10$ & $41.20\pm0.11$ & $1.9\pm0.1$ \\
HE 1254$-$0934 & $44.68\pm0.11$ & $42.53\pm0.12$ & $3.8\pm0.1$ \\
HE 1300$-$1325 & $43.46\pm0.09$ & $41.23\pm0.09$ & $<$1.4 \\
HE 1310$-$1051 & $43.55\pm0.10$ & $41.66\pm0.14$ & $<$1.9 \\
HE 1338$-$1423 & $43.73\pm0.10$ & $42.00\pm0.13$ & $2.0\pm0.0$ \\
HE 1405$-$1545 & $44.22\pm0.11$ & $42.03\pm0.12$ & $4.6\pm0.1$ \\
HE 1416$-$1256 & $44.13\pm0.12$ & $42.85\pm0.11$ & $5.6\pm0.5$ \\
\hline\end{tabular}

${}^{a}$QSO continuum luminosity at $5100\AA$ after removing the contribution from the host galaxy.
${}^{b}$Integrated \Ox\ luminosity from the unresolved QSO and extended ENLR.
${}^{c}$\Ox\ luminosity-weighted effective ENLR radius excluding the \HII-like regions. Upper limits are derived for those objects which do not show an ENLR and represent the shortest distance to an \HII-like region.

%% file: tab4.tex
\begin{tabular}{lcccccccc}\hline\hline\noalign{\smallskip}
Object${}^{a}$  &  $f$(\Ha)${}^{c}$ &  $f_\mathrm{cor}$(\Ha)${}^{d}$ &  $A_\mathrm{V}^{e}$ & $L_\mathrm{cor}$(\Ha)${}^{f}$ & SFR$_{\mathrm{H}\alpha}^{g}$ & SFR$_{\mathrm{FIR}}^{h}$ & SFR$_{\mathrm{1.4GHz}}^{i}$ & $\log \mathrm{sSFR}^{j}$ \\ 
   &  \multicolumn{2}{c}{$\left[10^{-16}\,\mathrm{erg}\,\mathrm{cm}^{-2}\,\mathrm{s}^{-1}\right]$} & $\left[\mathrm{mag}\right]$ & $\left[\mathrm{erg}\,\mathrm{s}^{-1}\right]$ &  \multicolumn{3}{c}{$\left[\mathrm{M}_{\sun}\,\mathrm{yr}^{-1}\right]$} & $\left[\mathrm{yr}^{-1}\right]$  \\ \noalign{\smallskip}\hline\noalign{\smallskip}
HE 0952$-$1552 & $16.2\pm1.3$ & $57\pm11$ & $1.7\pm0.1$ & $41.23\pm0.09$ & $2.3\pm1.0$ & $<$45.1 & 93.8 & $-10.8\pm0.3$\\
HE 1019$-$1414${}^{b}$ & $<23.7$ & $<58$ & ... & $<40.93$ & $<0.7$ & $<$20.5 & 39.9 & $<-10.9$\\
HE 1029$-$1401${}^{b}$ & $<74.9$ & $<118$ & ... & $<41.33$ & $<1.7$ & $<$25.8 & 149.2 & $<-10.9$\\
HE 1043$-$1346 & $144.7\pm16.1$ & $987\pm472$ & $1.2\pm0.1$ & $42.04\pm0.27$ & $7.3\pm3.4$ & 25.7 & $<$7.9 & $-10.0\pm0.4$\\
HE 1110$-$1910 & $10.0\pm1.8$ & $16\pm3$ & $1.1\pm0.8$ & $40.70\pm0.08$ & $0.8\pm0.4$ & $<$44.8 & $<$22.1 & $-10.9\pm0.5$\\
HE 1228$-$1637 & $176.2\pm14.3$ & $377\pm114$ & $0.8\pm0.2$ & $42.01\pm0.15$ & $6.7\pm2.5$ & $<$38.9 & $<$19.2 & $-10.0\pm0.3$\\
HE 1237$-$2252 & $58.1\pm3.0$ & $230\pm158$ & $0.8\pm0.1$ & $41.76\pm0.35$ & $4.5\pm2.7$ & 45.6 & $<$16.6 & $-10.5\pm0.2$\\
HE 1239$-$2426 & $196.9\pm7.9$ & $920\pm144$ & $1.5\pm0.1$ & $42.18\pm0.07$ & $11.4\pm4.3$ & 32.7 & 22.0 & $-10.1\pm0.2$\\
HE 1254$-$0934 & $79.4\pm8.6$ & $1698\pm227$ & $4.1\pm0.1$ & $42.94\pm0.06$ & $55.9\pm8.7$ & 161.4 & 432.6 & $-9.4\pm0.2$\\
HE 1300$-$1325 & $354.8\pm17.1$ & $1352\pm110$ & $1.4\pm0.1$ & $41.84\pm0.04$ & $4.7\pm0.5$ & 8.9 & $<$3.6 & $-10.1\pm0.4$\\
HE 1315$-$1028${}^{b}$ & $<2.8$ & $<7$ & ... & $<40.21$ & $<0.1$ & $<$34.8 & $<$17.1 & $<-11.6$\\
HE 1310$-$1051 & $134.9\pm6.7$ & $594\pm73$ & $1.7\pm0.1$ & $41.21\pm0.05$ & $1.5\pm0.5$ & $<$3.2 & $<$1.9 & $-10.0\pm0.3$\\
HE 1335$-$0847 & $120.2\pm39.8$ & $189\pm60$ & ... & $41.48\pm0.16$ & $7.0\pm3.1$ & $<$22.7 & $<$11.1 & $-9.6\pm0.3$\\
HE 1338$-$1423${}^{b}$ & $<229.8$ & $<567$ & ... & $<41.34$ & $<1.7$ & $<$4.6 & 10.7 & $<-10.8$\\
HE 1405$-$1545 & $153.5\pm7.9$ & $444\pm39$ & $1.3\pm0.1$ & $42.68\pm0.04$ & $37.7\pm8.6$ & $<$149.5 & $<$73.1 & $-9.7\pm0.4$\\
HE 1416$-$1256 & $61.9\pm11.5$ & $62\pm12$ & $<0.5$ & $41.44\pm0.09$ & $3.6\pm1.8$ & $<$61.1 & 49.5 & $-9.9\pm0.4$\\
\noalign{\smallskip}\hline\end{tabular}

${}^{a}$Companion galaxies are denoted with a (c) behind the name of the corresponding QSO.
${}^{b}$Galaxies for which the extended emission is dominated by the ENLR. Integrating the \Ha\ flux of the ENLR provides a firm upper limit on the possible contribution of \HII-like regions. We use the mean attentuation for each morphological type to roughly take dust extinction effects into account for the upper limit.
${}^{c}$Integrated \Ha\ flux of all \HII-like and intermediate regions of a galaxy.
${}^{d}$Attenuation-corrected \Ha\ flux.
${}^{e}$Mean attenuation of all specific regions of a galaxy taking into account the individual uncertainties as weights.
${}^{f}$Attenuation-corrected \Ha\ luminosity.
${}^{g}$SFR estimated from the attenuation-corrected \Ha\ luminosity following \citet{Kennicutt:1998}. Systematic offsets and uncertainties based on the results from extinsive simulations (Sect.~\ref{sect:simulations}) are already taken into account in the reported SFRs.
${}^{j}$\Ha-based specfic SFR are computed with the SED-based stellar masses (see Table~\ref{tbl:sample}).

%% file: tab5.tex
\begin{tabular}{lcc}\hline\hline
Object  &  $12+\log(\mathrm{O}/\mathrm{H})_\mathrm{T04}^{a}$ & Indicator${}^{b}$. \\ \hline
HE 0952$-$1552 & $(8.75/8.91)\pm 0.03 (\pm 0.07)$ & NLR \\
HE 1019$-$1414 & $(8.95/9.10)\pm 0.19 (\pm 0.08)$ & NLR \\
HE 1029$-$1401 & $(8.81/8.96)\pm 0.04 (\pm 0.05)$ & NLR \\
HE 1043$-$1346 & $(9.08/9.23)\pm 0.06 (\pm 0.04)$ & O3N2 \\
HE 1228$-$1637 & $(8.83/8.98)\pm 0.10 (\pm 0.04)$ & O3N2 \\
HE 1237$-$2252 & $(8.98/9.13)\pm 0.08 (\pm 0.05)$ & O3N2 \\
HE 1239$-$2426 & $(8.96/9.11)\pm 0.03 (\pm 0.05)$ & O3N2 \\
HE 1300$-$1325 & $(8.75/8.90)\pm 0.04 (\pm 0.02)$ & O3N2 \\
HE 1310$-$1051 & $(8.73/8.88)\pm 0.05 (\pm 0.04)$ & O3N2 \\
HE 1405$-$1545 & $(8.73/8.88)\pm 0.04 (\pm 0.05)$ & O3N2 \\
HE 1416$-$1256 & $(8.59/8.74)\pm 0.05 (\pm 0.13)$ & O3N2 \\
HE 1434$-$1600 & $(8.73/8.88)\pm 0.04 (\pm 0.08)$ & NLR \\
\hline\end{tabular}

${}^{a}$Two metallicities are given. The first one is measured at 1$R_e$ of the host galaxy from the data and the second is an extrapolated one towards the galaxy center assuming a slope of radius-metallicity as reported by \citet{Sanchez:2012b}. All metallicities and the radius-metallicity slope are converted to the \citet{Tremonti:2004} oxygen abundance scale using the conversion of \citet{Kewley:2008}. The measurement errors and the systematic errors are provided. The systematic errors are based on the extinsive simulation of the QSO-host deblending process presented in Sect.~\ref{sect:simulations}.
${}^{b}$Metallicities are either inferred using the O3N2 index of \citet{Pettini:2004} for \HII\ regions or using the calibration of \citet{Storchi-Bergmann:1998} for the ENLR.